\documentclass[10pt]{article}\usepackage[margin=1.4in]{geometry}

\usepackage{xspace} 
\def\url{url~}

\usepackage{amsmath,amssymb,listings}

\usepackage{caption,float}

\usepackage{epsfig,graphicx,epstopdf,eurosym} 
\usepackage{tikz}
\usetikzlibrary{positioning, shapes, fit,arrows}
\usepackage{multirow,chicago,cases}
\usepackage{eurosym} 

\usepackage{pgf} 
\usepackage{array,color,colortbl}
\usepackage{mathrsfs} 
\definecolor{midnightblue}{rgb}{0.10,0.10,0.44}
\definecolor{rltgreen}{rgb}{0,0.4,0}
\definecolor{rltred}{rgb}{0.75,0,0}
\definecolor{lightgreen}{rgb}{0,0.6,0}
\definecolor{LemonChiffon}{rgb}{1.,0.98,0.8}
\colorlet{mauve}{blue!70!red}

\usepackage{dsfont}
\definecolor{WildStrawberry}{rgb}{1.00,0.00,0.50}
\definecolor{ForestGreen}{rgb}{0.00,1.00,0.00}
\definecolor{Blue}{rgb}{0.00,0.00,1.00}
\definecolor{WildStrawberry}{rgb}{0.5,0.00,0.25}
\definecolor{ForestGreen}{rgb}{0.00,50,0.50}
\definecolor{Blue}{rgb}{0.00,0.00,1.00}

\def\r#1{\sref{#1}}

\def\hat{\widehat}
\def\tilde{\widetilde}

\def\ben{\begin{enumerate}}
\def\een{\end{enumerate}}
\def\bit{\begin{itemize}}
\def\eit{\end{itemize}}
\def\bqo{\begin{quotation}}
\def\eqo{\end{quotation}}
\def\bdi{\begin{description}}
\def\edi{\end{description}} 
\def\bb{\begin{block}}
\def\eb{\end{block}} 

\def\bal{\begin{aligned}}
\def\eal{\end{aligned}}

\newtheorem{lem}{Lemma}[section]
\newtheorem{pro}{Proposition}[section]
\newtheorem{cor}{Corollary}[section]
\newtheorem{conj}{Conjecture}[section]

\newtheorem{rem}{Remark}[section]
\newtheorem{com}{Comments}[section]
\newtheorem{ex}{Example}[section]
\newtheorem{nota}{Notation}[section]
\newtheorem{defi}{Definition}[section]
\newtheorem{hyp}{Assumption}[section]

\newcommand{\bt}{\begin{pro}}\newcommand{\et}{\end{pro}}
\newcommand{\bl}{\begin{lem}}\newcommand{\el}{\end{lem}}
\newcommand{\bp}{\begin{pro}}\newcommand{\ep}{\end{pro}}
\newcommand{\bcor}{\begin{cor}}\newcommand{\ecor}{\end{cor}}
\newcommand{\bconj}{\begin{conj}}\newcommand{\econj}{\end{conj}}

\newcommand{\bd}{\begin{defi} \rm }\newcommand{\ed}{\finproof\end{defi} }
\newcommand{\eds}{\end{defi} }
\newcommand{\brem }{\begin{rem} \rm }\newcommand{\erem }{\finproof\end{rem}}
\newcommand{\bcom}{\begin{com} \rm }\newcommand{\ecom }{\end{com}}
\newcommand{\brems }{\begin{rem} \rm }\newcommand{\erems }{\end{rem}}
\newcommand{\bex}{\begin{ex} \rm }\newcommand{\eex}{\finproof\end{ex}}
\newcommand{\bno}{\begin{nota} \rm }\newcommand{\eno}{\finproof\end{nota}}\newcommand{\enos}{\end{nota}}
\newcommand{\eexs}{\end{ex} }
\newcommand{\bhyp}{\begin{hyp} \rm }\newcommand{\ehyp}{\end{hyp}}

\newcommand{\bP}{\begin{proof}}
\newcommand{\eP}{\end{proof}}
\def\proof{\noindent {\it Proof. $\, $}}\def\proof{\noindent{\it\textbf{Proof}. $\, $}}
\def\finproof {\hfill $\Box$ \vskip 5 pt }

\def\bdm{\begin{displaymath}}
\def\edm{\end{displaymath}}
\def\be{\begin{enumerate}}
\def\ee{\end{enumerate}}
\newcommand{\beqa}{\begin{eqnarray}}
\newcommand{\eeqa}{\end{eqnarray}}
\newcommand{\beqe}{\beqa\begin{aligned}}
\newcommand{\eeqe}{\end{aligned}\eeqa}

\def\bea{\begin{eqnarray*}}
\def\eea{\end{eqnarray*}}
\def\bi{\vskip -7 pt\begin{itemize}\vskip -7 pt}
\def\ei{\vskip -7 pt\end{itemize}\vskip -7 pt}

\newcommand{\bde}{\begin{displaymath}}
\newcommand{\ede}{\end{displaymath}}
\newcommand{\bel }{\left\{\begin{array}{ll}}
\newcommand{\eel}{\cr \end{array} \right.}

\def\t{\tilde}

\def\to{\rightarrow}

\def\R{\mathbb{R}}

\def\cL{\mathcal{L}}

\def\cF{{\cal F}}

\def\cP{{\cal P}}

\def\cH{{\cal H}}
\def\cC{{\cal C}}

\def\cL{{\cal L}}

\def\b{\beta}

\def\ca{{\check{a}}}

\def\ff{{\tk\in{\cal F}}}

\def\I{\mathds{1}}
\def\sp{\,,\;}
\def\cdots{\ldots}

\def\eee{\end{document}}
\def\eef{\end{frame}\end{document}}

\renewcommand\be{\begin{equation}}
\renewcommand{\ee}{\end{equation}} 

\def\r#1{(\ref{#1})}

\def\Xi{Y}

\def\cL{L}

\def\cP{{\cal P}}

\def\cP{{\cal P}}

\def\cY{{\cal Y}}
\def\hat{\widehat}
\def\tilde{\widetilde}

\def\t0{0}
\def\0{0}

\def\ff{{\mathbb F}}

\def\R{{\mathbb R}}

\def\I{\mathds{1}}

\def\td{\theta}

\def\tb{\bar{\sigma}}

\def\HH1{{\cal H}^{2}(\ff)}

\def\L2{ L^2(0,T)}

\def\b{\,^\beta\! }

\def\M2{M^2}

\def\sp{,\;}

\def\sp{\;,\;}

\def\cY{{\cal Y}}

\def\bal{\begin{aligned}}
\def\eal{\end{aligned}}

\def\z{z}

\def\cL{{\cal Z}}

\def\be{\begin{equation}}
\def\ee{\end{equation}}

\def\ff{{\mathbb F}}

\def\I{\mathds{1}}
\def\R{\mathbb{R}}

\def\td{\theta}

\usepackage{float}
\usepackage{epsfig}
\usepackage{graphicx}
\usepackage{amsfonts}

\definecolor{GrapeFruit}{rgb}{0.92,0.33,0.08}
\definecolor{Wine}{rgb}{0.5,0,0.25}

\newcommand{\A}{A}

\def\t{{(2)}}

\def\t{{\{2\}}}

\def\sp{\,,\ \,  }
\def\ind{\I}
\def\r#1{(\ref{#1})}

\def\xitau2{\xi_{(\tau_2)}}
\def\Ltau2{\phi(\tau_1,\tau_2)}\def\Ltau2{\tilde{\xi}(\tau_1,\tau_2)}
\def\chitau2{\chi_{(\tau_2)}}

\def\ebb{\eb\end{document}}

\def\qqq{\quad\quad\quad}

\renewcommand{\bea}{\begin{eqnarray*}}
\renewcommand{\beqa }{\bea}
\renewcommand{\eeqa }{\eea}

\def\bll#1{
\beqa\label{#1}\bal
}\def\lel{\eal\eeqa}
\newcommand{\beql}[1]{\bll{#1}}
\newcommand{\eeql}{\lel}

\newcommand{\COMM}[1]{}
 
\newcommand{\indi}[1]{\I_{\{{#1}\}}}

\usepackage{listings}

\def\chitau2{P_{\tau_2}}

\def\ff{{\mathbb F}}

\def\xitau2{\xi}

\def\ff{{\cal G}}

\def\tb{\bar{\tau}}

\def\CVA{\mathcal{A}}\def\CVA{\Theta}

\def\tilde{\widetilde}

\def\Rf{\mathfrak{R}}\def\Rf{\mathfrak{r}}

\def\thisR{R}\def\thisR{\rho}

\def\tb{{\bar{\tau}}}\def\tb{\thetau}

\def\gg{{\mathbb G}}\def\gg{{\mathbb F}}

\def\ff{\mathbb{F}}

\def\ff{\mathbb{F}^*}

\def\gg{{\cal G}}\def\gg{\mathbb{G}}

\def\ff{{\cal F}}\def\ff{\mathbb{F}}

\def\Ts{\bar{T}}\def\Ts{\mathcal{T}}\def\Ts{T}

\def\tb{{\bar{\tau}}}

\def\Ts{\mathcal{T}}\def\Ts{\bar{T}}
\def\Ts{\bar{T}}\def\Ts{T}

\def\iota{\mathbf{l}}

\def\Ep{{E_{b}}}

\def\chitau{\chi}

\renewcommand{\bel}{\bea\bal}
\renewcommand{\eel}{\end{aligned}\eea}

\def\R{\mathbb{R}}

\def\Rf{\mathfrak{R}}\def\Rf{\mathfrak{r}}

\def\thisR{R}\def\thisR{\rho}

\def\tb{{\bar{\tau}}}\def\tb{\thetau}

\def\gg{{\mathbb F}}\def\gg{{\cal G}}\def\gg{{\mathbb G}}

\def\ff{\mathbb{F}}

\def\Ts{\mathcal{T}}\def\Ts{\mathsf{T}}\def\Ts{\bar{T}}\def\Ts{T}

\def\Ts{\bar{T}}\def\Ts{\mathcal{T}}\def\Ts{T}

\def\tb{{\bar{\tau}}}

\def\Ts{\mathcal{T}}\def\Ts{\mathsf{T}}
\def\Ts{\bar{T}}\def\Ts{T}

\def\kappa{S^{\star}}

\def\Rf{\mathfrak{R}}\def\Rf{\Rf}\def\Rf{R_{f}}\def\Rf{\bar{R}_{b}}\def\Rf{\bar{R}}

\def\thisR{R}\def\thisR{\thisR }\def\thisR{R_b}

\def\tb{\thetau}\def\tb{{\bar{\tau}}}

\def\td{\tau^{\delta}}

\def\thec{c}

\def\chitau{\chi}

\def\kappa{k}

\def\qr#1{\eqref{#1}}

\def\thea{a}

\def\theapim{\theapim}\def\theapim{\thea}
\def\theaim{\theaim}\def\theaim{\thea}

\def\RM{R}\def\RM{\mbox{ES}}\def\RM{\ES}

\def\FBSDE{\FBSDE\xspace}\def\FBSDE{BSDE\xspace}

\def\href#1#2{#2}

\def\CR{{\rm CR}\xspace}
\def\L{\tilde{L}}
\def\TRC{\Theta\xspace}

\def\CR{{\rm FV}\xspace}
\def\CL{{\rm CL}\xspace}
\def\CA {{\rm CA}\xspace}
\def\FVA {{\rm FVA}\xspace}

\def\ES{R}\def\ES{\mbox{ES}}\def\ES{\ES}

\renewcommand{\finproof}{\rule{4pt}{6pt}}
\renewcommand{\bea}{\begin{eqnarray}}
\renewcommand{\eea}{\end{eqnarray}}
\renewcommand{\bt}{\begin{pro}}\renewcommand{\et}{\end{pro}}

\def\KVA{\kva'}

\def\PIM{{\rm PIM}}
\def\RIM{{\rm RIM}}
\def\M{C}\def\M{\mbox{IM}}\def\M{\PIM}

\def\CA{{\rm CA}\xspace}

\def\CL{{\rm CL}\xspace}
\def\CVA{{\rm CVA}\xspace}
\def\FVA{{\rm FVA}\xspace}
\def\FDA{{\rm FDA}\xspace}
\def\DVA{{\rm DVA}\xspace}

\def\cH{{\cal H}}\def\cH{\mathcal{S}}\def\cH{\mathcal{L}}

\def\UCVA{{\rm \UCVA}\xspace}\def\UCVA{{\rm CVA}\xspace}
\def\UFVA{{\rm \UFVA}\xspace}\def\UFVA{{\rm FVA}\xspace}

\def\M{\mathbb{M}}\def\M{\mathbb{P}}
\def\eds{\end{defi}}
\def\cCst{\tilde{\mathcal{C}}^\star}
\def\cFst{\tilde{\mathcal{F}}^\star}

\def\RM{{\rm RM}\xspace}
\def\RC{{\rm RC}\xspace}

\def\FDA{{\rm FDA}\xspace}
\def\DVA{{\rm DVA}\xspace}
\def\FTP{{\rm FTP}\xspace}

\def\cH{{\cal H}}

\def\cCb{\mathcal{C}^\bullet}\def\cFb{\mathcal{F}^\bullet}
\def\cCs{{\mathcal{C}^\star}}\def\cFs{\mathcal{F}^\star}

\def\cCst{{\mathcal{C}}^\star}
\def\cFst{{\mathcal{F}}^\star}

\def\M{\mathbb{M}}\def\M{\mathbb{P}}

\def\Pf{P_1}\def\Pf{\cP}
\def\S{S}\def\S{J}

\def\cL{{\cal H}}\def\cL{\mathcal{L}}\def\cL{\mathcal{S}}
\def\L2{L^2}\def\L2{\cL^2}\def\L2{\cL_2}

\def\EC{{\ES }}\def\EC{{\rm CR}}
\def\ES{R}\def\ES{\mbox{ES}}\def\ES{\ES}\def\ES{{\rm EC
}\xspace}
\def\SCR{{\rm SCR}}

\def\sr#1{Sect.~\ref{#1}}\def\sr#1{Section~\ref{#1}}

\def\SHC{{\rm SHC}\xspace}
\def\UC{{\rm UC}\xspace}
\def\CET{{\rm CET1}\xspace} 
\def\SCR{{\rm SCR}}

\def\CVA{{\rm CVA}\xspace}
\def\FVA{{\rm FVA}\xspace}

\def\UCVA{{\rm \UCVA}\xspace}\def\UCVA{{\rm CVA}\xspace}
\def\UFVA{{\rm \UFVA}\xspace}\def\UFVA{{\rm FVA}\xspace}

\def\KVA{\kva'\xspace}

\def\CVA{{\rm CVA}^\circ}
\def\FVA{{\rm FVA}^\circ}

\def\UCVA{{\rm \UCVA}^\circ}\def\UCVA{{\rm CVA}^\circ}
\def\UFVA{{\rm \UFVA}^\circ}\def\UFVA{{\rm FVA}^\circ}

\def\KVA{{\rm KVA}^\circ}

\def\LOSS{L^\circ}
\def\CA{{\rm CA}^\circ\xspace} 

\def\XVA{TRC/FVA\xspace}\def\XVA{{\rm CA}\xspace}\def\XVA{{\rm XVA}\xspace}

\def\Theta{{\rm CA}}

\def\mtm{{\rm MtM} \xspace}
\def\kva{{\rm KVA} \xspace}
\def\cva{{\rm CVA} \xspace}
\def\fva{{\rm FVA} \xspace}
\def\mva{{\rm MVA} \xspace}
\def\ca{{\rm CA} \xspace}

\def\ca{{\rm CA}\xspace} 

\def\Q{{\mathbb P}}\def\Q{{\mathbb Q}}\def\Q{{\mathbb Q}^*}

\def\r{\textcolor{red}}\def\r{}

\def\b{\textcolor{blue}}\def\b{}
\def\tilde{}
 
\renewcommand{\bcor}{\brem}\renewcommand{\ecor}{\erem}
 
\def\VM{{\rm VM}}
\def\lambdabar{\lambda}

\def\mtm{P\xspace}\def\mtm{{\rm MtM} \xspace}
\def\tb{T}
\def\VaR{\mathbb{V}\mathrm{a}\mathbb{R}'}
\def\RM{\mathbb{ES}}
\def\theaim{a_{im}}
\def\arim{a_{rim}}
\def\apim{a_{pim}}

\usepackage{footnote}
\makesavenoteenv{tabular}
\makesavenoteenv{table}
\usepackage{booktabs,bigstrut}
\usepackage{algorithm}
\usepackage[english]{babel}
\usepackage{amsmath,amssymb,xcolor}

\def\A{\mathfrak{A}}
\def\CM{{\rm CM}\xspace}
\def\sr#1{Sect.~\ref{#1}}\def\sr#1{Section~\ref{#1}}

\def\LOSSt{\tilde{L}}

\def\kva{{\rm KVA} \xspace}
\def\cva{{\rm CVA} \xspace}
\def\fva{{\rm FVA} \xspace}
\def\ca{{\rm CA} \xspace}

\def\ca{{\rm CA}\xspace}

\def\cCst{\mathcal{C}^\circ}
\def\cFst{\mathcal{F}^\circ}

\def\cPs{\mathcal{P}^\circ}

\def\mtm{P\xspace}\def\mtm{{\rm MtM} \xspace}
\def\thec{c}\def\thec{1}

\def\theMtM{P}\def\theMtM{\mtm}

\def\RM{{\rm RM}\xspace}
\def\ES{\mathbb{ES}\xspace}
\def\EC{{\rm EC}\xspace}
\def\CR{{\rm CR}\xspace}
\def\FV{{\rm FV}\xspace}

\def\MVA{{\rm MVA} \xspace}
\def\MDA{{\rm MDA} \xspace}

\def\ES{\mathbb{ES}'\xspace}\def\ES{\mathbb{ES}\xspace}

\def\Ep{\mathbb{E}'\xspace}\def\Ep{\mathbb{E}\xspace}
\def\Qp{\mathbb{Q}'\xspace}\def\Qp{\mathbb{Q}\xspace}
\def\Qs{\mathbb{Q}^*\xspace} 
\def\VaR{\mathbb{V}\mathrm{a}\mathbb{R}'\xspace}\def\VaR{\mathbb{V}\mathrm{a}\mathbb{R}\xspace}
\def\Es{\mathbb{E}^*\xspace}  
\def\qpr{q'\xspace}\def\qpr{q\xspace}
\def\epr{e'\xspace}\def\epr{e\xspace}\def\epr{s\xspace}

\def\cCs{{\mathcal{C}^\circ}}\def\cFs{\mathcal{F}^\circ}

\begin{document}
\title{XVA Analysis From the Balance Sheet}


\author{Claudio Albanese\textsuperscript{1},  St\'ephane Cr\'epey\textsuperscript{2}, Rodney Hoskinson\textsuperscript{3}, and Bouazza Saadeddine\textsuperscript{2,4}}
{\let\thefootnote\relax\footnotetext{\textsuperscript{1} \textit{Global Valuation ltd, London, United Kingdom}}}

{\let\thefootnote\relax\footnotetext{\textsuperscript{2} \textit{LaMME, Univ Evry, CNRS, Universit\'e Paris-Saclay.}}}
{\let\thefootnote\relax\footnotetext{\textsuperscript{3} \textit{ANZ Banking Group, Singapore}}}

{\let\thefootnote\relax\footnotetext{\textsuperscript{4} \textit{Quantitative Research GMD/GMT Credit Agricole CIB, Paris}}}

{\let\thefootnote\relax\footnotetext{{{\it Acknowledgement:} This article has been accepted for publication in Quantitative Finance, published by Taylor \& Francis.
\newline
We are grateful for useful discussions to  
Lokman Abbas-Turki,
Agostino Capponi, 
Karl-Theodor Eisele, 
Chris Kenyon, 
Marek Rutkowski, and Michael Schmutz. The PhD thesis of Bouazza Saadeddine is funded by a CIFRE grant from CA-CIB and French
ANRT.
}}} 
{\let\thefootnote\relax\footnotetext{{{\it Disclaimers:}  
The views expressed herein by Rodney Hoskinson and co-authors are their personal views
 and do not reflect the views of ANZ Banking Group Limited ("ANZ"). No liability shall be accepted by ANZ whatsoever for any direct or consequential loss
 from any use of this paper and the information, opinions and materials contained herein.}}}  

{\let\thefootnote\relax\footnotetext{{{\it Email addresses:} 
claudio.albanese@global-valuation.com,  stephane.crepey@univ-evry.fr, rodney.hoskinson@edhec.com,  bouazza.saadeddine@univ-evry.fr}}}

\date{\today}

\maketitle

\rm
\begin{abstract}  
XVAs denote various counterparty risk related valuation adjustments that are applied to financial derivatives since the 2007--09 crisis.
We root a cost-of-capital XVA strategy in a
balance sheet perspective 
which is key in identifying the economic meaning of the XVA terms.
Our approach is first detailed in a static setup that is solved explicitly.
It is then plugged in the dynamic and trade incremental context of a real derivative banking portfolio.
The corresponding cost-of-capital XVA strategy
ensures to bank shareholders a submartingale equity process corresponding to a target hurdle rate on their capital at risk, consistently between and throughout deals. 
Set on a forward/backward SDE formulation, this strategy
can be solved efficiently using 
GPU computing combined with 
deep learning regression methods in a whole bank balance sheet context. 
A numerical case study emphasizes the workability and added value of 
the ensuing pathwise XVA computations.
\end{abstract}

\def\keywordname{{\bfseries Keywords:}}
\def\keywords#1{\par\addvspace\baselineskip\noindent\keywordname\enspace
\ignorespaces#1}
 
\vspace{2mm}



\section{Introduction}\label{s:intro}

XVAs, with X as C for credit, D for debt, F for funding, M for margin, or K for capital, are post-2007--09 crisis valuation adjustments for financial derivatives.  In broad terms to be
detailed later in the paper (cf.~Table
\ref{tab:accr} in Section \ref{ss:bscsm}), CVA is what the bank expects to lose due to \r{counterparty}
defaults in the future; DVA (irrelevant for pricing but material to bank creditors as we will see) is what the bank expects to gain due to its own default; FVA is the expected cost for the bank of having to raise variation margin (re-hypothecable collateral) ; MVA is the expected cost for the bank of having to raise initial margin (segregated collateral);  KVA is the expected cost for the bank of having to remunerate its shareholders through dividends 
for their capital at risk.

XVAs deeply affect the derivative pricing task by making it global, nonlinear, and entity dependent. However,
before these technical implications, the fundamental point is to understand what really deserves to be priced and what does not,
by rooting the pricing approach in a corresponding collateralization, 
accounting,
and dividend policy of the bank.


Coming after several papers on the valuation of defaultable assets
in the 90's, such as \citeN{DuffieHuang},
\citeN[Eq. (14.25) p.~448]{BieleckiRutkowski2002} obtained the formula 
\beql{e:cd}\cva-\DVA\eeql 
for the valuation of bilateral counterparty risk 
on a swap, assuming risk-free funding. 
This formula, rediscovered and generalized by others since the 2008--09 financial crisis (cf.~e.g.~\citeN{BrigoCapponi2010}), 
\b{is symmetrical, i.e.~it is the negative} of the analogous quantity considered from the point of view of the counterparty, consistent with the law of one price and the \citeN{MM1958} theorem.

Around 2010, 
the materiality of
the \DVA windfall benefit of a bank at its own default time became the topic of intense debates in the quant and academic communities.
At least, it seemed reasonable to admit that, if the own default risk of the bank was accounted for in the modeling, in the form of a DVA benefit, then the cost of funding (FVA) implication of this risk should be included as well, leading to the modified formula ($\cva-\DVA+\fva$). See for instance \citeANP{BurgardKjaer11} (\citeyearNP{BurgardKjaer11}, \citeyearNP{BurgardKjaer14}, \citeyearNP{BurgardKjaer17}), \citeN{Crepey2012bc}, 
\citeN{PallaviciniBrigo13bprel}, or \citeN{BichuchCapponiSturm16}. See also
\citeN{BieleckiCrepeyRutkowski11} for an abstract funding framework (without explicit reference to XVAs), generalizing \citeN{Piterbarg10} to a nonlinear setup.

Then \citeN{HullWhite13de} objected that the FVA was only the compensator of another windfall benefit of the bank at its own default, corresponding to the non-reimbursement by the bank of its funding debt. Accounting for the corresponding ``DVA2'' (akin to the FDA in this paper)
brings back to the original firm valuation formula:
$$\cva-\DVA+\fva-\FDA=\cva-\DVA,$$
as $\fva=\FDA$ (assuming risky funding fairly priced as we will see).

However, their argument implicitly assumes that the bank can perfectly hedge its own default: cf.~\citeN[end of Section 3.1]{BurgardKjaer14} and see \sr{ss:mcl} below. As a bank is an intrinsically leveraged entity, this is not the case in practice. 
One can mention the related corporate finance notion of debt overhang in \citeN{Myers77}, by which a project valuable for the firm as a whole may be rejected by shareholders because the project is mainly valuable to bondholders.
But, until recently, such considerations were hardly considered in the field of derivative pricing.

The first ones to recast the XVA debate in the perspective of the balance sheet of the bank were \citeN{BurgardKjaer11}, to explain
that an appropriately hedged derivative position has no impact on the dealer's funding costs. 
Also relying on balance sheet models of a dealer bank, \citeN{Castagna2014} and 
\citeN{AndersenDuffieSong2016} 
end up with conflicting conclusions, namely that the FVA should, respectively should not, be included in the valuation of financial derivatives.
Adding the KVA, but in a replication framework, \citeN{GreenKenyonDennis14}
conclude that
both the FVA and the KVA should be included
as add-ons in entry prices and as liabilities in the balance sheet.
 
\subsection{Contents}\label{ss:con}

Our key premise
is that
counterparty risk entails two distinct but intertwined sources of market incompleteness: 
\begin{itemize}
\item {A bank cannot perfectly hedge counterparty default losses, by lack of sufficiently liquid CDS markets;} 

\item {A bank can even less 
hedge
its own jump-to-default exposure, because this would mean selling protection on its own default, which is nonpractical and, under certain juridictions, even legally forbidden (see \sr{ss:bscsm}).} 
\end{itemize}
We specify 
the banking XVA metrics that align derivative entry prices to shareholder interest, given this impossibility for a bank to replicate the jump-to-default related cash flows. 
We develop a 
cost-of-capital XVA approach
consistent with
the accounting standards set out in 
IFRS 4 Phase II (see \citeN{IFRS4Phase2ED}),
 inspired from the Swiss solvency test and Solvency II insurance regulatory frameworks
(see \citeN{FOPI06} and \citeN{CEIOPS10}), which so far has no analogue in the banking domain. 
Under this approach, the valuation (CL) of the so-called contra-liabilities and the cost of capital (KVA) are sourced from clients 
at trade inceptions, on top of the ($\cva-\DVA$) complete market valuation of counterparty risk, in order to compensate bank shareholders for wealth transfer and risk on their capital. 

The cost of the corresponding collateralization, 
accounting,
and dividend policy  
is, by contrast with the complete market valuation ($\cva-\DVA$) of counterparty risk, 
\beql{e:cfk}\cva+\fva+\kva,\eeql  
computed unilaterally in a certain sense (even though we do crucially include the default of the bank itself in our modeling), and charged to clients on an incremental run-off basis at every new deal\footnote{See also Remark \ref{rem:mva} regarding the meaning of the FVA in \eqref{e:cfk}.}.

All in one,
our cost-of-capital XVA strategy makes 
shareholder equity a submartingale with drift corresponding to a hurdle rate $h$ on shareholder capital at risk, consistently between and throughout deals.
Thus we arrive at a sustainable strategy for profits retention, much like in the
above-mentioned
insurance regulation, but in a consistent continuous-time 
and banking framework.  

Last but not least, 
our approach can be solved efficiently using GPU computing combined with deep learning regression methods in a whole bank balance sheet context.

\subsection{Outline and Contributions}\label{ss:outl}

Section \ref{ss:bscsm} sets a financial stage where a bank is split across
several trading desks and entails different stakeholders.
\sr{s:theex} develops our cost-of-capital XVA approach in a one-period static setup.
Section \ref{s:dyn} revisits the approach at the dynamic and trade incremental level.
  \sr{s:num} is a numerical case study on large, multi-counterparty portfolios of interest rate swaps,
based on the continuous-time XVA equations for bilateral trade portfolios recalled in \sr{s:eqns}. 

The \textbf{\em main contributions} of the paper are:
\begin{itemize}

 \item The one-period static XVA model of \sr{s:theex}, with explicit formulas for all the quantities at hand, {offering a concrete grasp} on the related wealth transfer and risk premium issues;

\item Proposition \ref{p:wtprel},
which 
establishes the 
connections between XVAs and the core equity tier 1 capital of the bank, respectively bank shareholder equity; 

\item Proposition
 \ref{p:wrafinal}, which establishes that,
under the XVA policy represented by the balance conditions \qr{e:reset} between deals and the counterparty risk add-on
\qr{e:bsheet-entry-incr-delta-fin}
throughout deals,   
bank shareholder equity is
a submartingale with drift corresponding to a target hurdle rate $h$ on shareholder capital at risk.
This 
perspective 
solves the puzzle according to which, on the one hand, XVA computations are performed on a run-off portfolio basis, while, on the other hand, they are used for computing pricing add-ons to new deals;


\item The XVA deep learning (quantile) regression computational strategy of \sr{ss:xvaframew};


\item The numerical case study of \sr{s:num}, which emphasizes the materiality of 
refined, pathwise XVA computations, 
as compared to 
more simplistic XVA approaches.

\end{itemize}
From a broader point of view, this paper reflects a shift of paradigm 
regarding 
the pricing and risk management of financial derivatives, from 
hedging to balance sheet optimization, as quantified by relevant XVA metrics. In particular (compare with the last paragraph before Section \ref{ss:con}), our approach implies
 that the FVA (and also the MVA, see Remark \ref{rem:mva}) should be included
as an add-on in entry prices and as a liability in the balance sheet; the KVA should be included
as an add-on in entry prices, but not as a liability in the balance sheet.

From a computational point of view, this paper opens the way to
second generation XVA GPU implementation. The first generation consisted of nested Monte Carlo implemented by explicit CUDA programming on GPUs (see \citeN{AlbaneseCaenazzoCrepey17b},
\citeN{AbbasturkiCrepeyDiallo17}).
The second generation takes advantage of GPUs leveraging via pre-coded CUDA/AAD deep learning packages that are used for the XVA embedded regression and quantile regression task. Compared to a regulatory capital based KVA approach, an economic capital based KVA approach is then not only conceptually more satisfying, but 
also simpler to implement.

%
 
\section{Balance Sheet 
and Capital Structure 
Model of the Bank}
\label{ss:bscsm} 


We consider a dealer bank, which is a market maker involved in bilateral derivative portfolios. For simplicity, we only consider European derivatives.
The bank has two 
kinds of stakeholders, \textbf{\em shareholders} and  \textbf{\em bondholders}.
The shareholders have the control of the bank and are solely responsible for investment decisions before bank default. The
bondholders represent the \b{senior} 
creditors
of the bank, who have no decision power until bank default, but are protected by laws, of the pari-passu type, forbidding trades 
that would trigger value
away from them to shareholders during the default resolution process of the bank.
The bank also
has \b{junior} creditors, represented in our framework by an \textbf{\em external funder}, who can lend unsecured to the bank and is assumed to \b{suffer an exogenously given loss-given-default}
 in case of default of the bank.

\begin{figure}[h!]
\begin{center}
\includegraphics{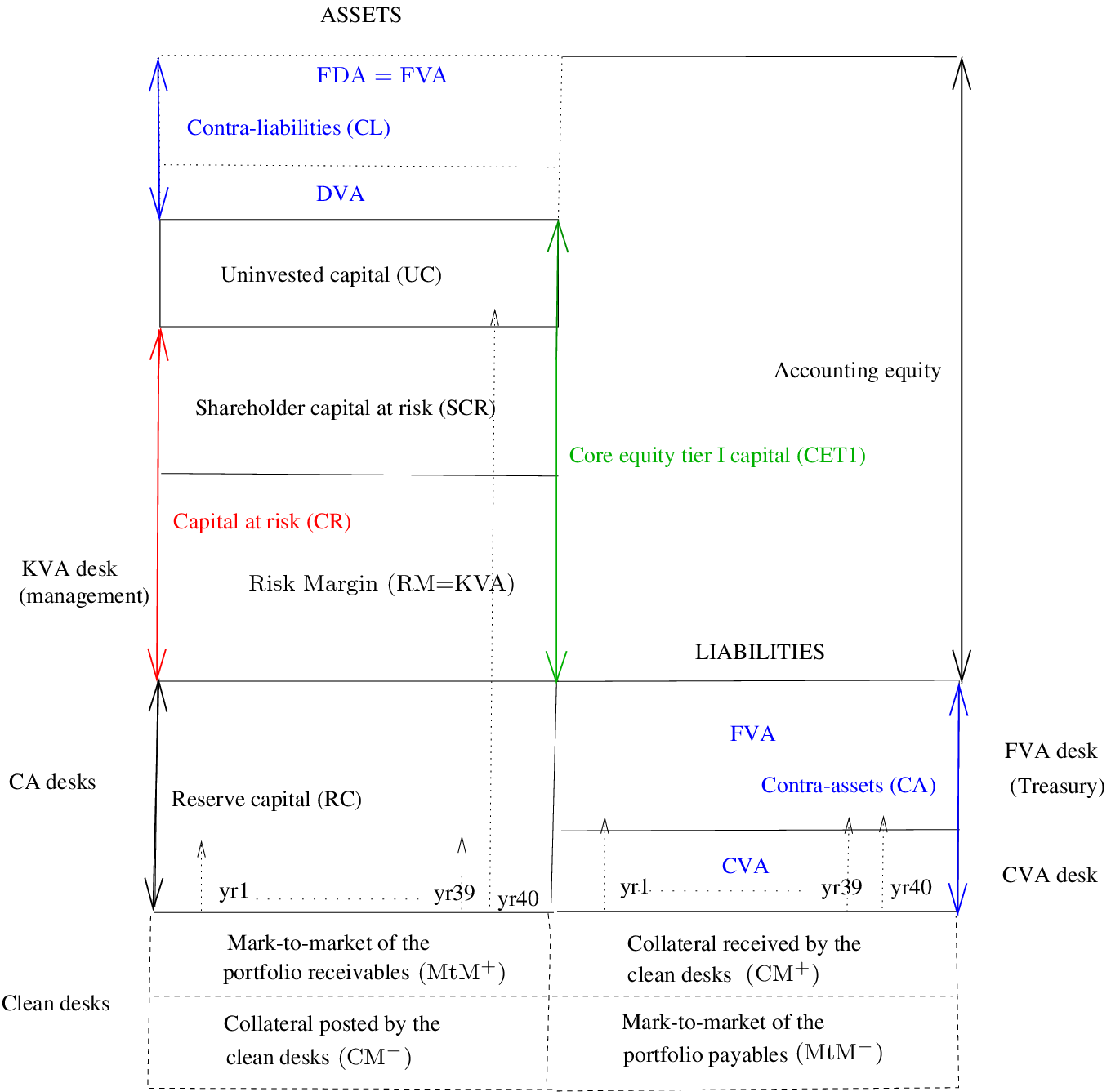}%
\end{center}
\caption{Balance sheet of a dealer bank. Contra-liability valuation (CL) at the top is shown in dotted boxes because 
it is only value to the bondholders (see \sr{ss:mcl}).
Mark-to-market valuation (MtM)
of the \r{derivative} portfolio of the bank
by the clean desks, as well as the corresponding collateral (clean margin CM), are shown in dashed boxes at the bottom. Their role will essentially vanish in our setup, where we assume a perfect
clean hedge by the bank. The arrows in the left column represent trading losses of the CA desks in ``normal years 1 to 39'' 
and in an ``exceptional year 40'' with full depletion 
(i.e.~refill via UC, under Assumption \ref{h:selfi}.ii) 
of RC, RM, and SCR.
The numberings yr1 to yr40 are fictitious yearly scenarios in line with a 97.5\% expected shortfall of the one-year-ahead trading losses of the bank that we use for defining its economic capital.
The arrows in the right column symbolize the average depreciation in time of contra-assets between deals. \b{The collateral between the bank and its
\r{counterparties}
 is not shown to alleviate the picture.}}
\label{fig:bs}
\end{figure}

We consider three kinds of business units within the bank (see Figure \ref{fig:bs} for the corresponding picture of the bank balance sheet and refer to Table \ref{tab:accr}
for a list of the main financial acronyms used in the paper): 
the \textbf{\em CA desks}, i.e.~the CVA desk and the FVA desk (or Treasury) of the bank, in charge of contra-assets, i.e.~of counterparty risk and its funding implications for the bank; the \textbf{\em clean desks}, who focus on the market risk of the contracts in their respective business lines;
the \textbf{\em management} of the bank, in charge of  
the dividend release policy of the bank.

\begin{table}[!htbp]
\begin{center}
\begin{tabular}{|p{1cm}|p{5.7cm}|p{6.8cm}|} 
\multicolumn{3}{c}{\textbf{\em Amounts on dedicated cash accounts of the bank:}}\\\hline
\bf CM & Clean margin&Definition \ref{d:shc} 
and Assumption \ref{h:selfi} \\
\bf RC &   Reserve capital&Definition \ref{d:shc} 
and Assumption \ref{h:selfi} \\
\bf RM &   Risk margin&Definition \ref{d:shc} 
and Assumption \ref{h:selfi} \\ 
\bf UC &   Uninvested capital& Definition \ref{d:shc} 
and Assumption \ref{h:selfi} \\  \hline
\multicolumn{3}{c}{}\\
\multicolumn{3}{c}{\textbf{\em Valuations:}}\\\hline
\bf CA &   Contra-assets valuation&\qr{e:shccet},
\qr{e:cadi},
and
\qr{e:rcfm}\\
\bf CL &   Contra-liabilities valuation&Definition \ref{d:shc} and \qr{e:cl},
\qr{e:bsheet-pedagodecompo}, and
\qr{e:bsheet-entry-incr-delta-fin} \\
\bf CVA &   Credit valuation adjustment& \qr{e:cvadi},
\qr{e:cadi}, \qr{e:L}, and  \qr{e:instrunocsa}--\qr{e:instrucsa}\\ 
\bf DVA &   Debt valuation adjustment& \qr{e:cl} and \qr{e:cvadi}\\  
\bf FDA &   Funding debt adjustment& \qr{e:cl} and \qr{e:fdadi}\\
\bf FV &   Firm valuation of counterparty risk& \qr{e:fvdi}  and 
\qr{e:fdadi}\\
\bf FVA &   Funding valuation adjustment &
Remark \ref{rem:mva}, \qr{e:cvadi},
\qr{e:cadi}, and \qr{e:L}\\
\bf KVA &  Capital valuation adjustment &  \qr{e:reset},
\qr{e:kvadiscr}, and 
\qr{e:kconsuite}\\ 
\bf MtM &   Mark-to-market& \qr{e:reset} and \qr{e:mtmdi}\\ 
\bf MVA &   Margin valuation adjustment & Remark \ref{rem:mva},  \qr{e:mvda}, 
\qr{e:L}, and 
 \qr{e:instrumva}
\\
\bf XVA &   Generic ``X'' valuation adjustment& First paragraph \\\hline
\multicolumn{3}{c}{}\\
\multicolumn{3}{c}{\textbf{\em Also:}}\\\hline
\bf CR &   Capital at risk& \qr{e:ecpt}\\ 
\bf CET1 &   Core equity tier I capital&  \qr{e:shccet}
and \qr{e:cL}\\ 
\bf EC &   Economic capital& Definitions \ref{d:ccrbis}  
and \ref{defi:ec}\\ 
\bf FTP &   Funds transfer price&\qr{e:bsheet-entry-incr-delta-fin} \\ 
\bf SHC &  Shareholder capital (or equity)& \qr{e:shccet} and \qr{e:lalafirst}\\ 
\bf SCR &   Shareholder capital at risk&Assumption \ref{h:selfi} and \qr{e:scr}\\ 
\hline
\end{tabular}
\end{center}
\caption{Main financial acronyms and place where they are introduced conceptually and/or specified mathematically in the paper, as relevant.}
\label{tab:accr}
\end{table}

Collateral means cash or liquid assets that are posted to guarantee a netted set of transactions 
against defaults. 
It comes in two forms: variation margin, which is
re-hypothecableotecable, i.e.~fungible across netting sets, and initial margin, which is segregated. We assume
cash only collateral.
Posted collateral is supposed to be remunerated at the risk-free rate (assumed to exist, with overnight index swap rates as a best market proxy). 

\brem\label{rem:mva} To alleviate the notation, in this conceptual section of the paper, we only consider an FVA as the global cost of raising collateral for the bank, as opposed to a distinction, in the industry and in later sections in the paper, 
between an FVA, in the strict sense of the cost of raising variation margin, and an MVA for the cost of raising initial margin.~\erem
 
The CA desks guarantee the trading of the clean desks against \r{counterparty}
defaults, through a \textbf{\em clean margin account}, which can be seen as (re-hypothecable) 
collateral exchanged between the CA desks and the clean desks. The corresponding clean margin amount (CM) also plays the role of the funding debt of the clean desks put at their disposal at a risk-free cost by the Treasury of the bank. This is at least the case when $\CM>0$ (clean desks clean margin receivers). In the case when $\CM<0$ (clean desks clean margin posters), ($-\CM$) corresponds to excess cash generated by the trading of the clean desks, usable by the Treasury for its other funding purposes. See the bottom, dashed boxes in Figure \ref{fig:bs}.

In addition, the CA desks value the contra-assets (future counterparty default losses and funding expenditures),
charge them to the (corporate) clients at deal inception, deposit the corresponding payments in a \textbf{\em reserve capital account}, and then are exposed to the corresponding payoffs. 
As time proceeds, contra-assets realize
and are covered by the CA desks with the reserve capital account. 

On top of reserve capital, the so-called
risk margin is sourced by the management of the bank from the clients at deal inception, deposited into a \textbf{\em risk margin account},
and then gradually released as KVA payments into the shareholder dividend stream. 

Another account contains the \textbf{\em shareholder capital at risk} earmarked by the bank to deal with exceptional trading losses (beyond the expected losses that are already accounted for by reserve capital). 
 
Last, there is one more \b{bank} account
with shareholder \textbf{\em uninvested 
capital}.


All cash accounts 
are remunerated at the risk-free rate.

\bd\label{d:shc} We write CM,
RC, RM, SCR, and UC for the respective {(risk-free discounted)} amounts on the clean margin, reserve capital, risk margin, shareholder capital at risk, and uninvested capital accounts of the bank. We also define
\beql{e:shccet}\SHC=\SCR+\UC\sp\CET=\RM+\SCR+\UC.~\finproof\eeql \eds

\noindent
From a financial interpretation point of view, before bank default, $\SHC$ corresponds to shareholder capital (or equity);
CET1 is the {\textbf{\em core equity tier I capital} of the bank, representing the financial strength of the bank assessed from a regulatory, structural solvency point of view, i.e.~the sum between shareholder capital and the risk margin (which is also loss-absorbing), but excluding
the value CL of the so-called contra-liabilities (see Figure \ref{fig:bs}). Indeed, the latter only benefits the bondholders  (cf.~\sr{ss:mcl}), hence
it only enters accounting equity. \b{Before the default of the bank, \textbf{\em shareholder wealth} and \textbf{\em bondholder wealth} are respectively given by 
$\SHC+\RM^{sh}$ and $\CL+\RM^{bh}$, for shareholder and bondholder components of RM to be detailed in Remark \ref{rem:share}; shareholder and bondholder wealths sum up to the accounting equity $\RM+\SCR+\UC+\CL$, i.e. the wealth of the firm as a whole (see Figure \ref{fig:bs}).}

\brem\label{rem:int} The purpose of our capital structure model of the bank 
is {\em not} to model the default of the bank, like in a \citeN{Merton1974} model, as the point of negative equity (i.e. $\CET<0$). In the case of a bank, such a default model would be unrealistic. 
For instance, at the time of its collapse in April 2008, Bear Stearns had billions of capital. In fact, the legal definition of default is an unpaid coupon or cash flow, which is a liquidity (as opposed to solvency) issue. Eventually we will model the default of the bank as a totally unpredictable event at some exogenous time $\tau$ calibrated to the credit default swap (CDS) curve referencing the bank. Indeed 
 we view the latter as the most reliable and informative credit data regarding anticipations of markets participants about future recapitalization, government intervention, bail-in,
and other bank failure resolution policies.

The aim of our capital structure model, instead, is to put in a balance sheet perspective the contra-assets and contra-liabilities of a dealer bank, items which are not present in the Merton model and play a key role in our XVA analysis.
\erem

 In line with the Volcker rule banning proprietary trading for a bank, we assume
a perfect
market hedge of the derivative portfolio of the bank by the clean desks, in a sense to be specified below in the respective static and continuous-time setups. 
By contrast, as jump-to-default exposures (own jump-to-default exposure, in particular) cannot be hedged by the bank (cf.~\sr{ss:con}), 
we conservatively assume no XVA hedge.

We work on a measurable space $(\Omega,\A)$ endowed with a probability measure $\Qs$, with $\Qs$ expectation denoted by $\Es$, which is used for the  linear valuation task, using the risk-free asset as our num\'eraire everywhere.

\b{\brem\label{rem:dyb} Regarding the nature of our reference probability measure $\Q$, ``physical or risk-neutral'', one should view it as a blend between the two. For instance,
even if we do not use this explicitly in the paper, one could conceptually think of $\Qs$
as the probability measure introduced by \citeN{Dybvig92} 
to deal with incomplete markets that are a mix of financial traded risk factors and unhedgeable ones (jumps to default, in our setup), recently revisited in a finance and insurance context by \citeN{ArtznerEiseleSchmidt20}. Namely, one could think of $\Q$ as the
unique probability measure on $\A$\footnote{See \citeN[Proposition 2.1]{ArtznerEiseleSchmidt20} for a proof.} 
that coincides {\em (i)} with a given
risk-neutral pricing measure  
{\em on} the financial $\sigma$ algebra $\subseteq \A$, and {\em (ii)}
with the physical probability measure {\em conditional} on the financial $\sigma$ algebra (the risk-neutral and physical measures being assumed equivalent on the financial $\sigma$ algebra). The risk-neutral pricing measure (hence, in view of {\em (i)}, $\Qs$ itself) is calibrated to prices of fully collateralized transaction for which counterparty risk is immaterial. The physical probability measure expresses user views on the unhedgeable risk factors. The uncertainty about $\Qs$ can be dealt with by a Bayesian variation on our baseline XVA approach, whereby paths of alternative, co-calibrated models are combined in a global simulation (cf. \citeN{Hoeting99bayesianmodel}).~\erem}
 

\subsection{Run-Off Portfolio}\label{ss:runoffprel}

Until \sr{ss:incrx}, we consider the case of a portfolio held on a run-off basis, 
i.e.~set up at time 0 and such that no new unplanned trades enter the portfolio in the future.

The trading cash flows of the bank (cumulative cash flow streams starting from 0 at time 0) then consist of
\begin{itemize}
\item the contractually promised cash flows $\cP$ from \r{counterparties},
\item the counterparty credit cash flows $\cC$ to \r{counterparties}
(i.e., because of counterparty risk, the effective cash flows from \r{counterparties}
are $\cP-\cC$), 
\item the risky funding cash flows $\cF$ to the external funder, and
\item the hedging cash flows $\cH$ of the clean desks to financial hedging markets
\end{itemize}
 (note that all cash flow differentials can be positive or negative). See \sr{ss:cfop} and 
\qr{e:cmmtm}--\qr{e:fundi}
for concrete specifications in respective one-period and continuous-time setups.

 \bhyp\label{h:selfi} \begin{enumerate}
\item {\it \textbf{(Self-financing condition)}}
$\RC+\RM+\SCR+\UC-\CM$ evolves like the 
received
trading cash flows $\cP-\cC-\cF-\cH$.

\item
 {\it \textbf{(Mark-to-model)}}
The amounts on all the accounts but UC are marked-to-model (hence the last, residual amount, UC, plays the role of an adjustment variable).
Specifically,
we assume that
the following \textbf{\em shareholder balance conditions} hold at all times:  
\beql{e:reset}
{\CM}=\mtm\sp
{\RC}=\ca
\sp
{\rm RM}=\kva, \eeql  
for theoretical target levels $\mtm$, $\ca$, and $\kva$ to be specified in later sections of the paper (which will also determine the theoretical target level for SCR).

\item
 {\it \textbf{(Agents)}}
The initial amounts $\mtm_0$, $\ca_0$, and $\kva_0$  are provided by the clients at portfolio inception time 0. Resets between time 0 and the bank default time $\tau$ (excluded) are on bank shareholders. At the (positive) bank default time $\tau$,
the property of the residual amount on 
the reserve capital and risk margin accounts 
is transferred from the shareholders to the bondholders of the bank.~\finproof\end{enumerate}\ehyp 
\b{ \brem \label{rem:nodelta} In an asymmetric setup with a price maker and a price taker, 
the price maker passes his costs to the price taker. Accordingly, in our setup, the (corporate) clients provide all the amounts to the clean margin,
reserve capital, and risk margin accounts of the 
bank required for resetting the accounts to their theoretical target levels \qr{e:reset} corresponding to the updated portfolio.~\erem} 


Under a cost-of-capital XVA approach, we define valuation so as to make shareholder trading losses (that include marked-to-model liability fluctuations) centered, then we add a KVA risk premium in order to ensure to bank shareholders some positive hurdle rate $h$ on their capital at risk.  

In what follows, such an approach is developed, first, in a static setup, which can be solved explicitly, and then,
in a dynamic and trade incremental setup, as suitable for dealing with a real derivative banking portfolio.

\section{XVA Analysis in a Static Setup}\label{s:theex}
 
{

\def\CVA{\cva} 
\def\UCVA{\cva}
\def\FVA{\fva} 
\def\UFVA{\fva}
\def\CA{\ca} 
\def\KVA{\kva}
\def\cC{\mathcal{C}^\circ}
\def\cF{\mathcal{F}^\circ}
\def\EC{{\rm EC}}
\def\FTP{{\rm FTP}}

In this section, we apply the cost-of-capital XVA approach to a portfolio made of {a single deal, $\cP$  (random variable promised to the bank)}, between a bank and a client, without prior endowment, in an elementary one-period (one year) setup. 
All the trading cash flows $\cP$, $\mathcal{C}$, $\mathcal{F}$, and $\cH$ are then random variables (as opposed to processes in a multi-period setup later in the paper).  
We first assume no collateral exchanged between the bank and its client (but collateral exchanged as always between the CA and the clean desks as well as collateral on the market hedge of the bank, 
the way explained after the respective Remarks \ref{rem:mva} and
\ref{rem:int}).
Risky funding 
assets are assumed fairly priced by the market, in the sense that~$\Es\mathcal{F}=0$.

The bank and client are both default prone with zero recovery to each other.
The bank also has zero recovery to its external funder.
We denote by $\S$
and $\S_\thec $ the survival indicators (random variables) of the bank and client at time 1, 
with default probability of the bank
$\Qs(\S=0)=\gamma$. 


Since prices and XVAs only matter at time 0 in a one-period setup,
we identify all the XVA processes, as well as the 
mark-to-market  (valuation 
by the clean desks) $\mtm$
 of the deal, 
 with their values at time 0.

For any random variable $\cY$,
we define 
\beql{e:decbi}
\cY^{\circ}=J\cY\mbox{ and } \cY^{\bullet}=-(1-J)\cY\sp \mbox{ hence } \cY=\cY^{\circ}-\cY^{\bullet}.\eeql
Let $\Ep$ denote the expectation with respect to the bank survival measure, say $\Qp $, associated with $\Qs$, i.e., for any random variable $\cY$,
\beql{e:survdi}&\Ep\cY =(1-\gamma)^{-1}\Es( \cY^\circ)\eeql
(which is also equal to $\Ep\cY^\circ$).
The notion of bank survival measure was introduced in greater generality by \citeN{Schoenbucher04}.
In the present static setup, \qr{e:survdi} is nothing but the $\Qs$ expectation of $\cY$ conditional on the survival of the bank
(note that,  whenever $\cY$ is independent from $J$, the right-hand-side in \qr{e:survdi} coincides with $\Es \cY $).

\bl \label{e:thelem} For any random variable $\cY$ and constant $Y$, we have 
\bel
Y =\Es (\cY^\circ  + (1-J)Y)  \Longleftrightarrow Y=\Ep   \cY.
\eel
\el
\proof Indeed, 
\bel
Y =\Es (J\cY  + (1-J)Y)  \Longleftrightarrow  \Es (J (\cY -Y))=0  \Longleftrightarrow    \Ep    (\cY -Y)=0  \Longleftrightarrow Y=\Ep   \cY,
\eel 
 where the equivalence in the middle is justified  by \qr{e:survdi}.~\finproof

\brem\label{rem:fun}
For simplicity in a first stage, we will ignore 
the possibility of using capital at risk 
for funding purposes, only considering in this respect reserve capital ${\RC}=\ca$ (cf.~\qr{e:reset}). The additional free funding source provided by capital at risk will be introduced later, as well as \r{collateral between bank and client}, in Section \ref{rem:compl}.~\erem

\subsection{Cash Flows}\label{ss:cfop}

\bl\label{l:cfoneper} Given the (to be specified) $\mtm$ and $\CA$ amounts (cf.~Assumption \ref{h:selfi}.ii),
the credit and funding cash flows $\mathcal{C}$ and $\mathcal{F}$ of the bank and its trading loss (and profit) $L$ are such that
\beql{e:CFpedago}  
&\cCs =J(1-\S_\thec)\Pf^{+}  
\sp \cFs=J\gamma (\theMtM-\ca)^+ \\
&\cCb=(1-J)  \big(\cP^-  - (1-J_\thec)\cP ^{+} \big)  \sp\cFb=(1-J) \big((\theMtM-\ca)^+ -\gamma (\theMtM-\ca)^+  \big) \\
&\LOSS= 
\cCs +\cFs  -J\ca\sp  L^\bullet=\cCb + \cFb +(1-J)\ca \sp L=\mathcal{C}+\mathcal{F}-\ca .
\eeql
 \el

\proof For the deal to occur,
the bank
needs to
borrow $(\theMtM-\ca)^+$ unsecured or invest $(\theMtM-\ca)^-$ risk-free (cf.~Remark \ref{rem:fun}). 
Having assumed zero recovery to the external funder,
unsecured borrowing is fairly priced as $\gamma~\times$ the amount borrowed by the bank (in line with our assumption that $\Es\mathcal{F}=0$),
i.e.~the bank must pay for its risky funding the amount
$$\gamma (\theMtM-\ca)^+ .$$
Moreover, at time 1, under zero recovery upon defaults:
\begin{itemize} 
\item If the bank is not in default (i.e.~$\S=1$), then the bank closes its position with the client while receiving $\Pf$ from its client if the latter is not in default (i.e.~$\S_\thec =1$), whereas the bank
pays $\Pf^{-}$ to its client if the latter is in default (i.e.~$\S_\thec =0$). 
In addition, the bank reimburses its funding debt $(\theMtM-\ca)^+$ or receives back the amount $(\theMtM-\ca)^-$ it had lent at time 0;
\item If the bank is in default (i.e.~$\S=0$), then the bank receives back $J_\thec \Pf^+$ on the derivative as well as the amount $(\theMtM-\ca)^-$ it had lent at time 0. 
\end{itemize}
Also accounting for the hedging loss $\cH$, the trading loss of the bank over the year is
\beql{e:recapprel}
& 
L=  \gamma (\theMtM-\ca)^+ 
- \S \big( \S_\thec  \Pf-(1-\S_\thec)\Pf^{-} - (\theMtM-\ca)^+ + (\theMtM-\ca)^- \big)\\ & \qqq 
- (1-\S)\big(J_\thec  \Pf^+ + (\theMtM-\ca)^- \big) +\cH.
\eeql
In the static setup, the perfect clean hedge condition (see after Remark \ref{rem:int}) 
writes $\cH= \cP-\mtm $. Inserting this into the above yields
\beql{e:recap}
& 
L
=   (1-J_\thec)\cP ^{+}+ \gamma (\theMtM-\ca)^+ -\ca   - (1-J) (\cP ^- + (\theMtM-\ca)^+),
\eeql
as easily checked for each of the four possible values of the pair $(\S,\S_\thec)$.
That is, 
\beql{e:theg}
 &\LOSS= \underbrace{J   (1-J_\thec)\cP ^{+}   }_{\cCs} 
+\underbrace{J   \gamma (\theMtM-\ca)^+  }_{\cFs}- J \CA   \\
&L^\bullet= \underbrace{(1-J)  \big(\cP^-  - (1-J_\thec)\cP ^{+} \big) }_{\mathcal{C}^\bullet}+
 \underbrace{(1-J) \big((\theMtM-\ca)^+ -\gamma (\theMtM-\ca)^+  \big)}_{\mathcal{F}^\bullet} +(1-J)\ca,
\eeql 
where the identification of the different terms 
as part of $\mathcal{C}$ or $\mathcal{F}$
follows from their financial interpretation.~\finproof\\ 
\brem\label{e:nged} 
The derivation \qr{e:recapprel} implicitly allows for negative equity (that arises whenever $\LOSS>{\CET}$,  cf.~\qr{e:shccet}),
which is interpreted as recapitalization. 
In a variant of the model excluding both recapitalization and negative equity,
the default of the bank would be modeled in a structural fashion as the event $\{\LOSSt=\CET$\},
where
\beql{e:Lstat}
\LOSSt
= \big(    
(1-J_\thec)\cP ^{+}   
+   \gamma (\theMtM-\ca)^+
-\ca\big)\wedge {\CET}, \eeql
and we would obtain, instead of \qr{e:recap}, the following trading loss for the bank:
\bel
&
\indi{{\CET}> \LOSSt}  \LOSSt  + \indi{{\CET}= \LOSSt}\big({\CET} -\Pf^-  -   (\theMtM-\ca)^+ \big).
\eel 
In this paper we consider a model with recapitalization for the reasons explained in Remark \ref{rem:int}.

Structural XVA approaches in a static setup have been proposed in \citeN{AndersenDuffieSong2016} (without KVA) and 
\citeN{Kjaer2019} (including the KVA). Their marginal, limiting results as a new deal size goes to zero are comparable to some of the results that we have here.  But then, 
 instead of developing a continuous time version of their corporate finance model and
taking the small trade limit, these papers 
start the development of the continuous time model
from the single period small trade limit model. By contrast, in our framework, 
we 
have end to end development in the continuous time model of \sr{s:dyn} and in the present single period model.~\finproof\erems

%

\subsection{Contra-assets and Contra-liabilities}

To make shareholder trading losses centered (cf.~the next-to-last paragraph of Section \ref{ss:bscsm}), 
clean and CA desks value by $\Qs$ expectation their shareholder sensitive cash flows. These include, in case of default of the bank, 
 the transfer of property from the CA desks to the clean desks of the collateral amount MTM on the clean margin account, as well as (cf.~Assumptions \ref{h:selfi}.ii and iii) the
 transfer from shareholders to bondholders of the residual value $\RC=\CA$ on the reserve capital account. Accordingly:

\bd\label{d:ccr} We let \beql{e:mtmdi}\mtm=\Es\big( \cPs  +(1-J) \mtm \big) \eeql and 
\beql{e:cadi}&\CA= \CVA+\FVA,\eeql
 where
\beql{e:cvadi}
   & \CVA=\Es\big( \cC  +(1-J)\CVA \big) \\
 &\FVA=\Es\big( \cF  +(1-J)\FVA \big), 
\eeql  
hence $ \CA=\Es\big( \cC+ \cF +(1-J)\CA \big)$.
We also define the contra-liabilities value \beql{e:cl}\CL=\DVA+\FDA,\eeql where
\begin{eqnarray}
 &&\label{e:dvadi}\DVA= \Es\big(\mathcal{C}^\bullet+(1-J)\CVA\big)
 \\\label{e:fvadi}
&&\FDA=\Es\big( \mathcal{F}^\bullet  +(1-J)\FVA \big)  .
\end{eqnarray}  
Finally we define the firm valuation of counterparty risk,
\begin{eqnarray}
 &&\label{e:fvdi}\FV=\Es(\mathcal{C}+\mathcal{F}) .~\finproof
\end{eqnarray}  
\eds

The definitions of $\mtm, \cva,$ and $\fva$ are in fact fix-point equations. However, the following result shows that these equations are well-posed and yields explicit formulas for all the quantities at hand.
\bp\label{p:ccr}
We have 
\beql{L:discrs}
&\theMtM
=\Ep\cPs 
\\ &  \CVA
= \Ep \big( (1-J_\thec)\cP ^{+} \big)\\
&\FVA
=\gamma (\theMtM-\ca)^+= \frac{\gamma}{1+\gamma}(\theMtM- 
\cva)^+ 
\eeql
and
\beql{e:fdadi}  
&\Es  L^\circ  
=\Ep L=0  \\
&\FDA
=\FVA
\\& \FV
=\Es\mathcal{C} =\cva-\DVA =\ca-\CL  .
\eeql
\ep
\proof The first identities in each line of \qr{L:discrs} follow from Definition \ref{d:ccr}  
by 
Lemma \ref{e:thelem} 
and definition of the involved cash flows in Lemma \ref{l:cfoneper}.  
Given \eqref{e:cadi},
the formula $\FVA=\gamma (\theMtM-\ca)^+$ in \qr{L:discrs}
 is in fact a semi-linear equation
\beql{e:thfva}\FVA= \gamma (\theMtM-\cva-\FVA)^+ .\eeql
But, as $\gamma$ (a probability) is nonnegative, this equation has the unique solution given by the right-hand side in the third line of \qr{L:discrs}.  

Regarding \qr{e:fdadi}, we have
$$\Es  L^\circ  = (1-\gamma)\Ep \big( (1-J_\thec)\cP ^{+} + \gamma (\theMtM-\ca)^+- \CA\big)=0,$$
 by application of \qr{e:survdi}, the first line in \qr{e:theg}, \qr{L:discrs}, and \qr{e:cadi}. Hence, using \qr{e:survdi} again, 
$$\Ep L=(1-\gamma)^{-1}\Es L^\circ  =0.$$
This is the first line in \qr{e:fdadi}, which implies
the following ones by definition of the involved quantities and from the assumption that $\Es\mathcal{F}=0$.~\finproof\\

\noindent
\b{Note that $\theMtM=\Ep\cPs$  also coincides with $\Ep\cP$ (cf.~\qr{L:discrs} and the parenthesis following \qr{e:survdi}). In  practice 
$\cPs$ has less terms than $\cP$ (that also includes cash flows from bank default onward), which is why we favor the formulation $\Ep\cPs$ in \qr{L:discrs}.
The alternative formulation $\Ep\cP$ may seem more in line with the intuition of MtM as value deprived from any credit/funding considerations. However, as the
measure underlying $\Ep$ is the survival one  
(see before Lemma \ref{e:thelem}), 
this intuition is in fact simplistic and only strictly correct in the case without wrong way risk between credit and market (cf. the 
parenthesis preceding Lemma \ref{e:thelem}).}

\subsection{Capital Valuation Adjustment}

Economic capital (\EC) is the level of capital at risk that a regulator would like to see on an economic, structural basis. 
Risk calculations are typically performed by banks
``on a going concern'', i.e.~assuming that the bank itself does not default.
Accordingly:
\bd\label{d:ccrbis}  
The economic capital ($\EC$)  of the bank 
is given by
 the 97.5\% expected shortfall\footnote{See e.g.~\citeN[Section 4.4]{foellmerSchied2016}.} of the 
\b{bank trading loss $L$ under $\Qp $, 
which
we
denote by\footnote{Note that, by definition of $\Qp $, this quantity does not depend on $L^\bullet$.} 
 $\mathbb{ES}(\LOSS)$.}~\finproof\eds

The risk margin (sized by the to-be-defined KVA in our setup) is also loss-absorbing, i.e.~part of capital at risk, and the KVA is originally sourced from the client  (see Assumption \ref{h:selfi}.iii). Hence,
{\it shareholder} capital at risk only consists of the {\it difference} between the (total) capital at risk and the KVA.
Accordingly (and also accounting, regarding \qr{e:kvadiscr}, for the last part in Assumption \ref{h:selfi}.iii):
\bd\label{d:ccrter}   
The capital at the risk (CR) of the bank is given by $\max(\EC,\kva)$  and the ensuing shareholder capital at risk  (SCR) by
\beql{e:scr}\SCR=\max(\EC,\kva)-\kva=(\EC-\kva)^+,\eeql 
where, given some hurdle rate (target return-on-equity) $h$,   
\begin{eqnarray} \label{e:kvadiscr}
&& \kva=\Es\big(h   \SCR^\circ +(1-J)\KVA\big) .~\finproof
\end{eqnarray}
\eds
\b{\brem\label{rem:share} In view of \eqref{e:kvadiscr} and of the last balance condition in \qr{e:reset}, we have
\beql{e:share}   \RM^{sh}=\Es\big(h   \SCR^\circ)  \sp \RM^{bh}=\Es\big( (1-J)\KVA\big).\eeql
We refer the reader to the last bullet point in \citeN[Definition A.1]{AlbaneseCaenazzoCrepey16abx} for the analogous split of RM between shareholder and bondholder wealth in a dynamic, continuous-time setup.~\erem}
\bp\label{p:ccrbis}
We have 
\begin{eqnarray}
\label{e:thiskva}
&& \kva
 =  h \SCR = \frac{h}{1+h}  \EC = \frac{h}{1+h}  \mathbb{ES}(\LOSS)  . 
\end{eqnarray}  
\ep
\proof The first identity  follows  
from Lemma \ref{e:thelem}.  
The resulting KVA semi-linear equation (in view of \qr{e:scr}) is solved similarly to the FVA equation
 \qr{e:thfva}.~\finproof \\

\noindent
The KVA formula \qr{e:thiskva} (as well as its continuous-time analog \qr{e:kconsuite}) can be used either in the direct mode, for computing the KVA corresponding to a given $h$, or in the reverse-engineering mode, 
for defining the 
``implied hurdle rate'' associated with the 
actual 
level on the risk margin account of the bank.
 Cost of capital proxies have always been used to estimate return-on-equity.
The KVA is a refinement, 
fine-tuned for derivative portfolios, but the base return-on-equity concept itself is far older than even the CVA. 
 In particular, the KVA is very useful in
the context of collateral and capital optimization.

\paragraph{KVA Risk Premium and Indifference Pricing Interpretation}\label{ss:kvaindiffp}

The $\ca$ component of the FTP 
corresponds to the expected costs for the 
shareholders
of concluding the deal. This CA component makes the shareholder trading loss $\LOSS$ 
centered (cf.~the first line in
\qr{e:fdadi}).
On top of expected shareholder costs,
the bank charges to the clients a risk margin ($\RM$). 
Assume the bank shareholders endowed with a 
utility function $U$ on $\R$ such that $U(0)=0$.
In a shareholder indifference pricing framework, the
risk margin arises as per the following equation:
\beql{e:bounds}
&
\Es U( J(  \RM  -L) )= \Es U ( 0)=0 
\eeql
(the expected utility of the bank shareholders without the deal),
where 
$$
\Es U(J(  \RM  -L))= \Es \big( JU(   \RM  -L  )\big)=(1-\gamma)\Ep U( \RM  -L),$$
by \qr{e:survdi}. Hence
\beql{e:boundsconcl}
&
\Ep U( \RM  -L)=0 .
\eeql
 The corresponding \RM is interpreted as the minimal admissible
risk margin for the deal to occur, seen from bank shareholders' perspective.

Taking for concreteness $U (-\ell)=\frac{1- e^{\rho \ell}}{\rho}$, for some risk aversion parameter $\rho$,
\qr{e:boundsconcl} yields
$
\RM=\rho^{-1} \ln \Ep e^{\rho L }=\rho^{-1} \ln \Ep e^{\rho L^\circ},
$
by the observation following \qr{e:survdi}.
In the limiting case where the shareholder risk aversion parameter $\rho\to 0$ and $\Ep U (-L)\to -\Ep (L)=0$ (by  the first line in \qr{e:fdadi}), then $\RM\to 0.$

In view of \qr{e:reset} and \qr{e:thiskva},
the corresponding implied KVA and hurdle rate $h$ are such that
\beql{e:endogkva}\kva= \rho^{-1}  \ln \Ep e^{\rho L^\circ} \sp \frac{h}{1+h}= \frac{ \rho^{-1}\ln \Ep e^{\rho L^\circ}}{  \mathbb{ES}(L^\circ)}. \eeql
Hence, ``for $h$ and $\rho$ small'',
\bel h\approx  \frac{{\rm \mathbb{V}ar}(L^\circ)}{2 \mathbb{ES}(L^\circ)} \, \rho\,  \eel
(as $\Ep(L^\circ)=0$), where ${\rm \mathbb{V}ar}$ is the $\Qp$ variance operator.
The hurdle rate $h$ in our KVA setup plays the role of a risk aversion parameter, 
like $\rho$ in the exponential utility framework. 
 
An indifference price has a competitive interpretation. Assume that the bank is competing for the client with other banks. 
Then, in the limit of a continuum of competing banks with a continuum of indifference prices, whenever a bank makes a deal, this can only be at its indifference price. 
Our stylized indifference pricing model of a KVA defined by a constant hurdle rate $h$ exogenizes (by comparison with the endogenous hurdle rate $h$ in \qr{e:endogkva}) the impact on pricing of the competition between banks. It does so in a way that generalizes smoothly to a dynamic setup (see \sr{s:dyn}), as required to deal with a real derivative banking portfolio. It then provides a refined notion of return-on-equity 
for derivative portfolios, where a full-fledged optimization approach would be impractical.

\subsection{Collateral With Clients and Fungibility of Capital at Risk as a Funding Source} \label{rem:compl} 

In case of variation margin (VM) that would be exchanged between the bank and its client, and of initial margin that would be received (RIM) and posted (PIM) by the bank, at the level of, say, some $\Qp $ value-at-risk 
of $\pm(\cP- \VM)$, then 
\begin{itemize}
\item $\cP$ needs be replaced by  $(\cP- \VM-\RIM)$ everywhere in the above, whence an accordingly modified (in principle: diminished) CVA, 
\item an additional initial margin related cash flow in $\mathcal{F}^\circ$ given as $J\gamma\PIM$,  triggering an additional adjustment MVA in CA, where
\beql{e:mvda} 
\MVA= \Es\big( J\gamma\PIM  +(1-J)\MVA \big)=\gamma \PIM;
\eeql
\item additional initial margin related cash flows in $\mathcal{F}^\bullet$ given as $(1-J)(\PIM-\gamma\PIM)$ and $(1-J)\MVA$,  triggering an additional adjustment $\MDA=\MVA$ in CL;
\item the second FVA formula in \qr{L:discrs} modified into $\FVA
= \frac{\gamma}{1+\gamma}(\theMtM- \VM-
\cva-\MVA)^+ .$
\end{itemize}

Accounting further for the additional free funding source provided by capital at risk (cf.~Remark \ref{rem:fun}), then, in view of the specification given in the first sentence of Definition \ref{d:ccrter} for the latter,  one needs replace $(\theMtM-\ca)^\pm$ by $(\theMtM- \VM-\ca-\max(\EC,\kva))^\pm$ everywhere before. This results in the same CVA and MVA as in the bullet points above, but in the following \textit{system} for the random variable $\LOSS$ and the FVA and the KVA numbers (cf.~the corresponding lines in \qr{e:theg}, \qr{L:discrs}, \qr{e:thiskva}, and recall \qr{e:cadi}):
\beql{e:thegmod}
 &\LOSS= J   (1-J_\thec)\cP ^{+}   
+J   \gamma (\theMtM- \VM-\ca-\max(\EC,\kva))^+  +J\gamma\PIM  - J \CA\\
&\FVA=\gamma (\theMtM- \VM-\ca-\max(\EC,\kva))^+\\
& \kva = \frac{h}{1+h}  \mathbb{ES}(\LOSS) .
\eeql 
This system entails a coupled dependence between, on the one hand, the FVA and KVA numbers and, on the other hand, the shareholder loss process $\LOSS$.  However, once CVA, PIM, RIM, and MVA computed as in the above,
the system \qr{e:thegmod}
can be addressed numerically by Picard iteration,
starting from, say, $L^{(0)}=\kva^{(0)}=0$ and 
$\FVA^{(0)}=\frac{\gamma}{1+\gamma}(\theMtM- \VM-\cva-\mva)^+ $ (cf.~the last line in \qr{L:discrs}),
and then iterating in \qr{e:thegmod} until numerical convergence.  
\brem\label{rem:fvanotmva} 
The rationale for funding FVA but not MVA from $\CA+\max(\EC,\kva)$ is set out before Equation (15) in \citeN{AlbaneseCaenazzoCrepey17b}.~\erem
\subsection{Funds Transfer Price}\label{ss:mcl}

The  funds transfer price (all-inclusive XVA rebate to MtM) aligning the deal to shareholder interest (in the sense of a given hurdle rate $h$, cf.~the next-to-last paragraph of Section \ref{ss:bscsm}) is
\beql{e:bsheet-pedagodecompo}
\FTP &= \underbrace{{\rm CVA} +{\rm FVA}}_{\mbox{Expected shareholder costs ${\rm CA}$}} 
+ \underbrace{\rm KVA}_{\mbox{Shareholder risk premium}} \\
& = \underbrace{{\rm {CVA}}-{\rm {DVA}}}_{\mbox{Firm valuation $\FV$}}
+ \underbrace{{\rm {DVA}}+{\rm FDA}}_{\mbox{Wealth transfer ${\rm CL}$}}
+\underbrace{{\rm KVA}}_{\mbox{Shareholder Risk premium}}, 
\eeql 
where all terms are explicitly given in Propositions \ref{p:ccr} and \ref{p:ccrbis} (or the corresponding 
variants of \sr{rem:compl} in the
refined setup considered there).

\paragraph{Wealth Transfer Analysis} The above results implicitly assumed that the bank cannot hedge jump-to-default cash flows (cf.~\sr{ss:con}).  To understand this,
let us temporarily suppose, for the sake of the argument, that the bank would be able to hedge its own jump-to-default through a further deal, whereby the bank would deliver a payment $L^\bullet$ at time 1 in exchange of a
fee fairly valued as
\beql{e:dd} \CL= \Es L^\bullet
=\DVA+\FDA,\eeql
deposited in the reserve capital account of the bank at time 0. 

We include this hedge and assume that the client would now contribute at the level of $\FV=\CA-\CL$ (cf.~\qr{e:fdadi}), instead of $\CA$ before, to the reserve capital account of the bank at time 0. Then the amount that needs be borrowed by the bank for implementing its strategy is still  $\gamma(\theMtM-\CA)^+$ as before (back to the baseline funding setup of Remark \ref{rem:fun}). But
the trading loss of the bank becomes, instead of $L$ before,
\beql{e:thesuite}
\mathcal{C}+\mathcal{F}-\FV+(L^\bullet-\CL)=\mathcal{C}+\mathcal{F}-\CA+L^\bullet =L+L^\bullet  
=\LOSS ,
\eeql
where the last line in \qr{e:fdadi} and the last identity in \qr{e:CFpedago} were used in the first and second equality.
By comparison with the situation from previous sections without own-default hedge by the bank:
\begin{itemize}
\item the shareholders are still indifferent to the deal in expected counterparty default and funding expenses terms, 
\item the recovery of the bondholders becomes zero,
\item the client is better off by the amount $\CA-\FV=\CL$.
\end{itemize}
The CL originating cash flow $L^\bullet$ has been hedged and monetized by the shareholders, who have passed the corresponding benefit to the client. 

Under a cost-of-capital pricing approach, the bank would still charge to its client a KVA add-on $\frac{h}{1+h}\mathbb{ES}(\LOSS) $, as risk compensation for the nonvanishing shareholder trading loss $\LOSS$ still triggered by the deal.
If, however, the bank could also hedge the (zero-valued, by the first line in \qr{e:fdadi}) loss $L^\circ$, hence the totality of $L=L^\circ-L^\bullet$ (instead of $L^\bullet$ only in the above),
then the trading loss 
and the KVA would vanish. As a result, the all-inclusive XVA add-on 
(rebate from MtM valuation)
would boil down to
$$ \FV=\CVA -\DVA$$
(cf. \qr{e:cd}),
the value of counterparty risk and funding to the bank as a whole.
 
%

}

\paragraph{Connection With the Modigliani-Miller Theory}

The Modigliani-Miller invariance result, with \citeN{MM1958} as a seminal reference, consists in various facets of a broad statement that 
the funding and capital structure policies of a firm are irrelevant to the profitability of its investment decisions.
Modigliani-Miller (MM) irrelevance, as we put it for brevity hereafter, was initially seen as a pure arbitrage result.
However, it was later understood that there may be market incompleteness issues with it. So quoting
\citeN[page 9]{DuffieSharer86}, ``generically,
shareholders find the span of incomplete markets a binding constraint [...] shareholders are not indifferent to the financial policy of the firm if
it can change the span of markets (which is typically the case in incomplete markets)''; or 
\citeN[page 197]{Gottardi95}:
``When there are derivative securities and markets are incomplete the
financial decisions of the firm have generally real effects''.

A 
situation where shareholders may ``find the span of incomplete markets a binding constraint'' is 
when market completion 
is legally forbidden. 
This corresponds to the XVA case, which is also at the crossing between market incompleteness and the presence of derivatives pointed out above as the MM non irrelevance case in \citeN{Gottardi95}.
Specifically, the contra-assets and contra-liabilities that emerge
endogenously from the impact of counterparty risk on the derivative portfolio of a bank 
cannot be ``undone'' by shareholders, because jump-to-default risk cannot be replicated by a bank.

As a consequence, MM irrelevance is expected to break down in the XVA setup. In fact, as visible on the trade incremental FTP (counterparty risk pricing) formula
\qr{e:bsheet-pedagodecompo} (cf.~also \qr{e:bsheet-entry-incr-delta-fin} and Proposition \ref{p:wrafinal} in a dynamic and trade incremental setup below), cost of funding and cost of capital are material to banks and need be reflected in entry prices for ensuring shareholder indifference to the trades, i.e.~preserving their hurdle rate throughout trades.

\section{XVA Analysis in a Dynamic Setup}\label{s:dyn}

We now consider a dynamic, continuous-time setup, with model filtration $\gg$ and a (positive) bank default time $\tau$ endowed with an intensity $\gamma$.
The bank
 survival probability measure 
 associated with the 
measure $\Qs$ 
is then the
 probability measure $\Qp $ with $(\gg,\Qs)$ density process $Je^{\int_0^{\cdot}\gamma_s ds}$ (assumed integrable), where $J=\ind_{[0, \tau)}$ is the bank survival indicator process  (cf.~\citeN{Schoenbucher04} and  \citeN{CollinDufresneGoldsteinHugonnier2004}).
In particular, writing
$Y^{\circ}=JY +(1-J)Y_{\tau-},$ 
for any left-limited process $Y$, we have by application of the results of \b{\citeN{CrepeySong15c} (cf.~the condition (A) there):}
\bl\label{e:lemQps}
For every $\Qp $ (resp. sub-, resp. resp. super-) martingale $Y$, the 
process $Y^{\circ}$
is a  $\Qs $ (resp. sub-, resp. resp. super-) martingale.~\finproof
\el
\brem In the dynamic setup, the survival measure formulation is a light presentation, sufficient for the purpose of the present paper (skipping the related integrability issues),
of an underlying reduction of filtration setup, which is detailed in the above-mentioned reference (regarding Lemma \ref{e:lemQps},  cf.~also \citeN[Lemma 1]{CollinDufresneGoldsteinHugonnier2004}).~\erem

\subsection{Case of a Run-Off Portfolio}\label{ss:runoff}

First, we consider the case of a portfolio held on a run-off basis (cf.~\sr{ss:runoffprel}). We denote by $T$ the
final maturity of the portfolio and we assume that all prices and XVAs vanish at time $T$ if $T<\tau$. Then
the results of \citeN{AlbaneseCaenazzoCrepey16abx} show that
all the qualitative insights provided by the one-period XVA analysis of \sr{s:theex}
are still valid. 
The trading loss of the bank
is now given by the process
\beql{e:clpinew}
& 
L
= \cC  +\cF    + \ca  - \ca_0    \eeql 
and the bank 
{\it shareholder} trading loss
by the 
\b{$\Qp$ (hence $\Qs $, by Lemma \ref{e:lemQps})}
martingale
\beql{e:clpinewstopped}
& 
\LOSS
= \cCst+\cFst  + \CA  - \ca_0. \eeql 
In \qr{e:clpinew}-\qr{e:clpinewstopped}, we have $\ca=\cva+\fva$ as in \qr{e:cadi}; the processes $\cC,$ $\cF$, $\cva,$ and $\fva$ are continuous-time processes analogs, detailed in the case of bilateral trade portfolios in \sr{s:cfs}-\ref{ss:basics},
of the eponymous 
quantities in \sr{s:theex} (which were constants or random variables there).
\bp\label{p:wtprel} The core equity tier 1 capital of the bank is given by
\begin{eqnarray} 
\label{e:cL}&&{\rm CET1} ={\rm CET1}_0-L. 
\end{eqnarray}
Shareholder equity
is given by
\beql{e:lalafirst}\SHC 
=\SHC_0-(L +\kva-\kva_0).\eeql\ep
  
\proof  
In the continuous-time setup, Assumption \ref{h:selfi}.i is written as
$$\RC+\RM+\SCR+\UC-\CM -( \RC+\RM+\SCR+\UC-\CM)_0=\cP-(\cC+ \cF+\cH). $$
Given the definition of CET1 in \qr{e:shccet},  the perfect clean hedge condition 
(see after Remark \ref{rem:int}) 
written in the dynamic setup as
$\cP+\mtm - \mtm_0 -\cH=0$, and the balance conditions \qr{e:reset}, this is equivalent to 
$$ \ca+\CET-(\ca+\CET)_0=-(\cC+\cF) .$$
In view of \qr{e:clpinew},
we obtain \qr{e:cL}. 

 As $\SHC=\CET -\RM$
(cf.~\qr{e:shccet}), we have by \qr{e:cL}:
$$
\SHC =\CET_0-L -\RM =\CET_0 -\RM_0-(L  +\RM -\RM_0),$$
which, by the third balance condition in \qr{e:reset}, yields \qr{e:lalafirst}.~\finproof\\

Moreover, by Lemma \ref{e:lemQps},
\b{the continuous-time process $\KVA$ that stems from \qr{e:ecpt}-\qr{e:kcon}}
is
a  $\Qs $
supermartingale with terminal condition $\KVA_{T}=0$ on $\{T<\tau\}$ and drift coefficient
$h\SCR$, 
where SCR is given as in
 \qr{e:scr}, but for $\EC$ there dynamically defined as the time-t conditional, 97.5\% expected shortfall  
 of $(\LOSS_{t+1}-\LOSS_t)$ under $\Qp $,  killed at $\tau$.

\brem\label{rem:beft} It is only before $\tau$ that the right-hand-sides in
the definitions \qr{e:shccet} really deserve the respective interpretations of shareholder equity of the bank and core equity tier 1 capital. Hence, it is only 
the parts of \qr{e:cL} and \qr{e:lalafirst}
stopped before $\tau$, i.e.
\beql{e:explpost}{\rm CET1}^\circ ={\rm CET1}_0-\LOSS\sp \SHC^\circ=\SHC_0-(\LOSS +\KVA-\kva_0),\eeql
which are interesting financially. ~\erems

\subsection{Trade Incremental Cost-of-Capital XVA Strategy}\label{ss:incrx}

In \citeN{AlbaneseCaenazzoCrepey16abx} and in \sr{ss:runoff} above, the derivative portfolio of the bank is assumed held on a run-off basis. By contrast, 
real-life derivative portfolios are incremental. 

Assume a new deal 
shows up at time $\theta \in (0,\tau)$.
We denote by $\Delta{\rm \cdot},$ for any portfolio related process, the difference between the time $\theta$ values of this process
for the run-off versions of the 
portfolio with and without the new deal.
 
\bd\label{p:incr} We apply the following trade incremental pricing and accounting policy:
\begin{itemize}
\item The clean desks pay $\Delta\mtm $ to the client and the CA desks add an amount $\Delta\mtm $ on\footnote{i.e.~remove $(-\Delta\mtm)$ from, if $\Delta\mtm <0$.} the clean margin account;
\item The CA desks charge to the client an amount $\Delta\ca$ and add it on\footnote{i.e.~remove $(-\Delta\ca)$ from, if $\Delta\ca<0$.} the reserve capital account; 
\item The management of the bank charges the amount $\Delta\kva$ to the client and adds it on\footnote{i.e.~removes $(-\Delta\kva)$ from, if $\Delta\kva<0$.}
the risk margin account.~\finproof
\end{itemize} 
\eds
 
%

The funds transfer price of a deal is the all-inclusive XVA add-on charged by the bank to the client in the form of a rebate with respect to the mark-to-market $\Delta\mtm$ of the deal.
Under the above scheme,
the overall price charged to the client for the deal is $\Delta\mtm -\Delta\Theta -\Delta\kva $, i.e.
\beql{e:bsheet-entry-incr-delta-fin} 
\FTP &= \Delta\Theta + \Delta\kva=\Delta\cva+\Delta\fva + \Delta\kva
\\&=\Delta\FV+\Delta\CL + \Delta\kva,
\eeql 
by \qr{e:cadi} and the last line in
\qr{e:fdadi} (which still hold in continuous time, see \b{\citeN[Equations (1) and (66)]{AlbaneseCaenazzoCrepey16abx}})
 applied to the portfolios with and without the new deal. 
 \brem As opposed to the $\Delta\XVA$ terms, which entail portfolio-wide computations, 
$\Delta\mtm$ reduces to the
so-called clean valuation of the new deal,
by trade-additivity of $\mtm$ 
(as follows from 
\b{\citeN[Equations (25) and (37)]{AlbaneseCaenazzoCrepey16abx}}).
\erem

Obviously, the legacy portfolio of the bank
has a key impact on the $\FTP$. It may very well happen that the new deal is risk-reducing with respect to the portfolio, in which case $\FTP <0$, i.e.~the overall, XVA-inclusive price charged by the bank to the client would be $\Delta\mtm -\FTP>\Delta\mtm $
(subject of course to the commercial attitude adopted by the bank under such circumstance).

In order to exclude for simplicity jumps of our $L$ and KVA processes at $\theta$
(the ones related to the initial portfolio, but also those, starting at time $\theta$, corresponding to the augmented portfolio),
we assume a quasi-left continuous model filtration $\gg$ and  
 a $\gg$ predictable stopping time $\theta$. The first assumption excludes that martingales can jump
at predictable times. It is satisfied in all practical models and, in particular, in all models with L\'evy or Markov chain driven jumps. The second assumption
is reasonable regarding the time at which a financial contract is concluded. Note that it was actually already assumed regarding
the (fixed) time 0 at which the portfolio of the bank is supposed to have been set up in the first place.

\bl\label{p:wra}
Assuming the new trade at time $\theta$ handled by the trade incremental policy of Definition \ref{p:incr} after that the balance conditions \qr{e:reset} have been held before $\theta$, 
then shareholder equity $\SHC^\circ$ (see Remark \ref{rem:beft})
is a $\Qs $
submartingale on $[0,\theta]\cap\R_+$,
with 
drift coefficient 
$ h \SCR$ killed at $\tau$. 
\el

\proof   
In the case of a trade incremental portfolio, a priori,
the second identity in \qr{e:explpost} is only guaranteed to hold {\it before} $\theta$.
However, in view of the observation made in Remark \ref{rem:nodelta} and because, under our (harmless) technical assumptions, there can be no dividends arising from the portfolio expanded with the new deal (i.e. jumps in the related processes $L$ and KVA, defined on $[\theta,+\infty)$)
at time $\theta$ itself, the process $\SHC$ does not jump at $\theta$. The process $L$ and KVA related to the legacy portfolio cannot jump at $\theta$ either. As a result, the second identity in \qr{e:explpost} still holds at $\theta$. It is therefore valid
on $[0,\theta]\cap\R_+$.
The result then follows from the respective martingale and supermartingale properties of the (original) processes $\LOSS$ and $\KVA$ recalled before and after Proposition \ref{p:wtprel}.~\finproof\\

The above XVA strategy 
can be iterated between and throughout every new trade. We call this approach the
\textbf{\em trade incremental cost-of-capital XVA strategy}. By an iterated application of Lemma \ref{p:wra} at every new trade, we obtain the following:
\bp\label{p:wrafinal}
Under a dynamic and trade incremental cost-of-capital XVA strategy,  
shareholder equity $\SHC^\circ$ is
a $\Qs $ submartingale on $\R_+$,
with 
drift coefficient 
$ h \SCR$ killed at $\tau$.~\finproof
\ep
 
\noindent
Thus, a trade incremental cost-of-capital XVA strategy results in a sustainable strategy for profits retention, both between and throughout deals, which
was already the key principle behind Solvency II (see \sr{ss:con}).
Note that, without the KVA (i.e.~for $h=0$), the (risk-free discounted) shareholder equity process $\SHC^\circ$  would only be \b{a $\Qs $
martingale,
which could only be acceptable to shareholders without risk aversion (cf.~\sr{ss:kvaindiffp})}.

\subsection{Computational Challenges}

Figure \ref{XVANMC}
yields a picturesque representation, in the form of a corresponding XVA dependence tree, of the continuous-time XVA equations. 
\begin{figure}[h!]
\centering
\includegraphics[scale=0.5]{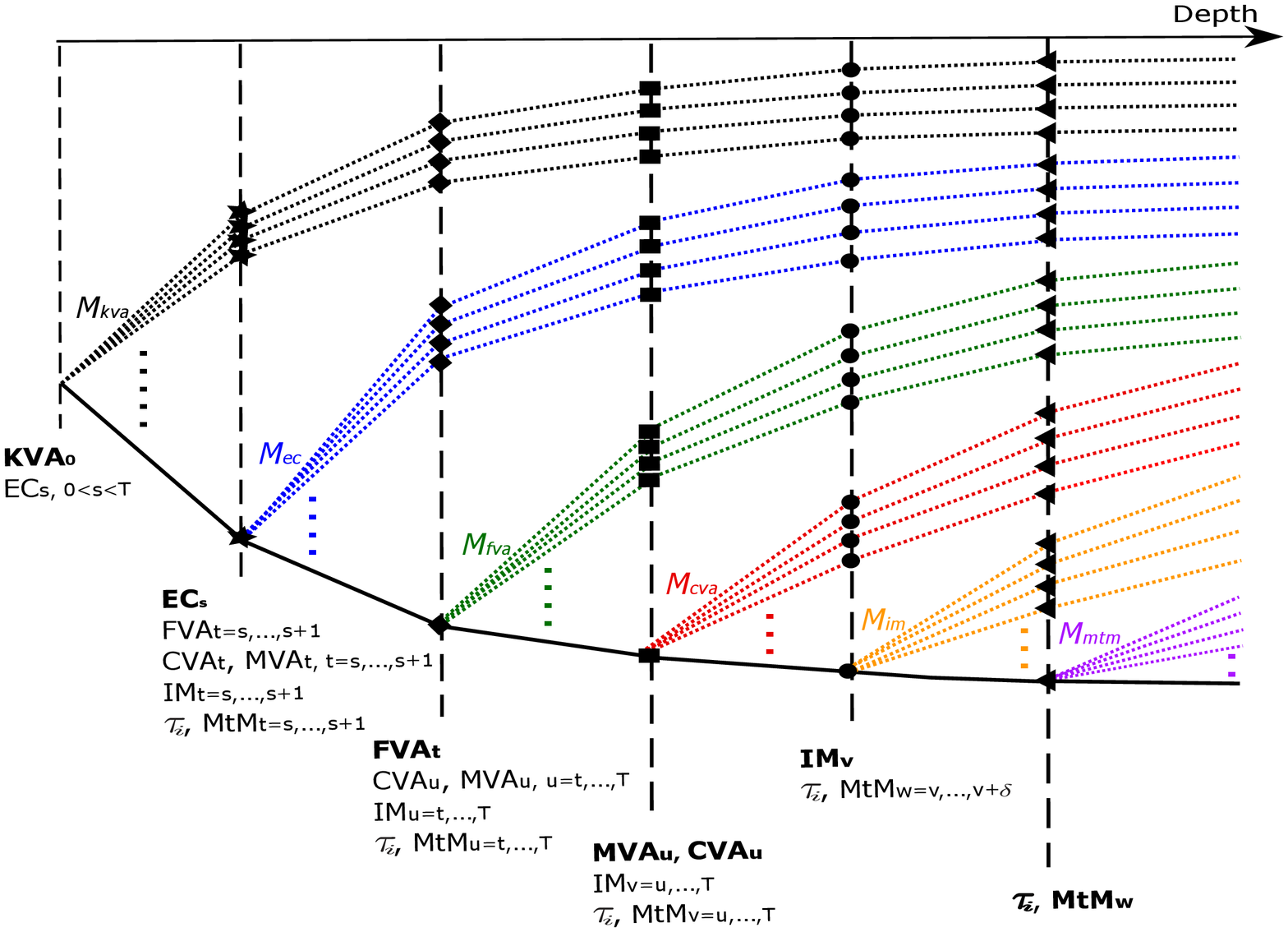}
\caption{The XVA equations dependence tree ({\it Source}: \protect\citeN{AbbasturkiCrepeyDiallo17}).
\label{XVANMC}}
\end{figure}

For concreteness, we restrict ourselves to the case of bilateral trading  in what follows, referring the reader to 
 \citeN[Section 6.2]{ArmentiCrepey16} for the more general and realistic situation of a bank also involved in centrally cleared trading. 
As visible from the corresponding equations in \sr{s:eqns},
the CVA of the bank
 can then be computed as the sum of its CVAs restricted to each netting set (or 
\r{counterparty} 
$i$ of the bank, with default time denoted by $\tau_i$ in Figure \ref{XVANMC}).
The initial margins and the MVA are also most accurately calculated at each netting set level. By contrast, the FVA is defined in terms of a
{semilinear equation} that can only be solved at the level of the overall derivative portfolio of the bank.
{The KVA} can only be computed at the level of the overall portfolio and relies on conditional risk measures
of future fluctuations of the shareholder trading loss process $\LOSS$, which itself {involves future fluctuations of the other XVA processes} (as these are part of the bank liabilities).
  
Moreover, the fungibility of capital at risk with variation margin (cf.~Remark \ref{rem:fvanotmva}) induces a  coupling between, on the one hand, the ``backward'' FVA and KVA processes and, on the other hand, the ``forward'' shareholder loss process $\LOSS$.
As in the static case of  \sr{rem:compl} (cf.~the last paragraph there), the ensuing forward backward system can be decoupled by Picard iteration.

These are heavy computations encompassing all the derivative contracts of the bank. Yet these computations
require accuracy so that trade incremental XVA computations, which are required as XVA add-ons to derivative entry prices (cf.~\sr{ss:incrx}),
are not in the numerical noise of the machinery.

As developed in \citeN[Section 3.2]{AbbasturkiCrepeyDiallo17}, computational strategies for (each Picard iteration of) the XVA equations involve a mix of nested Monte Carlo (NMC) and of simulation/regression schemes, optimally implemented on GPUs.
In view of Figure \ref{XVANMC}, a pure NMC approach would involve five nested layers of simulation (with respective numbers of paths $M_{xva}\sim \sqrt{M_{mtm}}$, see \citeN[Section 3.3]{AbbasturkiCrepeyDiallo17}). 
Moreover, nested Monte Carlo implies intensive repricing of the mark-to-market cube, i.e. pathwise MtM
valuation for each netting set,  or/and high dimensional interpolation. 
In this work, we use
no nested Monte Carlo or conditional repricing of future MtM cubes: beyond the base MtM layer in the XVA dependence tree, each successive layer  (from right to left in Figure \ref{XVANMC}, at each Picard iteration) will be ``learned" instead.


\subsection{Deep (Quantile) Regression XVA Framework}\label{ss:xvaframew}

We denote by  $\Ep_t$,  $\VaR_t$, and $\ES_t$
(and simply, in case $t=0$, $\Ep$, $\VaR$, and $\ES$) 
the time-$t$ conditional expectation, value-at-risk, and expected shortfall
with respect to the bank survival measure $\Qp$.

We compute the mark-to-market cube 
using CUDA routines. 
The pathwise XVAs are obtained by deep learning regression, i.e. extension of \citeN{long} kind of schemes to deep neural network regression bases as also considered in \citeN{HurePhamWarin19} or \citeN{BeckBeckerCheriditoJentzenNeufeld19}, 
based on the classical quadratic (also known as mean square error, MSE) loss function.
The conditional value-at-risks and expected shortfalls involved in the embedded 
pathwise EC and IM computations are obtained by deep quantile regression, as follows.

Given features $X$ and labels $Y$ (random variables),
we  want to compute the conditional value-at-risk and expected shortfall functions $\qpr (\cdot)$ and $\epr  (\cdot)$ such that $\VaR(Y|X)=\qpr (X)$ and $\ES(Y|X)=\epr  (X)$.
Recall from \citeN{fiss:zieg:gnei:15} 
and \citeN{fissler2016} that value-at-risk is $\textit{elicitable}$, expected shortfall is not, but their pair is \textit{jointly elicitable}. Specifically,
 we consider loss 
functions $\rho$ of the form 
(where in our notation $Y$ is a signed loss, whereas it is a signed gain  in their paper)
{\def\z{\theq}\def\hun{f}\def\hde{g}\def\hun{f}\def\theq{q(X)}\def\thee{e(X)}\def\thee{s(X)}
\beql{e:lossvares}
&\rho_\alpha ( q(\cdot),s(\cdot);X,Y)= (1-\alpha)^{-1}\left(\hun(Y) -\hun(\z) \right)^+ + \hun(\z)  +\\
& \qqq   \hde  (\thee)   -  \dot{\hde}(\thee) \left( \thee-\theq  - (1-\alpha)^{-1} (Y - \theq)^{+}   \right) .
\eeql}

One can show (cf.~also \citeN{dimitriadis2017})  that, for a suitable choice of the functions $f$, $g$  including 
$f(z)=z$ and $g=-\ln(1+e^{-z})$
 (our choice in our numerics), the pair of the conditional value-at-risk and expected shortfall functions
%
is 
the minimizer, over all measurable pair-functions $(q(\cdot),\epr(\cdot))$, of the error
\beql{e:theopt} \Ep\rho{}( q(\cdot),\epr(\cdot);X,Y).\eeql 
  
In practice, 
one minimizes numerically the error
\qr{e:theopt}, based on $m$ independent simulated values of $(X,Y)$,
over a 
parametrized family of functions $(q,\epr)(x)\equiv (q,\epr)_{\theta}(x)$.
\citeN{dimitriadis2017} restrict themselves to multilinear functions.
In our case we  use a feedforward neural network
parameterization  (see e.g.~\citeN{GoodfellowBengioCourville2017}).
The minimizing pair $(q,\epr)_{\hat\theta}$ then
represents the two scalar 
neural network 
approximations of the conditional value-at-risk and expected shortfall functions pair.

The left and right panels of
Figure \ref{fig:XVANNES} show the respective deep neural networks for pathwise 
value-at-risk/expected shortfall (with error \qr{e:theopt})
and pathwise XVAs (with classical quadratic norm
 error). 
Deep learning methods often show
particularly good generalization and scalability performances (cf.~\sr{s:timings}).
In the case of conditional value-at-risk and expected shortfall computations, deep learning quantile regression is also easier to implement 
than more naive methods, such as the
resimulation and sort-based scheme of \citeN{BarreraCrepeyDialloFortGobetStazhynski17}  for the value-at-risk and expected shorfall at each outer node of a nested Monte Carlo simulation.
\begin{figure}[h!]
\centering
\includegraphics[viewport=50 100 750 500,width=6.625cm,clip]{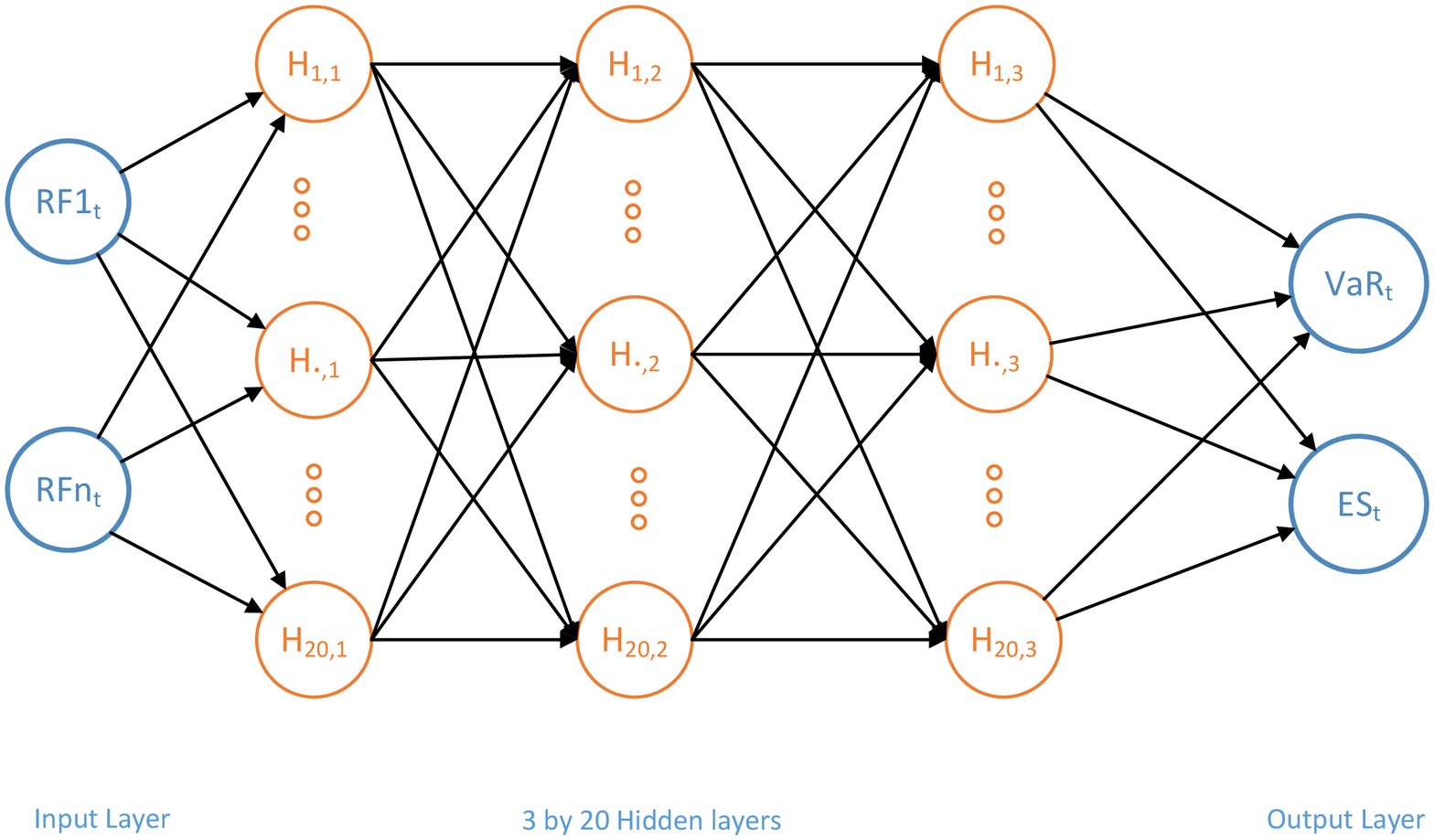}\includegraphics[viewport=50 100 750 500,width=6.625cm,clip]{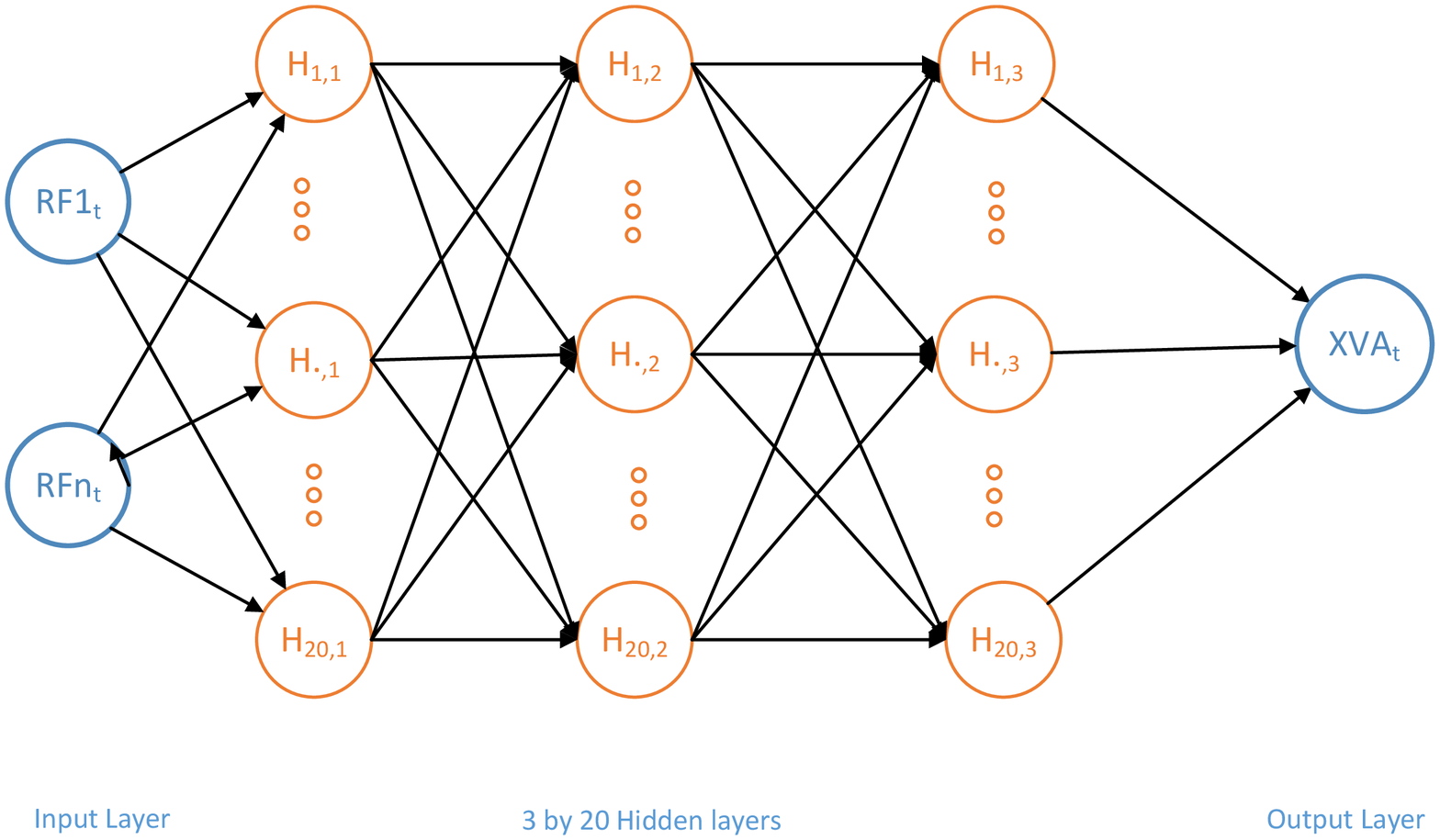}
\caption{Neural networks with state variables (realizations of the risk factors at the considered pricing time) as features. \textit{(Left)} Joint value-at-risk/expected shortfall neural network: 
output is joint estimate of pathwise conditional 
value-at-risk and expected shorfall, at a selected confidence level,
of the label (inputs to initial margin or economic capital)
given the features. 
\textit{(Right)} XVAs
neural network: 
output is estimate of pathwise conditional mean of the label (XVA generating cash flows) given the features.}
\label{fig:XVANNES}
\end{figure}


The neural network topology and hyper-parameters used \r{by default} in our examples are detailed in Table \ref{tab:ANNhypers}. We use hyperbolic tangent activation functions in all cases.
 \begin{table}[!htbp]
\footnotesize
  \centering
    \begin{tabular}{|llllllll|}
    \hline
          & CVA   & FVA   & IM & MVA   & Gap CVA\footnote{See \sr{s:num}.} & EC & KVA \bigstrut\\
    \hline
    Hidden Layers & \multicolumn{1}{l}{3} & \multicolumn{1}{l}{5} & \multicolumn{1}{l}{3} & \multicolumn{1}{l}{3} & \multicolumn{1}{l}{3} & \multicolumn{1}{l}{3} & \multicolumn{1}{l|}{3} \bigstrut[t]\\
    Hidden Layer Size & \multicolumn{1}{l}{20} & \multicolumn{1}{l}{6} & \multicolumn{1}{l}{20} & \multicolumn{1}{l}{20} & \multicolumn{1}{l}{20} & \multicolumn{1}{l}{20} & \multicolumn{1}{l|}{20} \\
    Learning Rate & \multicolumn{1}{l}{0.025} & \multicolumn{1}{l}{0.025} & \multicolumn{1}{l}{0.05} & \multicolumn{1}{l}{0.1} & \multicolumn{1}{l}{0.1} & \multicolumn{1}{l}{0.025} & \multicolumn{1}{l|}{0.1} \\
    Momentum & \multicolumn{1}{l}{0.95} & \multicolumn{1}{l}{0.95} & \multicolumn{1}{l}{0.5} & \multicolumn{1}{l}{0.5} & \multicolumn{1}{l}{0.5} & \multicolumn{1}{l}{0.95} & \multicolumn{1}{l|}{0.5} \\
    Iterations & \multicolumn{1}{l}{100} & \multicolumn{1}{l}{50} & \multicolumn{1}{l}{150} & \multicolumn{1}{l}{100} & \multicolumn{1}{l}{100} & \multicolumn{1}{l}{100} & \multicolumn{1}{l|}{100} \\
	Loss Function & MSE& MSE & \eqref{e:lossvares} & MSE & \eqref{e:lossvares} & \eqref{e:lossvares} & MSE \\
    Application & netting set & portf. & netting set & netting set & netting set & portf. & portf. \bigstrut[b]\\
    \hline
    \end{tabular}
  \caption{Neural network topology and learning parameters used by default in our numerics (portf. $\equiv$ overall derivative portfolio of the bank).}
  \label{tab:ANNhypers}
\end{table} 
Algorithm \ref{algo:case1} yields our fully (time and space) discrete scheme for simulating the Picard iteration
 \qr{e:prat-csa} until numerical convergence to the XVA processes. Note that, as opposed to more rudimentary, expected exposure
 based 
XVA computational approaches (see Section 1 in \citeN{AbbasturkiCrepeyDiallo17}), this algorithm requires the simulation of the 
\r{counterparty}
 defaults.
\begin{algorithm}[h!]
\begin{itemize} 
\item Simulate forward $m$ realizations (Euler paths) of the  market risk factor processes and of the 
\r{counterparty}
 survival indicator processes (i.e. default times) on a refined time grid;
\item 
 For each pricing time $t=t_i$ of a pricing time grid, with coarser time step denoted by $h$, and for each \r{counterparty}
 $c$:
\begin{itemize} 
\item Learn the corresponding $\VaR_{t}$ and $\ES_{t}$ terms visible in 
\qr{e:gapri} or (under the time-discretized outer integral in) \qr{e:instrucsa}; 
\item Learn the corresponding $\Ep_{t}$ terms visible in 
\qr{e:instrunocsa} through \qr{e:instrumva};
\item Compute the ensuing pathwise CVA and MVA as per \qr{e:instrunocsa}--\qr{e:instrumva};
\end{itemize}
\item 
For FVA$^{(0)}$, consider the following time discretization of \qr{e:fvainit} (in which $\lambda$ is the risky funding spread process of the bank) with time step $h$:
\beql{e:fvaz}
&   {\rm {FVA}}^{(0)}_t \approx\Ep_t[
{\rm {FVA}}^{(0)}_{t+h}] +h \lambdabar_t\Big(\sum_c J^c_t  (P^c_t-\VM^c_t) 
-\cva_t -\mva_t-\fva_t^{(0)}
  \Big)^+
\eeql
and, for each $t=t_i$,
 learn the corresponding $\Ep_{t}$ in \qr{e:fvaz},
then solve the semi-linear equation for ${\rm {FVA}}^{(0)}_t$;
\item 
For each Picard iteration $k$ (until numerical convergence), simulate forward $L^{(k)}$ as per the first line in \qr{e:prat-csa} (which only uses known or already learned quantities), and:
\begin{itemize}
\item For economic capital EC$^{(k)}$,  for each $t=t_i$, learn 
$\ES_{t} \big( (L^{(k)})^\circ_{t+1}-  (L^{(k)})^\circ_{t} \big)$
(cf.~Definition \ref{defi:ec});
\item KVA$^{(k)}$ and FVA$^{(k)}$ then require a backward recursion solved by deep learning approximation much like the one for FVA$^{(0)}$ above.
\end{itemize}
\end{itemize}
\caption{Deep XVAs algorithm. \label{algo:case1}}
\end{algorithm}

\section{Swap Portfolio Case Study}\label{s:num}

We consider an interest rate swap portfolio case study with 
\r{counterparties}
 in different economies, first involving 
10 one-factor Hull White interest-rates, 9 Black-Scholes exchange rates,  and 11
Cox-Ingersoll-Ross default intensity processes.
The default times of the 
\r{counterparties}
and the bank itself are jointly modeled 
by a ``common shock'' or dynamic Marshall-Olkin copula model
as per
\citeN[Chapt.~8--10]{BieleckiBrigoCrepeyHerbertsson13} and 
\citeN{CrepeySong15FS} (see also \citeANP{Elouerkhaoui07} (\citeyearNP{Elouerkhaoui07}, \citeyearNP{Elouerkhaoui17})).
This whole setup results in about 40 risk factors used as deep learning features (including the 
\r{counterparty}
 default indicators).

In this model we consider a bank portfolio of 10K randomly generated swap trades, with
\begin{itemize}  
  \item trade currency and counterparty both uniform on $[1,2,3\ldots,10]$,
  \item notional uniform on $[10K,20K,\ldots,100K]$,
   \item collateralization  (cf.~\sr{ss:collat}): either ``no CSA \r{counterparty}''
without initial margin (IM) nor variation margin (VM), or 
``CSA \r{counterparty}''
with $\VM=\mtm$ and posted initial margin
  (PIM) pledged at $99\%$ gap risk value-at-risk,
received initial margin (RIM) covering $75\%$ gap risk and leaving excess as residual gap CVA,
\item for economic capital, $97.5\%$ expected shortfall of 1-year ahead trading loss of the bank shareholders.  
\end{itemize}
By default we use
Monte Carlo simulation with 
50K paths of 16 coarse (pricing) and 32 fine (risk factors) time steps per year.

\subsection{Validation Results}

The validation of our deep learning methodology is done in the setup of a portfolio of swaps issued at par, with final maturity $T=10$ years, 
without initial margin (IM) nor variation margin (VM). 

We  first focus on the CVA, as the latter is amenable to
validation by a standard nested Monte Carlo (``NMC'') methodology.
Figures \ref{fig:nmc_hist}, \ref{fig:nmc_qqplot} and \ref{fig:mse_nmc_vs_learned} show that the learned CVA is consistent with that obtained from a nested Monte Carlo simulation. 
Regarding Figure \ref{fig:mse_nmc_vs_learned} (and also later below), note the equivalence of optimising the mean quadratic error
\begin{itemize}
  \item  between the ANN learned estimator $h\left(X\right)$ and the labels $Y$ (``MSE''),
$\mathbb{E}\left[\left(h\left(X\right)-Y\right)^2\right]$, and
\item between the ANN learned estimator and the conditional expectation $\mathbb{E}\left[Y\left|X\right.\right]$ (in our case estimated by NMC), $\mathbb{E}\left[\left(h\left(X\right)-\mathbb{E}\left[Y\left|X\right.\right]\right)^2\right]$.
\end{itemize}
The equivalence stems from the following identities, which hold 
for any random variables $X$, $Y$ and hypothesis function $h$ such that $Y$ and $h\left(X\right)$ are  square integrable:
\begin{equation}
\begin{aligned}\label{eq:mseacc}
&\mathbb{E}\left[\left(h\left(X\right)-Y\right)^2\right]=\mathbb{E}\left[\left(h\left(X\right)-\mathbb{E}\left[Y\left|X\right.\right]\right)^2\right]+\mathbb{E}\left[\left( \mathbb{E}\left[Y\left|X\right.\right]-Y\right)^2\right]\\&\qqq\qqq\qqq\qqq+2\mathbb{E}\big[\left(h\left(X\right)-\mathbb{E}\left[Y\left|X\right.\right]\right) \left( \mathbb{E}\left[Y\left|X\right.\right]-Y\right)\big] \\ 
&\qqq=\mathbb{E}\left[\left(h\left(X\right)-\mathbb{E}\left[Y\left|X\right.\right]\right)^2\right]+\mathbb{E}\left[{\rm \mathbb{V}ar}\left(Y\left|X\right.\right)\right]
\end{aligned}
\end{equation}
(as the second line vanishes), where $\mathbb{E}\left[{\rm \mathbb{V}ar}\left(Y\left|X\right.\right)\right]$ does not depend on $h$.
\par
The CVA error profile on Figure \ref{fig:mse_nmc_vs_learned} reveals slightly more difficulty in learning the earlier CVAs. 
This is  because of a higher variance of the  corresponding cash flows (integrated over longer time frames) in conjunction with a lower variance of the features (risk factors diffused over shorter time horizons).

\begin{figure}[!htbp]
\centering
\includegraphics[width=6.625cm]{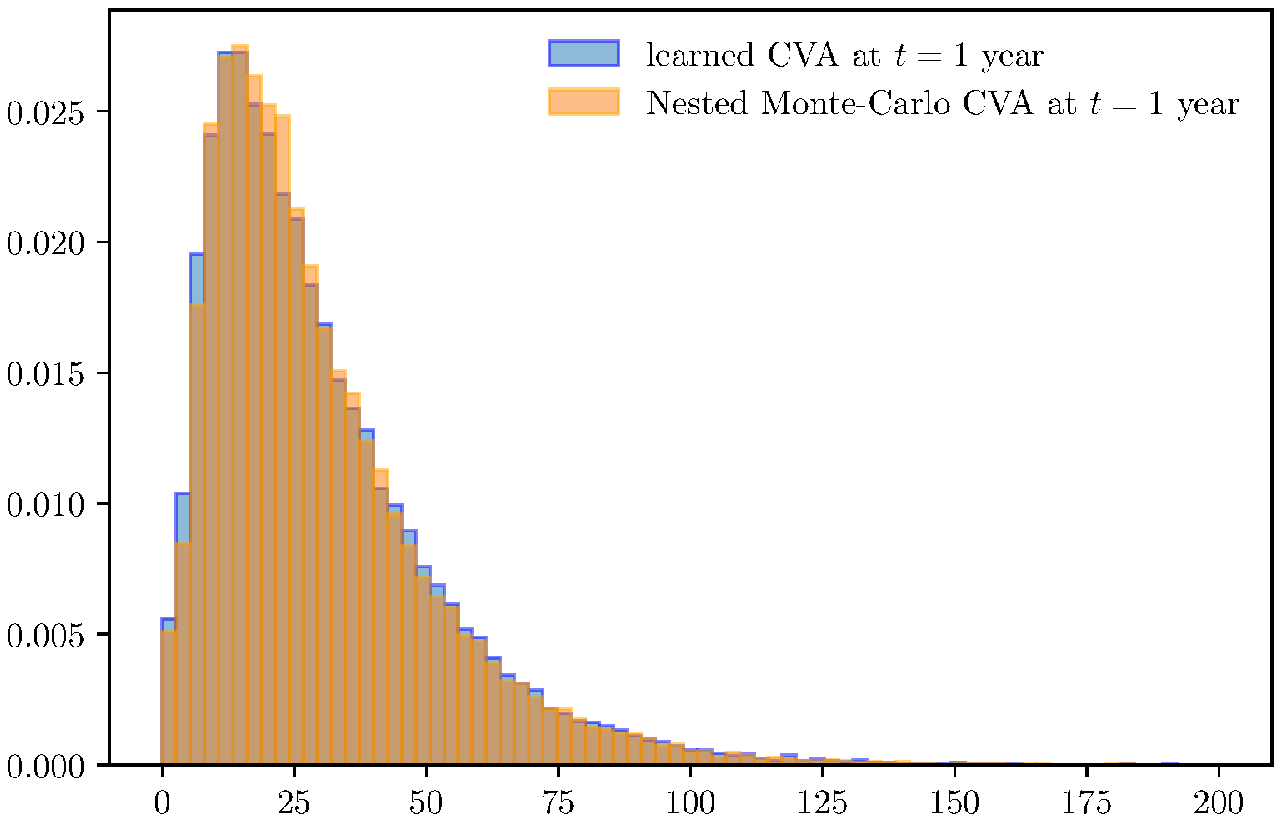}
\includegraphics[width=6.625cm]{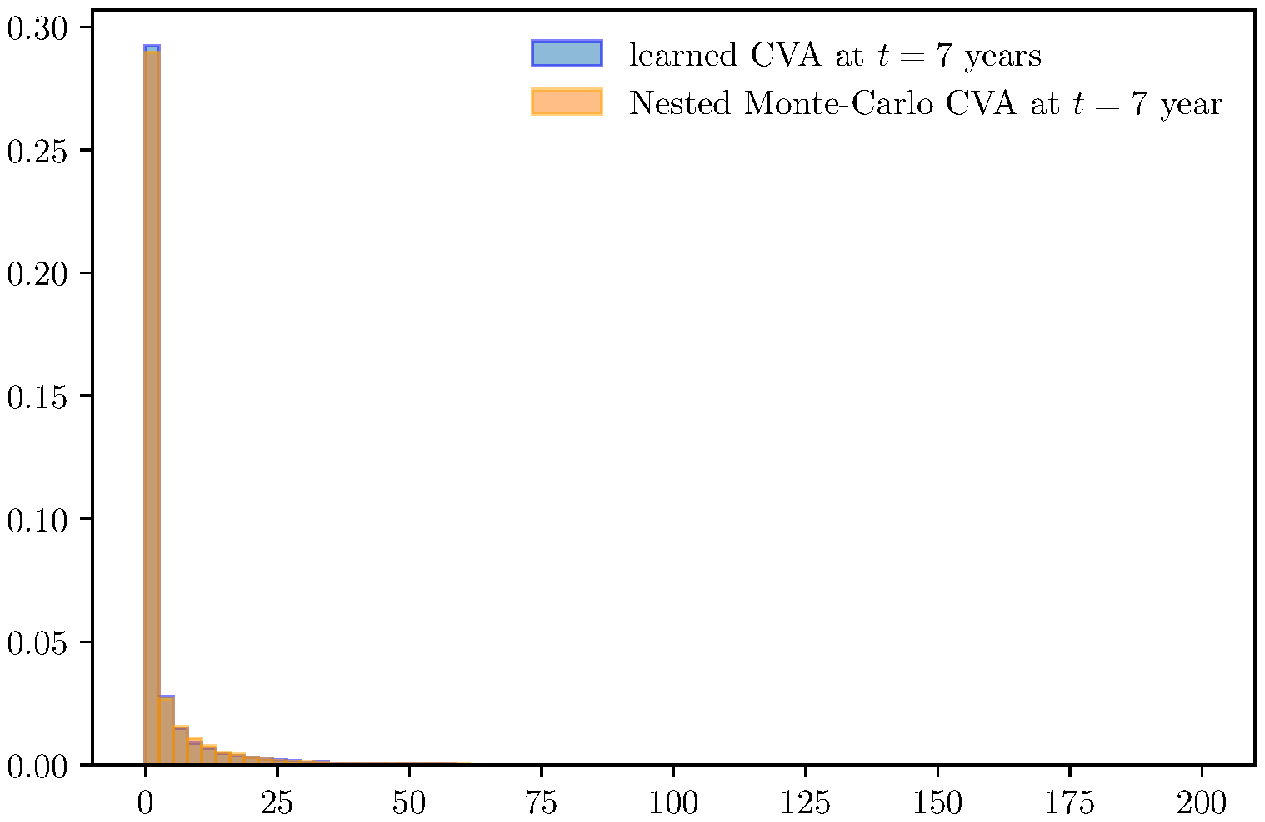} 
\caption{Random variables CVA$^c_1$ and CVA$^c_7$ (in the case of a no CSA netting set $c$, 
 respectively observed after 1 and 7 years) obtained by learning (blue histogram) versus nested Monte Carlo (orange histogram). All histograms are based on out-of-sample paths.}
\label{fig:nmc_hist}
\end{figure}

\begin{figure}[!htbp]
\centering
\includegraphics[width=6.625cm]{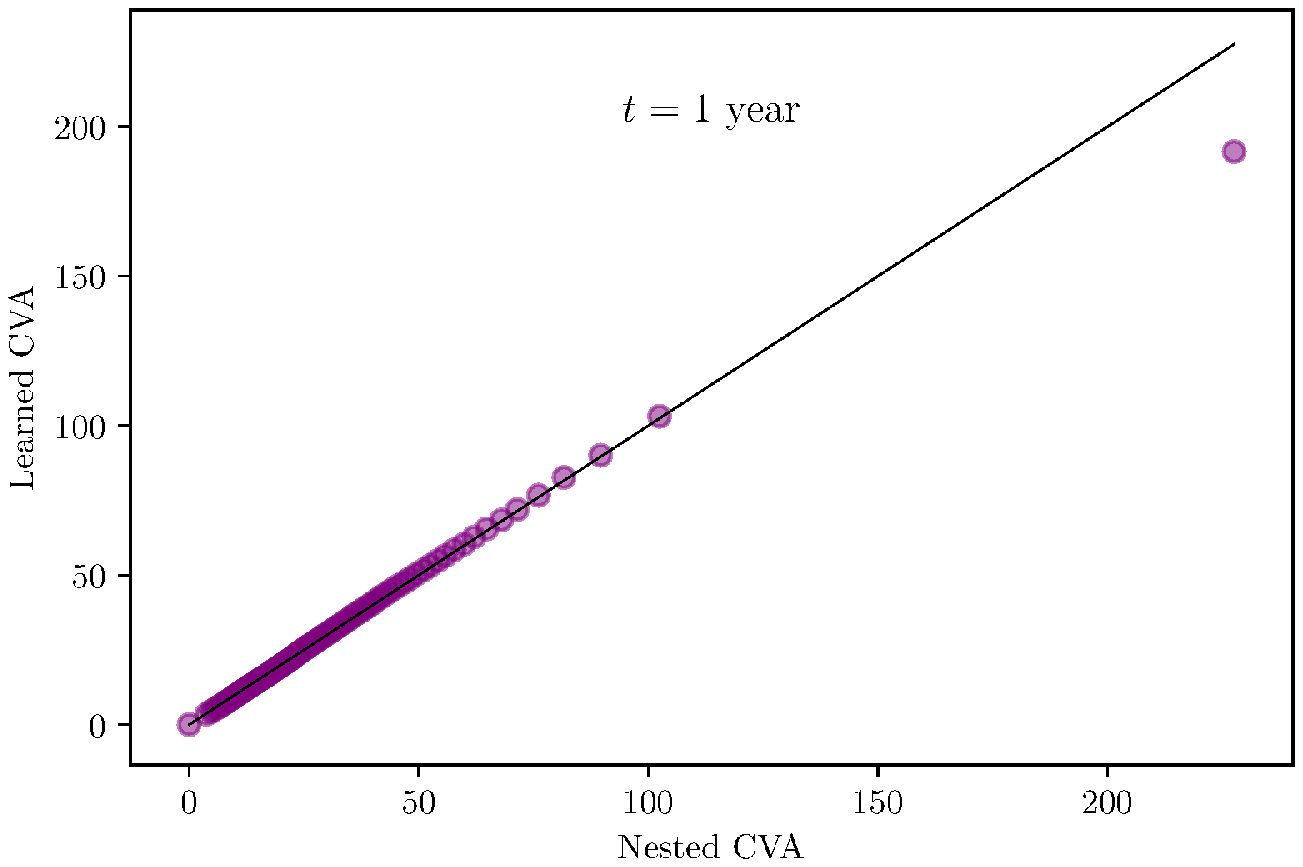}
\includegraphics[width=6.625cm]{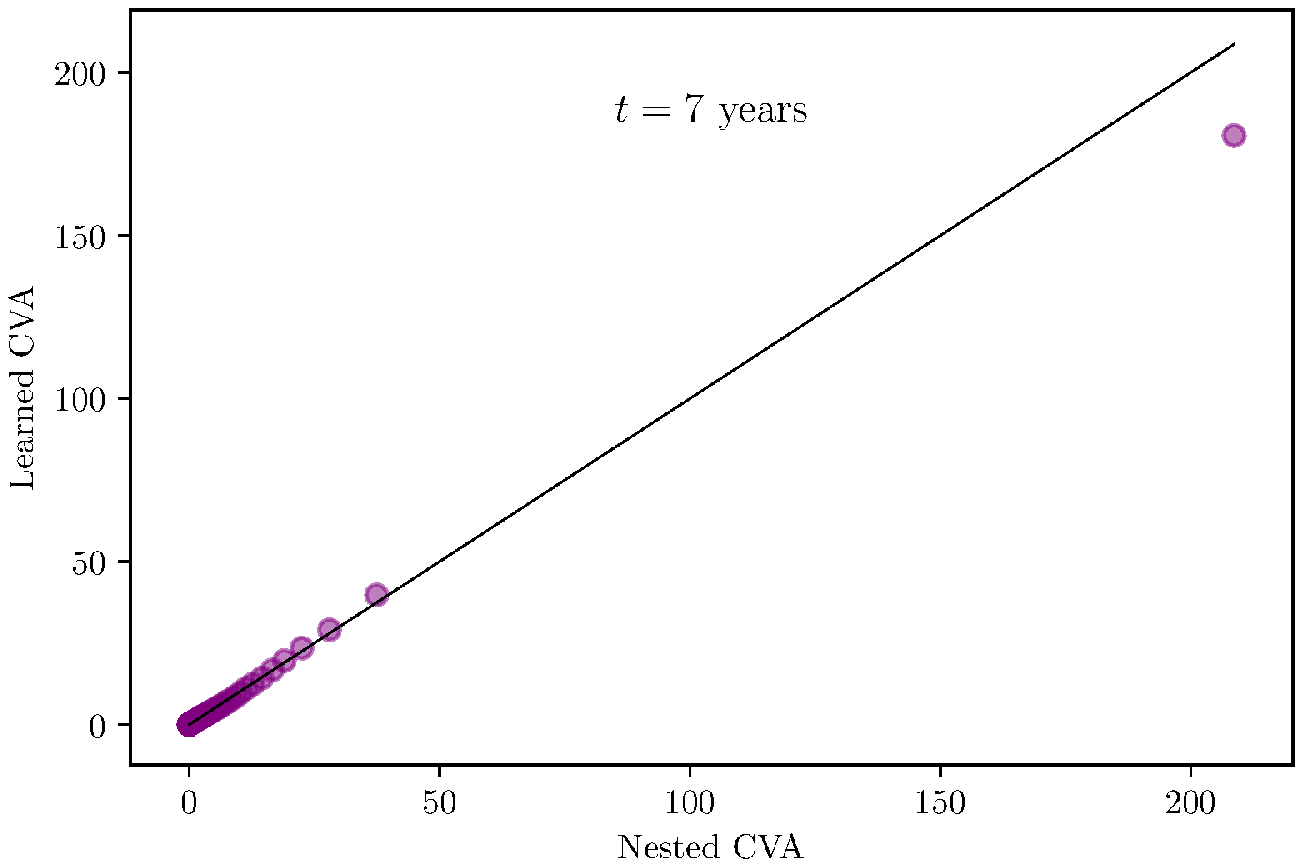} 
\caption{QQ-plot of learned versus nested Monte Carlo CVA for the random variables CVA$^c_1$ ({\em left}) and CVA$^c_7$ ({\em right}). Paths are out-of-sample.}
\label{fig:nmc_qqplot}
\end{figure}

\begin{figure}[!htbp]
\centering
\includegraphics[width=\textwidth]{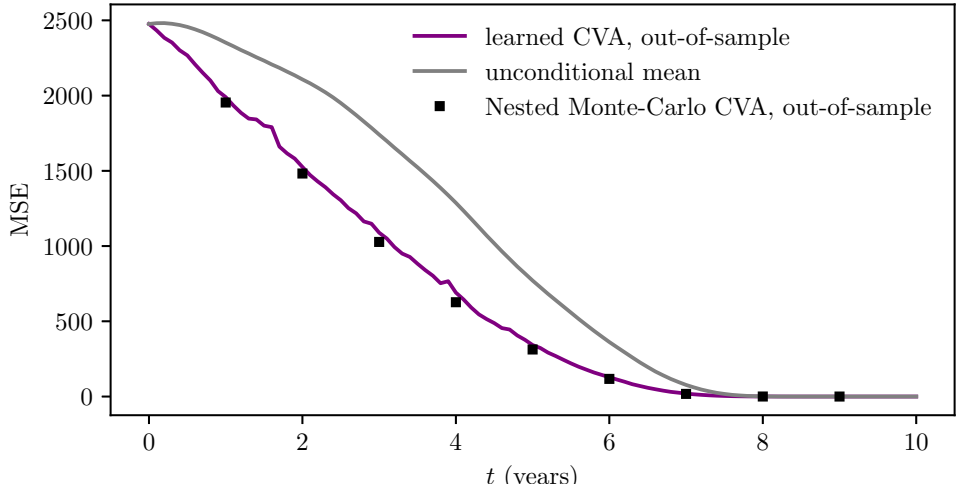}
\caption{Empirical quadratic loss of each CVA estimator at all coarse time-steps. The lower, the closer to the true conditional expectation (cf.~\eqref{eq:mseacc}). Since the nested Monte Carlo method is computationally expensive, it was carried out only once every 10 coarse time-steps.}
\label{fig:mse_nmc_vs_learned}
\end{figure}

Table \ref{tbl:acc_nmc} shows the computational cost and accuracy of the nested Monte Carlo method for different number of inner paths, using 32768 outer paths.
The convergence is already achieved for approximately 128 inner paths, in line with the NMC square root rule that is recalled in an XVA setup in \citeN[Section 3.3]{AbbasturkiCrepeyDiallo17}.
 Figure \ref{fig:speed_vs_acc} and Table \ref{tbl:acc_training_vs_nmc} show that a good accuracy can be achieved through learning at a lower computational cost than through nested Monte Carlo, while also enjoying the advantages of the approach being parametric. Indeed, once the CVA is learned, one would pay only the cost of inference later on, which is generally negligible compared to training time. By contrast, a nested Monte Carlo approach would require to relaunch the nested simulations every time the CVA estimator is needed on new paths. Early stopping could be used to help reduce training time further while improving regularization.
\begin{table}[!htbp]
\centering
\begin{tabular}{rrr}
\toprule
 \# of inner paths &    MSE (vs labels) &  Computational time (seconds) \\
\midrule
                     2 &  0.523 &                     37.562 \\
                     4 &  0.427 &                     37.815 \\
                     8 &  0.393 &                     37.819 \\
                    16 &  0.370 &                     38.988 \\
                    32 &  0.360 &                     40.707 \\
                    64 &  0.353 &                     57.875 \\
                   128 &  0.348 &                    157.536 \\
                   256 &  0.349 &                    301.406 \\
                   512 &  0.348 &                    584.475 \\
                  1024 &  0.348 &                   1213.756 \\
\bottomrule
\end{tabular}
\caption{Accuracy and computation times for the estimation of a CVA at a given coarse time-step using the nested Monte Carlo procedure. The MSE here is the mean quadratic error between the nested Monte Carlo estimator and the labels, and hence quantifies how well it is doing as a projection.}
\label{tbl:acc_nmc}
\end{table}

\begin{figure}[!htbp]
\centering
\includegraphics[width=\textwidth]{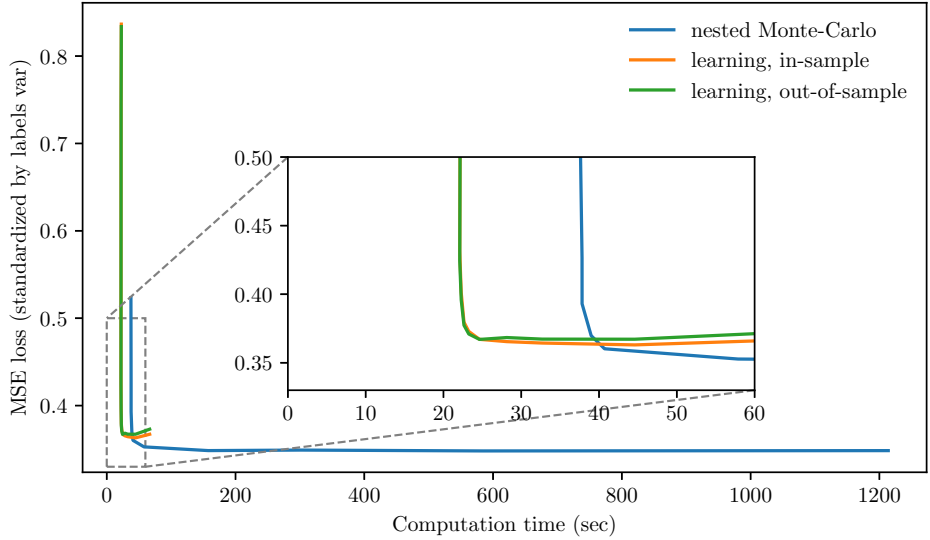}
\caption{Speed versus accuracy in the case of a CVA at a given pricing time. We kept varying the number of inner paths for the nested Monte Carlo estimator and the number of epochs for the learning approach and recorded the computation time and the empirical quadratic loss.}
\label{fig:speed_vs_acc}
\end{figure}

\begin{table}[!htbp]
\centering
\begin{tabular}{lrrrr}
\toprule
{} &  MSE (vs NMC CVA) &  MSE (vs labels) &  Simulation time &  Training time \\
\# of epochs &                      &                     &                               &                \\
\midrule
1           &                0.977 &               0.979 &                        21.992 &          0.880 \\
2           &                0.729 &               0.729 &                        21.992 &          0.434 \\
4           &                0.423 &               0.425 &                        21.992 &          0.524 \\
8           &                0.399 &               0.401 &                        21.992 &          0.719 \\
16          &                0.371 &               0.369 &                        21.992 &          1.088 \\
32          &                0.369 &               0.365 &                        21.992 &          1.800 \\
64          &                0.370 &               0.363 &                        21.992 &          3.243 \\
128         &                0.371 &               0.363 &                        21.992 &          6.227 \\
256         &                0.370 &               0.361 &                        21.992 &         10.883 \\
512         &                0.370 &               0.362 &                        21.992 &         20.096 \\
1024        &                0.371 &               0.362 &                        21.992 &         39.338 \\
\bottomrule
\end{tabular}
\caption{Accuracy and computation times (in sec) for the calculation of a CVA at a given coarse time-step using the learning approach. MSE against NMC CVA is the mean quadratic error between the learned CVA and a CVA obtained using a nested Monte Carlo with 512 inner paths, while MSE against labels designates the mean quadratic error between the learned CVA and the labels that were used during training and thus quantifies how well it is doing as a projection. Both errors are respectively normalized by the variances of the nested Monte Carlo estimator and of the labels. The paths used here are out-of-sample.}
\label{tbl:acc_training_vs_nmc}
\end{table}

More generally, in the presence of a multiple number of XVA layers (cf.~Figure \ref{XVANMC}), a purely nested Monte Carlo approach would require multiple layers of nested simulations, which would amount to a computational time that is exponential in the number of XVA layers, while the computational complexity for the learning approach is linear.

As with mainstream interpolation (as opposed to regression in our case) learning problems, a good architecture is key to better learning and hence better approximation of our XVA metrics. As expected, increasing the model capacity reduces the in-sample error as shown in the bottom panel of Figure \ref{fig:learning_curve}. Although fine-tuning in our case suggests a single layer yields the best out-of-sample performance for the CVA, a standard guess such as 3 layers can also be considered good enough as shown in the top panel. Of course such conclusions may depend on the complexity of the portfolio and the number of counterparties and risk factors.

\begin{figure}[!htbp]
\centering
\includegraphics[width=0.9\textwidth]{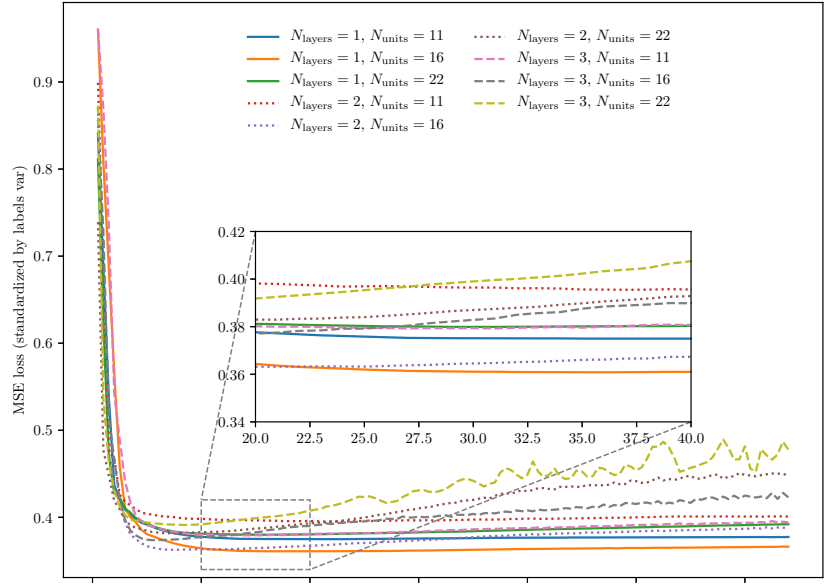}\\\vspace{-1.2cm}
\includegraphics[width=0.9\textwidth]{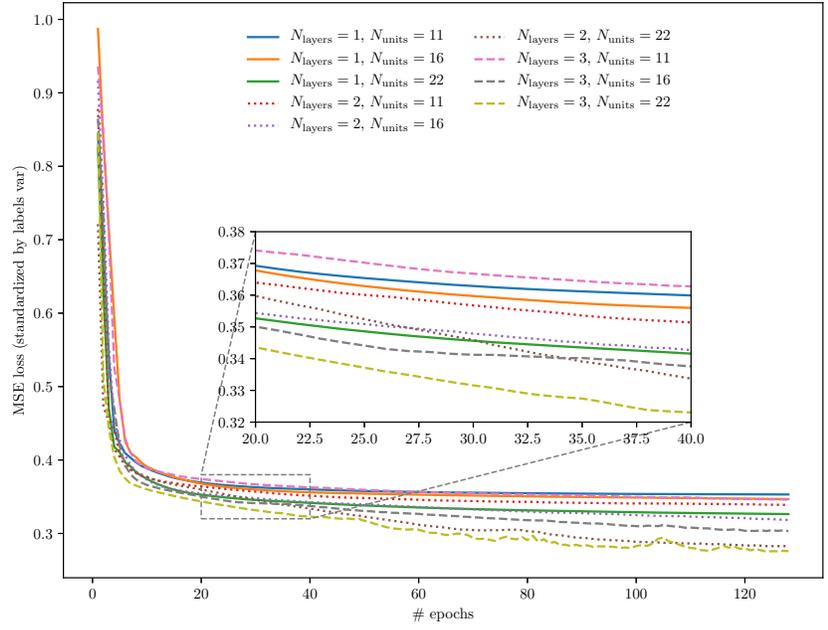}
\caption{Empirical quadratic loss during CVA learning at time-step $t=5\ \text{years}$, standardized by the variance of the labels. {\em (Bottom)} Paths are in-sample.  {\em (Top)} Paths are out-of-sample.}
\label{fig:learning_curve}
\end{figure}
%

 Figure \ref{fig:weightsreuse} shows the learned FVA$^{(0)}$ profile as per \qr{e:fvaz}.
 The orange FVA curve represents the mean FVA originating cash flows, which, in principle as on the picture, matches the blue mean FVA itself learned from these cash flows. The 5th and 95th percentiles FVA estimates are a bit less smooth in time then the mean profiles, as expected.  
 \begin{figure}[h!] 
 \centering 
 \includegraphics[width=\textwidth,clip]{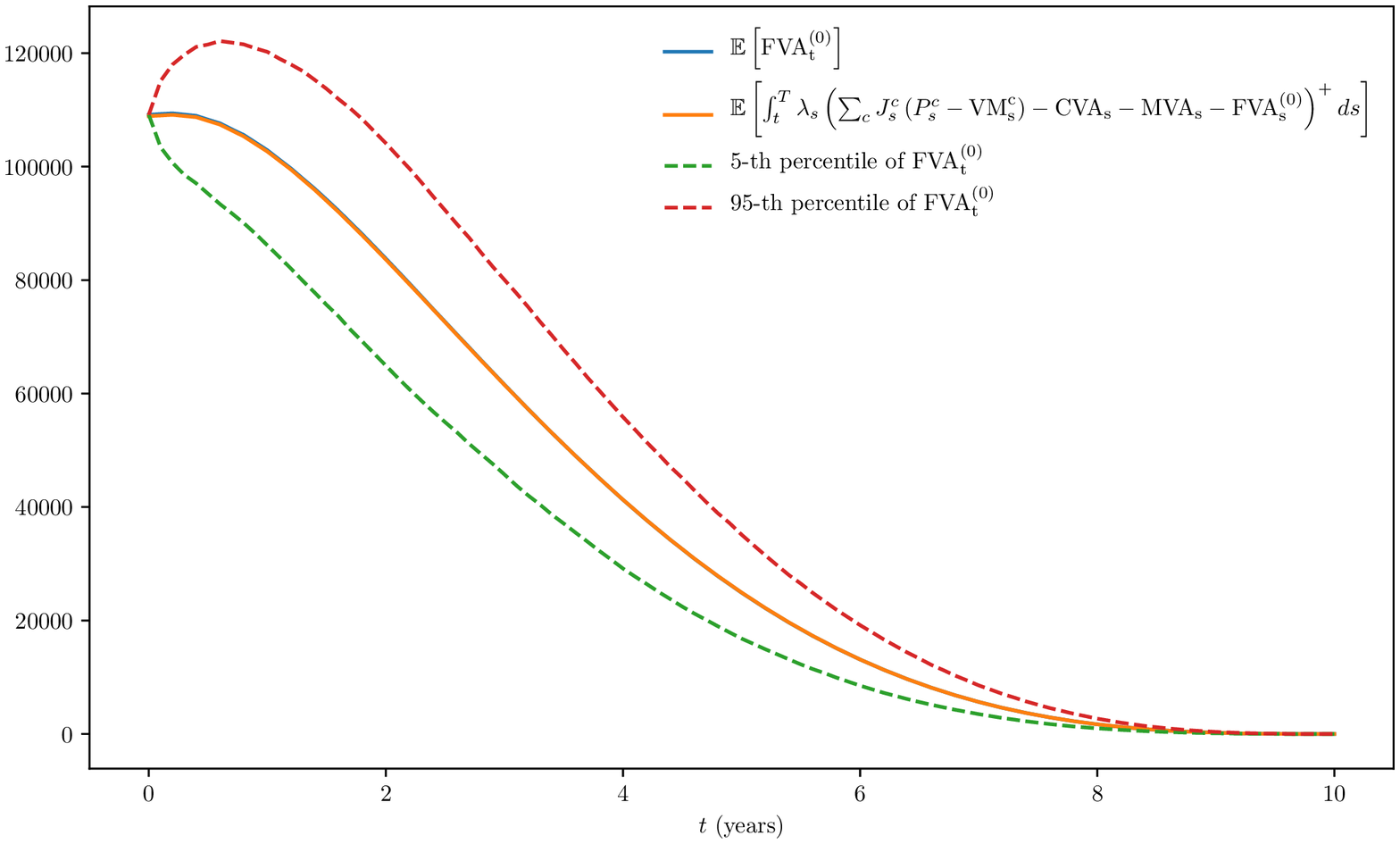} 
 \caption{Learned FVA$^{(0)}$.}
 \label{fig:weightsreuse} 
 \end{figure}

Figure \ref{fig:PicardSanity} \textit{(left)} is a sanity check that  the profiles of the successives iterates $L^{(k)}$ of the shareholder trading loss process $\LOSS$ in Algorithm \ref{algo:case1} \r{converge rapidly with $k$.}
Figure \ref{fig:PicardSanity} \textit{(right)} shows the loss process $L^{(3)}$,
displayed as its mean and mean
$\pm$ 2 stdev profiles. Consistent with its martingale property, the loss process $L^{(3)}$ appears numerically centered around zero.
The latter holds, at least, beyond $t\sim 5$ years. For earlier times, the regression errors, accumulated backward across pricing times since the final maturity of the portfolio, induce a non negligible bias (the corresponding  confidence intervals no longer contains 0). This is the reason why we use a coarser pricing time step than simulation time step in Algorithm \ref{algo:case1}.
\begin{figure}[!htbp]
\centering
\includegraphics[viewport=50 50 780 550,width=7.625cm,clip]{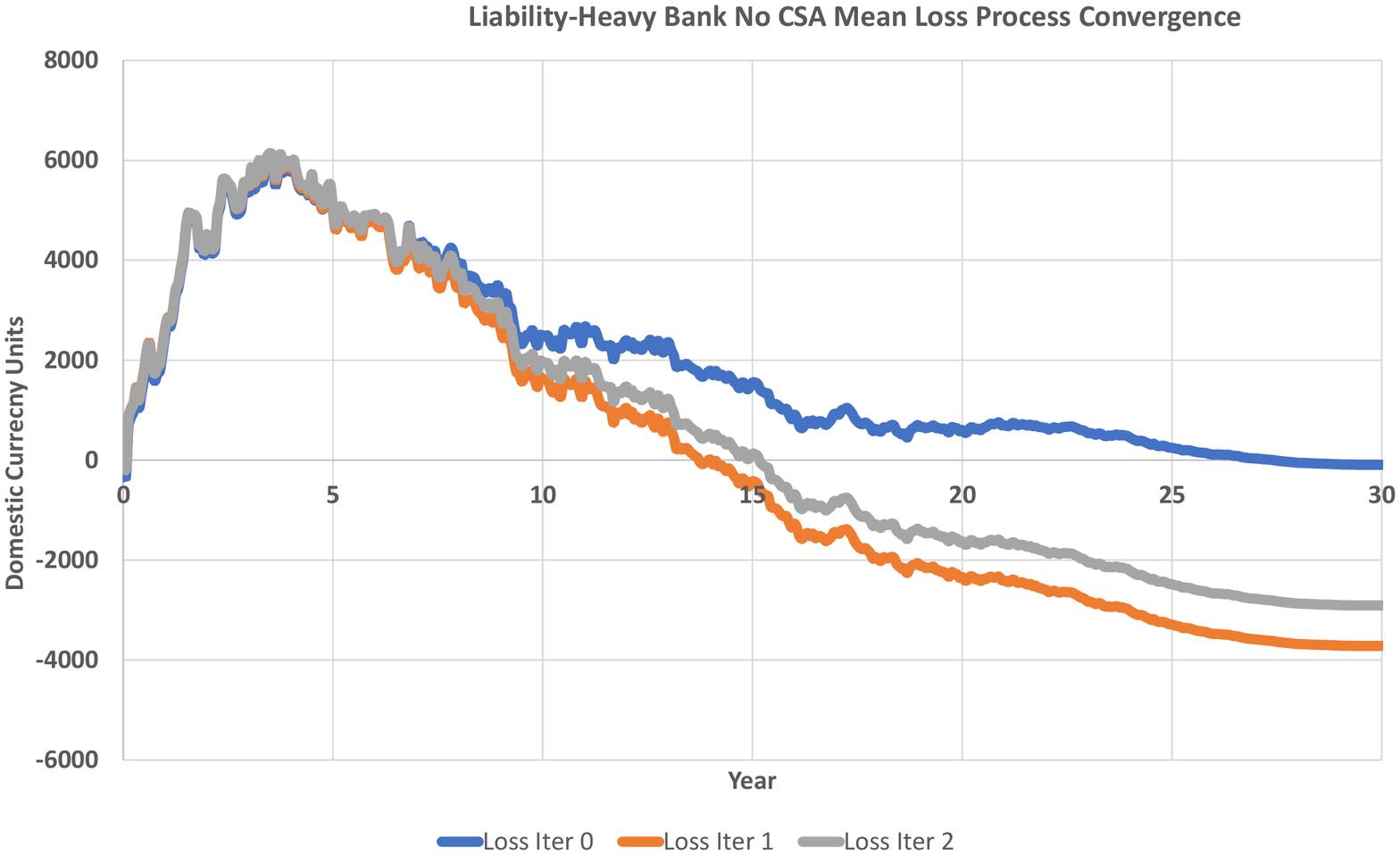}
\includegraphics[viewport=50 50 900 600,width=8.925cm,clip]{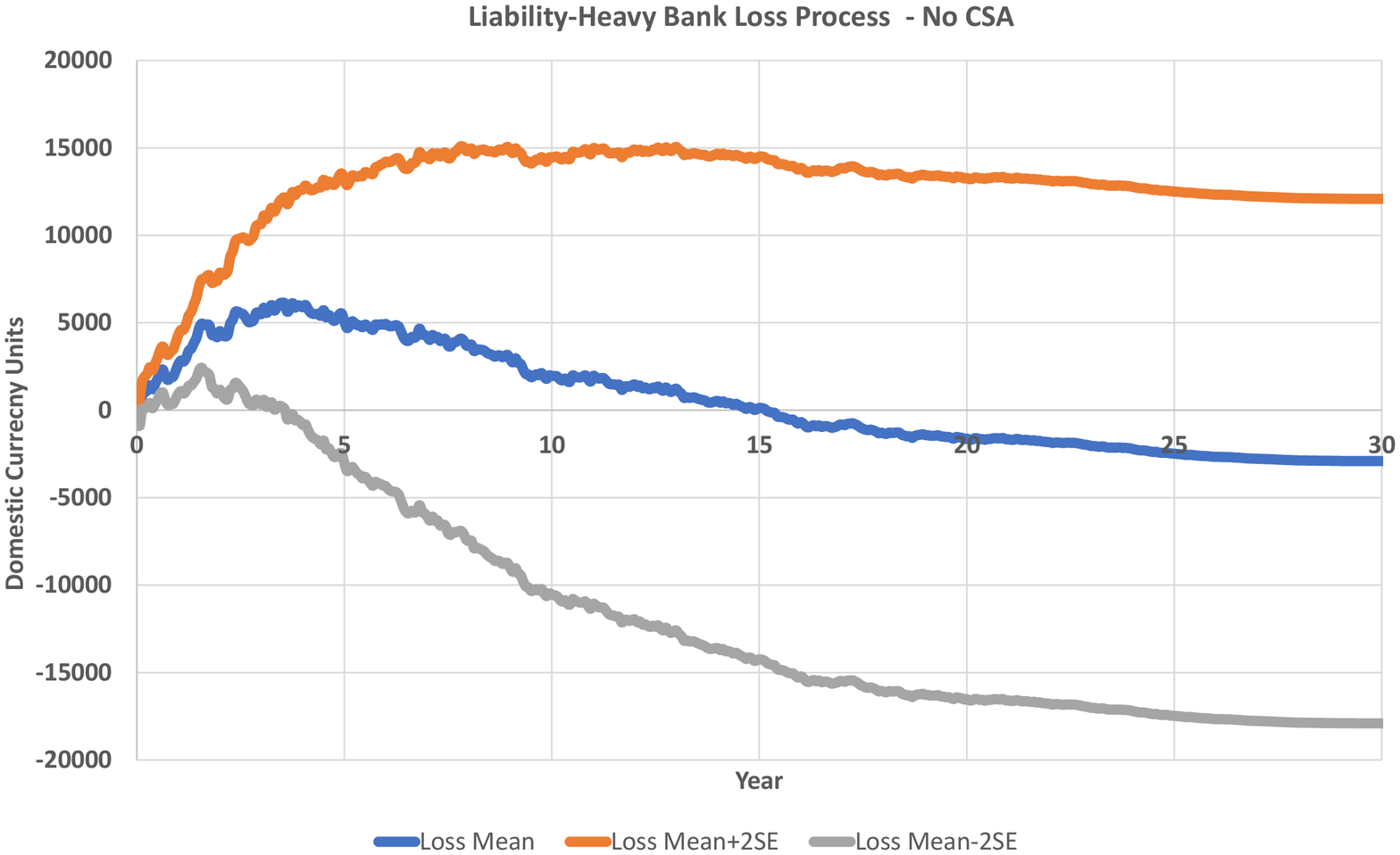}
\caption{\textit{\textit{(Left)}} Profiles of the processes $L^{(k)}$, for $k=1,2,3$; 
\textit{(Right)} Mean $\pm$ 2 stdev profiles of the process $L^{(3)}$.}
\label{fig:PicardSanity}
\end{figure}

\subsection{Portfolio-wide XVA Profiles}

For the financial case study that follows, we  consider
\begin{itemize}
  \item swap rates uniformly distributed on $[0.005,0.05]$
(hence swaps already in-the-money or out-of-the-money at time 0),
  \item number of six-monthly coupon resets uniform on $[5\cdots{}60]$  (final maturity of the portfolio $T=30$ years), 
  \item portfolio direction: either ``asset heavy'' bank mostly in the receivables in the future, or ``liability-heavy'' bank mostly in the payables in the future 
(respectively corresponding, with our data, to a bank $75\%$ likely to pay fixed in the swaps, or $75\%$ likely to receive fixed).
\end{itemize}

The figures that follow only display profiles, i.e.~term structures, that is, expectations as a function of time of the corresponding processes. But all these processes are computed pathwise,
based on the deep learning regression and  quantile regression methodology of \sr{ss:xvaframew},
allowing for all XVA inter-dependencies.
Of course, XVA profiles (or pathwise XVAs if wished) are much more informative for traders than the spot XVA values (or time 0 confidence intervals) returned by most XVA systems.

Assuming 10 counterparties,
Figure \ref{fig:LiabilityHeavyMtM} shows the GPU generated profiles of 
\beql{e:mtm}\mtm=\sum_{c} P^c\ind_{[0,\tau^\delta_c)}\eeql
in the case
of the asset-heavy portfolio  and of the
liability-heavy portfolio.
\begin{figure}[!htbp]
\centering
\includegraphics[viewport=50 250 580 550,width=6.625cm,clip]{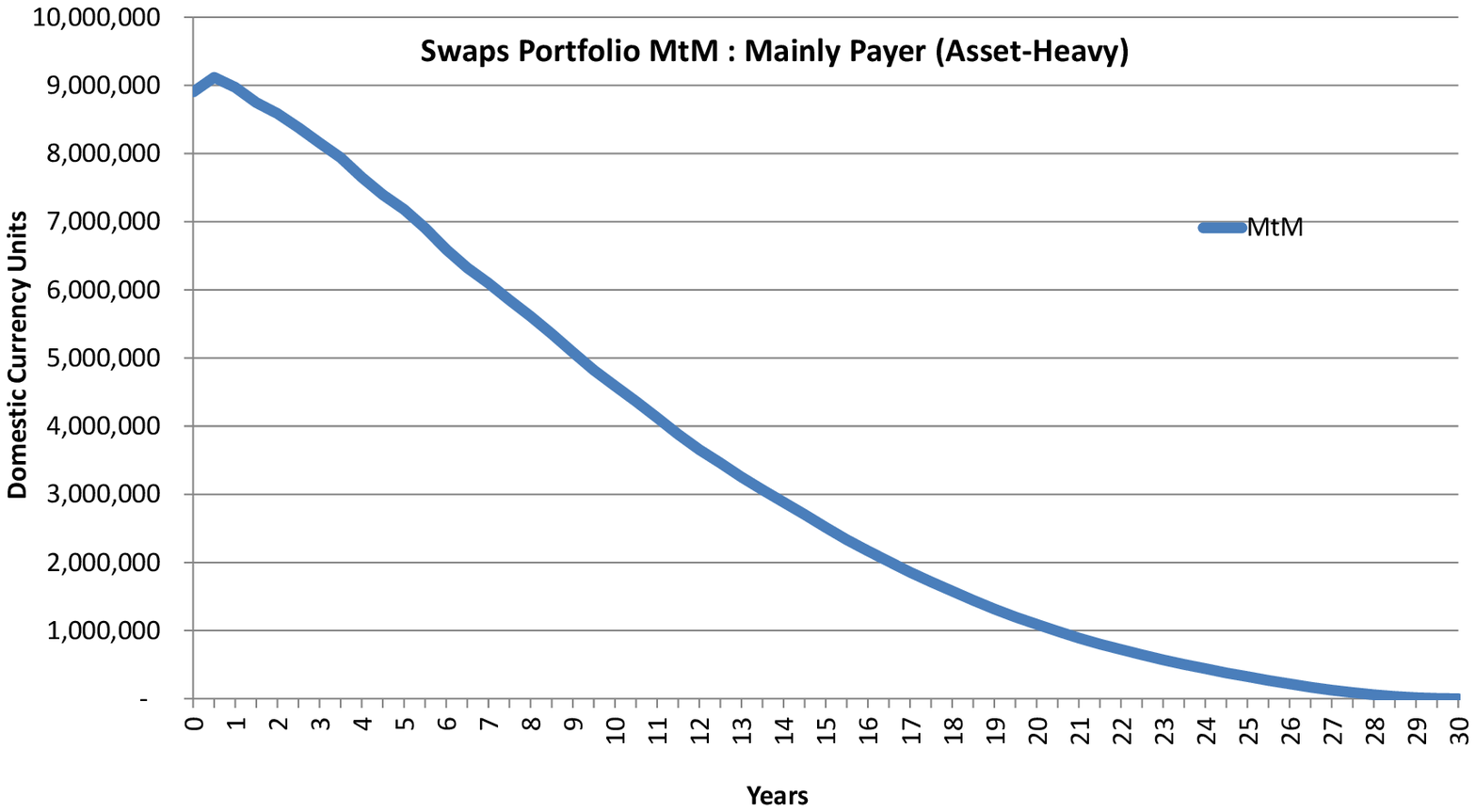}
\includegraphics[viewport=50 250 580 550,width=6.625cm,clip]{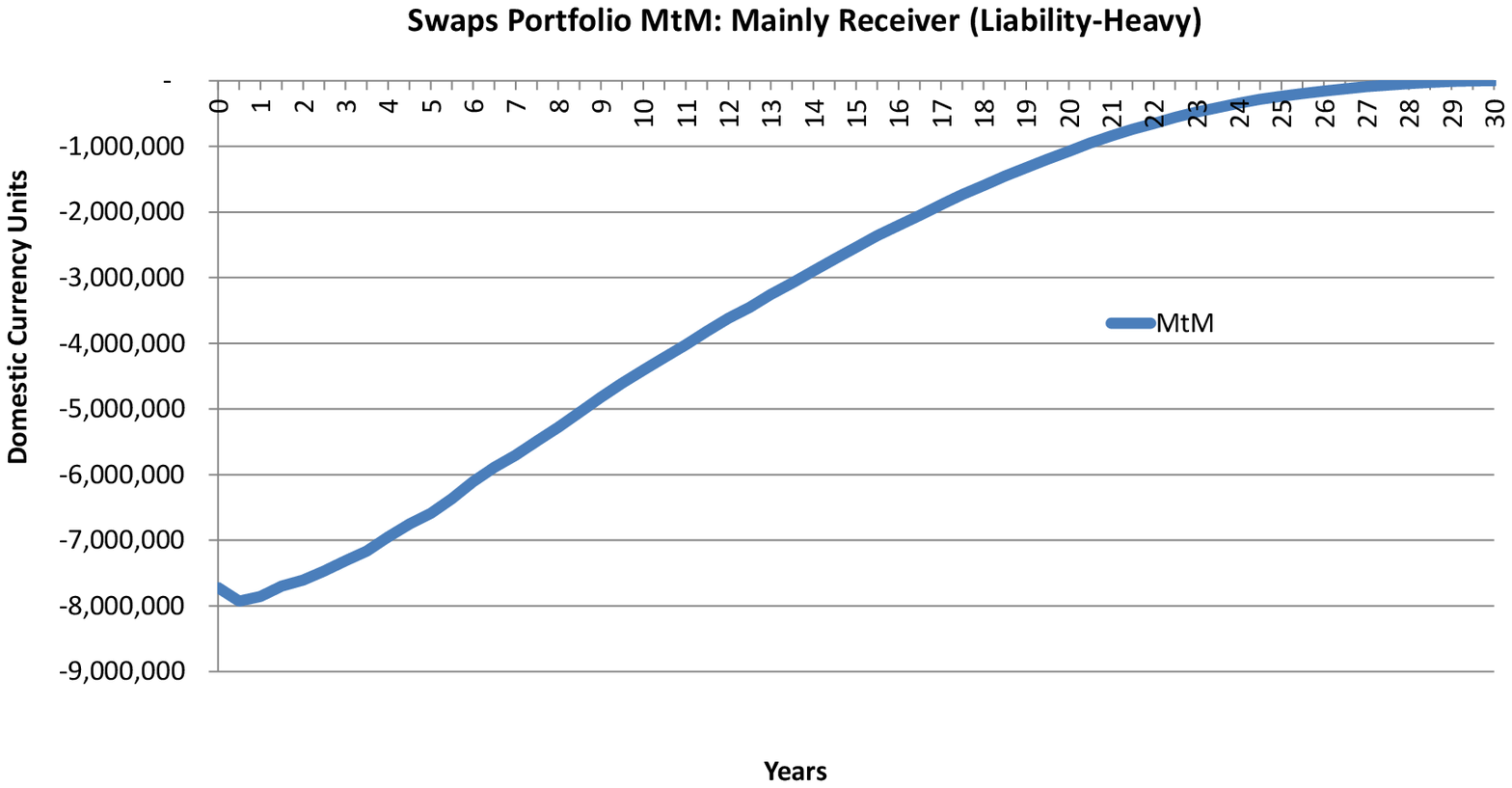}
\caption{MtM profiles. \textit{(Left)} Asset-heavy portfolio. 
\textit{(Right)} Liability-heavy portfolio.}\label{fig:LiabilityHeavyMtM}
 \end{figure}

Figure \ref{fig:LiabilityHeavyXVANoCSA} shows the porftolio-wide XVA profiles of the asset-heavy \textit{(top)}  vs.~liability--heavy \textit{(bottom)} portfolio
and of the no CSA \textit{(left)} vs.~CSA portfolio \textit{(right)}. Obviously, asset--heavy or no CSA means more CVA. The correponding curves also emphasize the transfer from counterparty credit into \b{liquidity funding risk}
prompted by extensive collateralisation. Yet FVA/MVA risk is ignored in current derivatives capital regulation. 
\begin{figure}[!htbp]
\centering
\includegraphics[viewport=50 250 580 550,width=6.625cm,clip]{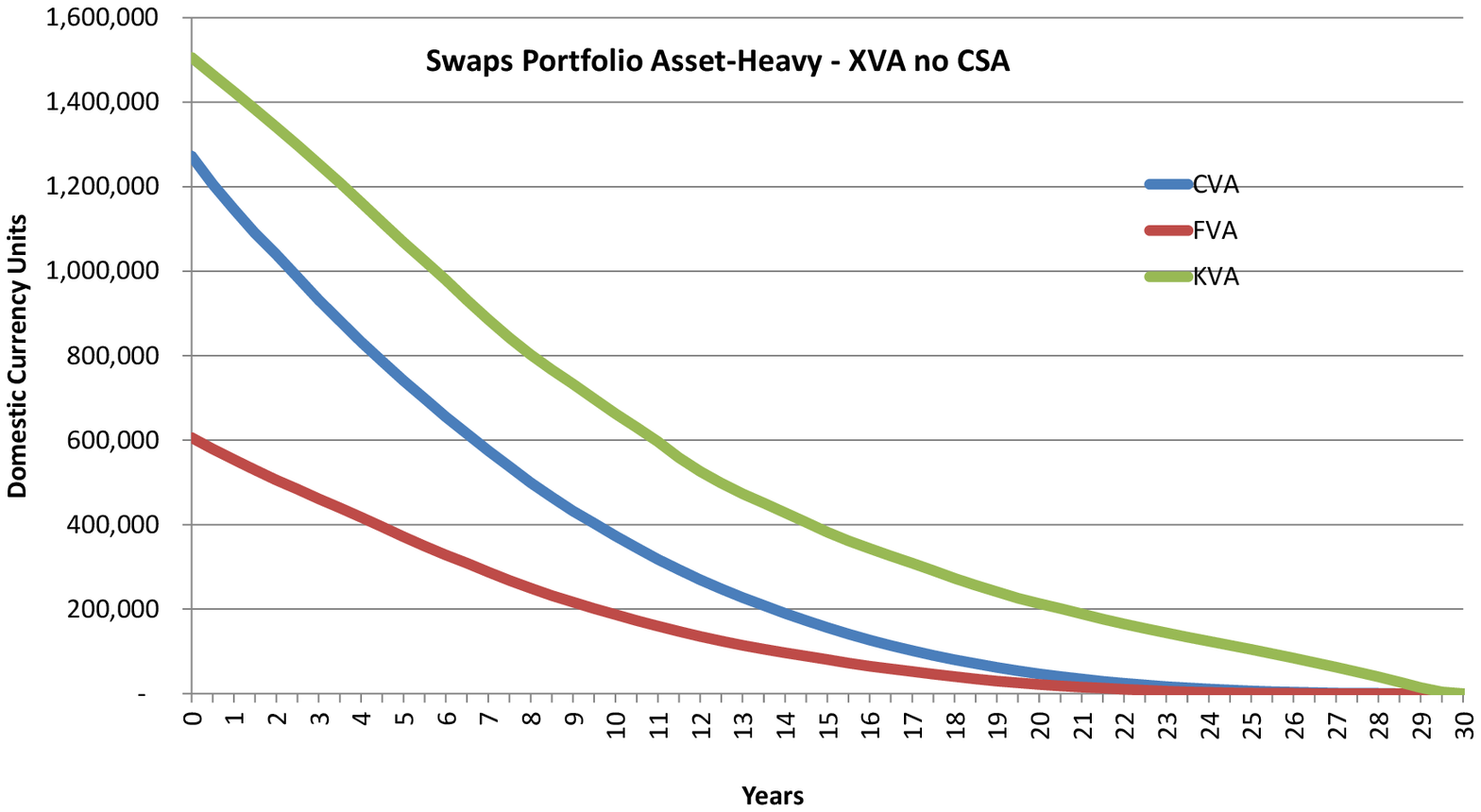}
\includegraphics[viewport=50 250 580 550,width=6.625cm,clip]{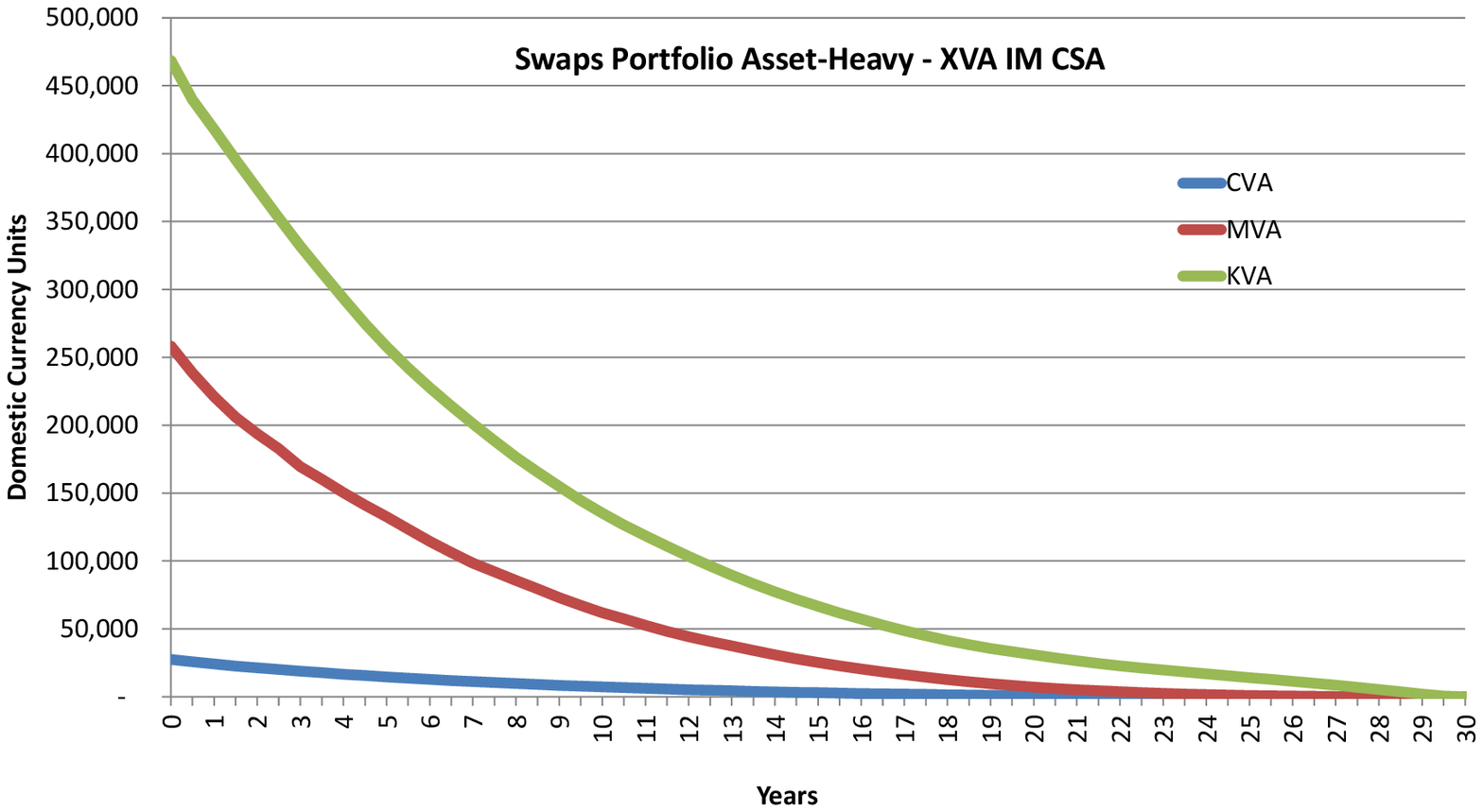}\\
\includegraphics[viewport=50 250 580 550,width=6.625cm,clip]{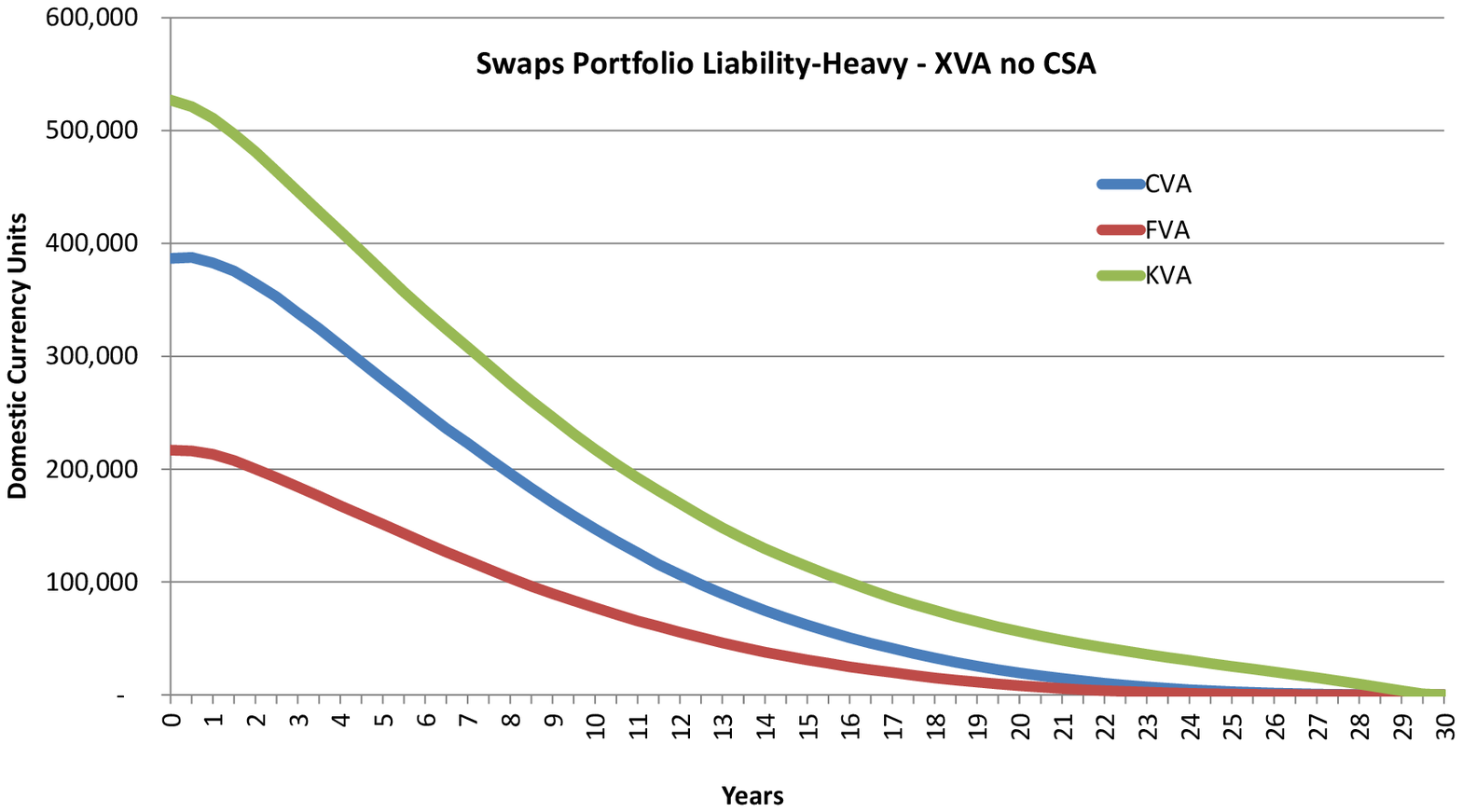}
\includegraphics[viewport=50 250 580 550,width=6.625cm,clip]{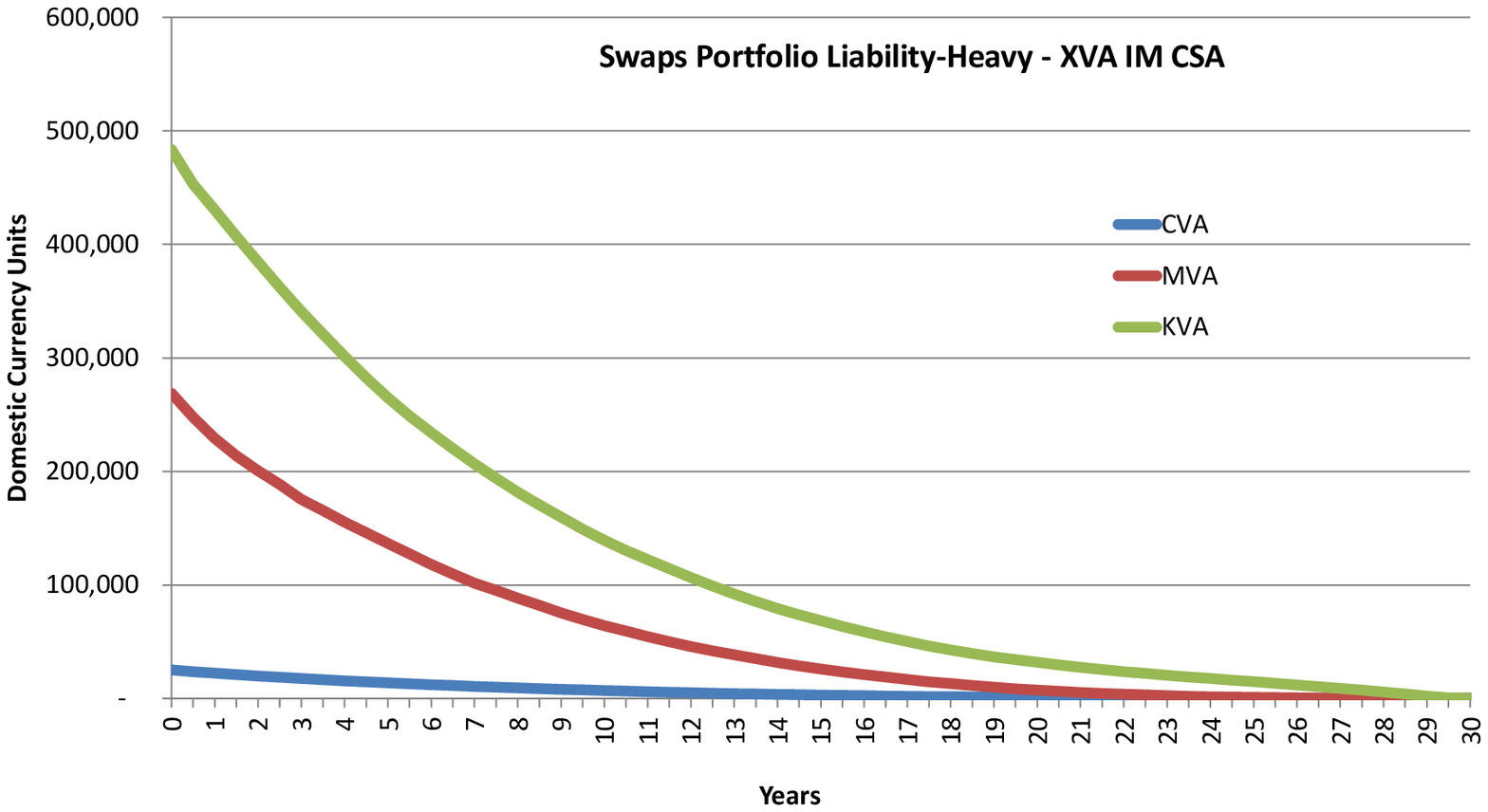}
\caption{\textit{(Top left)} Asset-heavy portfolio, no CSA.
\textit{(Top right)} Asset-heavy portfolio under CSA.
\textit{(Bottom left)} Liability--heavy portfolio, no CSA.
\textit{(Bottom right)} Liability-heavy portfolio under CSA.}
\label{fig:LiabilityHeavyXVANoCSA}
\end{figure}

Figure \ref{fig:AssetHeavyIMConvexityIMCSA} shows that \textit{(top left)} capital at risk as funding
(cf.~\sr{rem:compl})
 has a material impact on the already (reserve capital as funding) reduced FVA, 
\textit{(top right)} treating KVA as a risk margin (cf.~\qr{e:kvadiscr}) gives a huge discounting impact,
\textit{(bottom left)} deep learning detects material initial margin convexity in the asset-heavy CSA portfolio, and \textit{(bottom right)} deep learning detects material economic capital convexity in the asset-heavy no CSA portfolio. 
\begin{figure}[h!]
\centering
\includegraphics[viewport=50 250 580 550,width=6.625cm,clip]{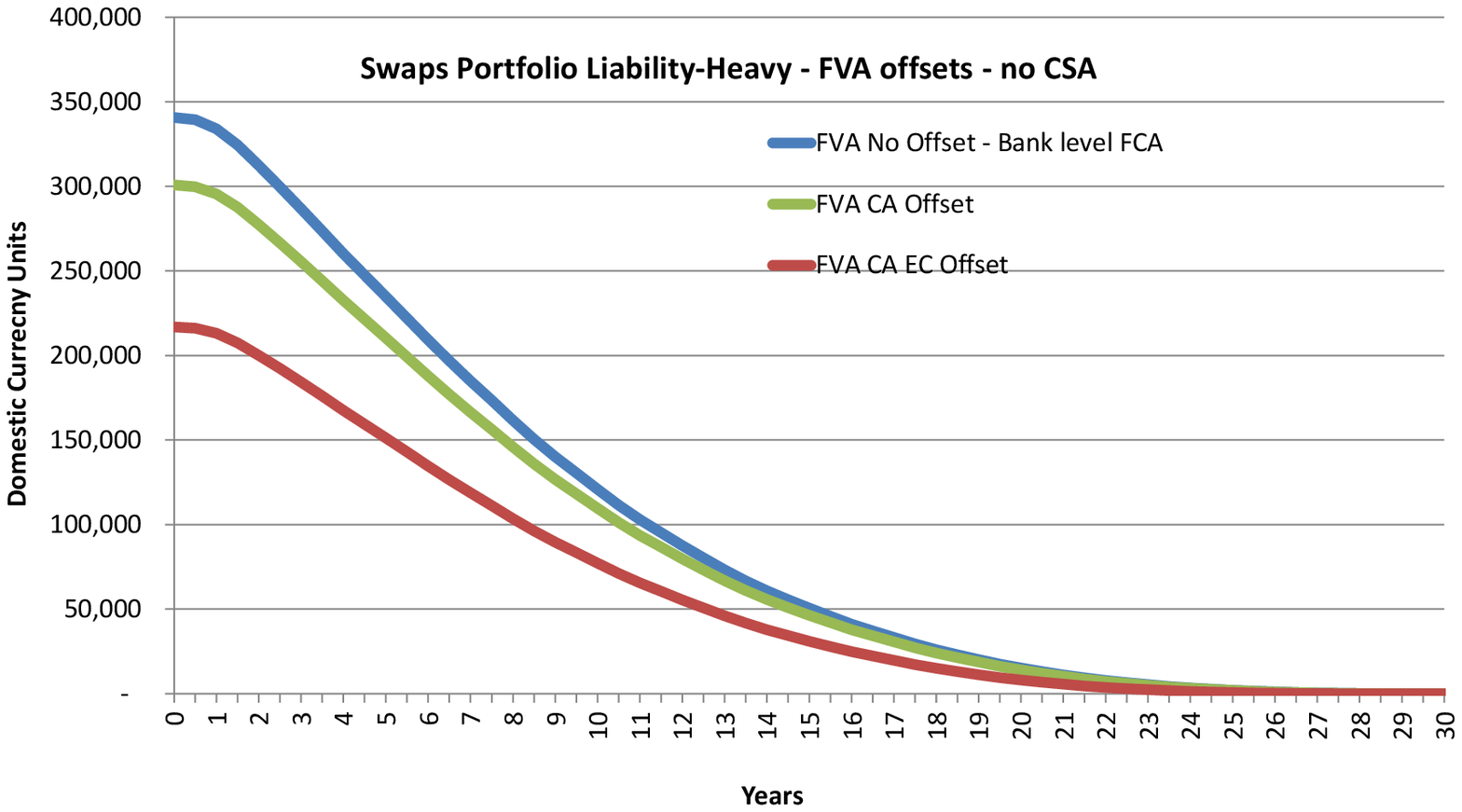} 
\includegraphics[viewport=50 250 580 550,width=6.625cm,clip]{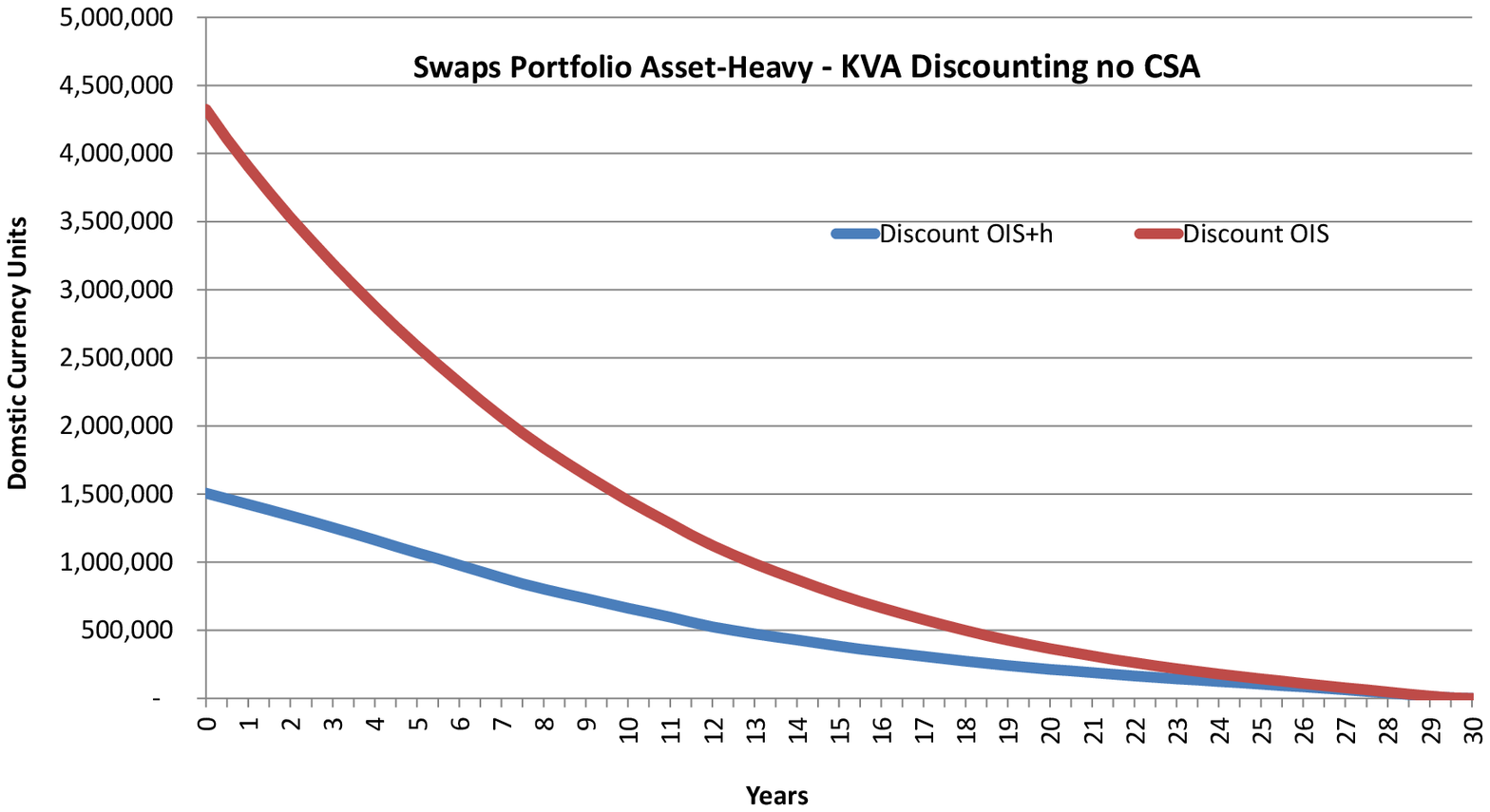}\\ 
\includegraphics[viewport=50 250 580 550,width=6.625cm,clip]{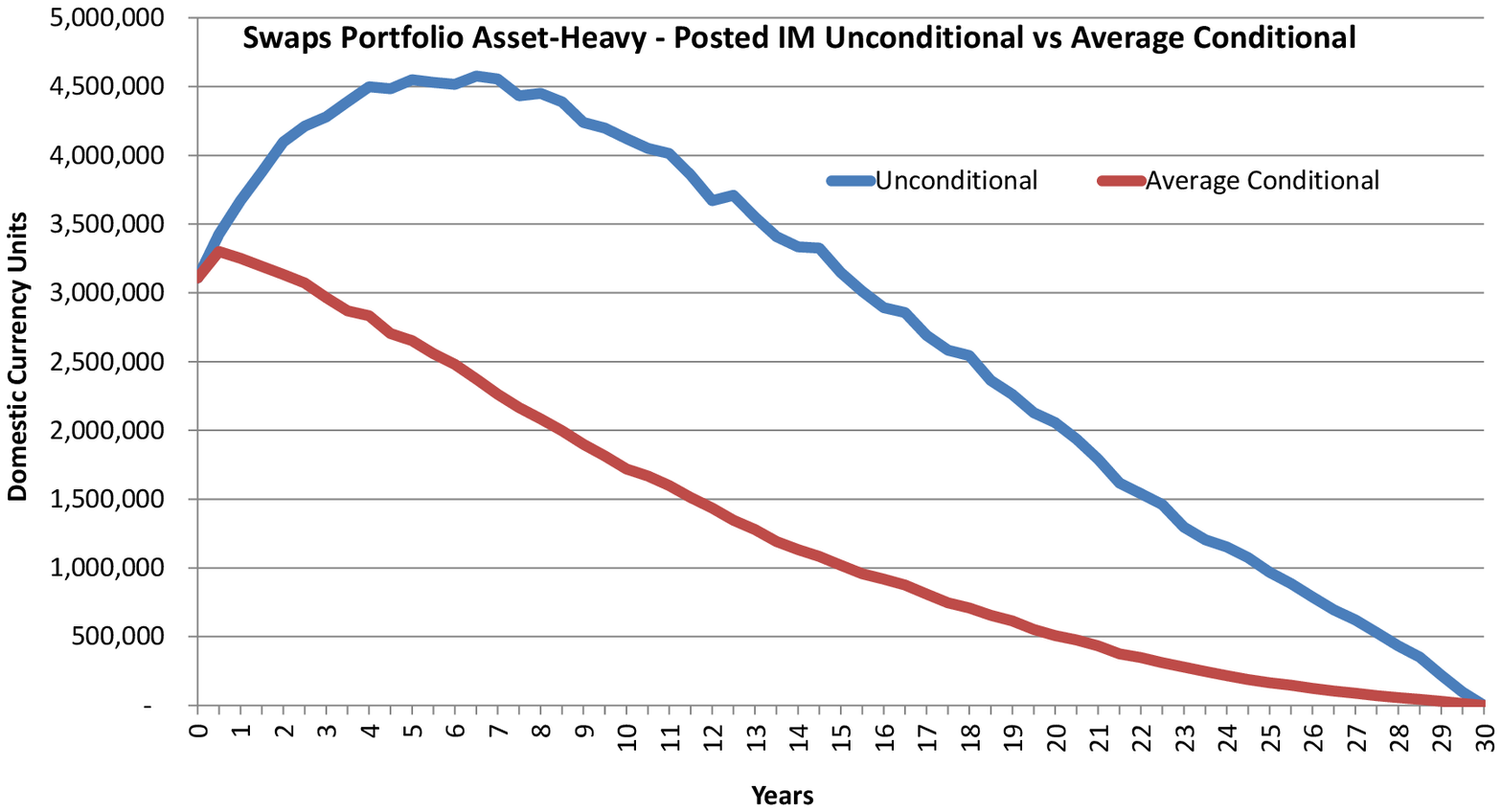}
\includegraphics[viewport=50 250 580 550,width=6.625cm,clip]{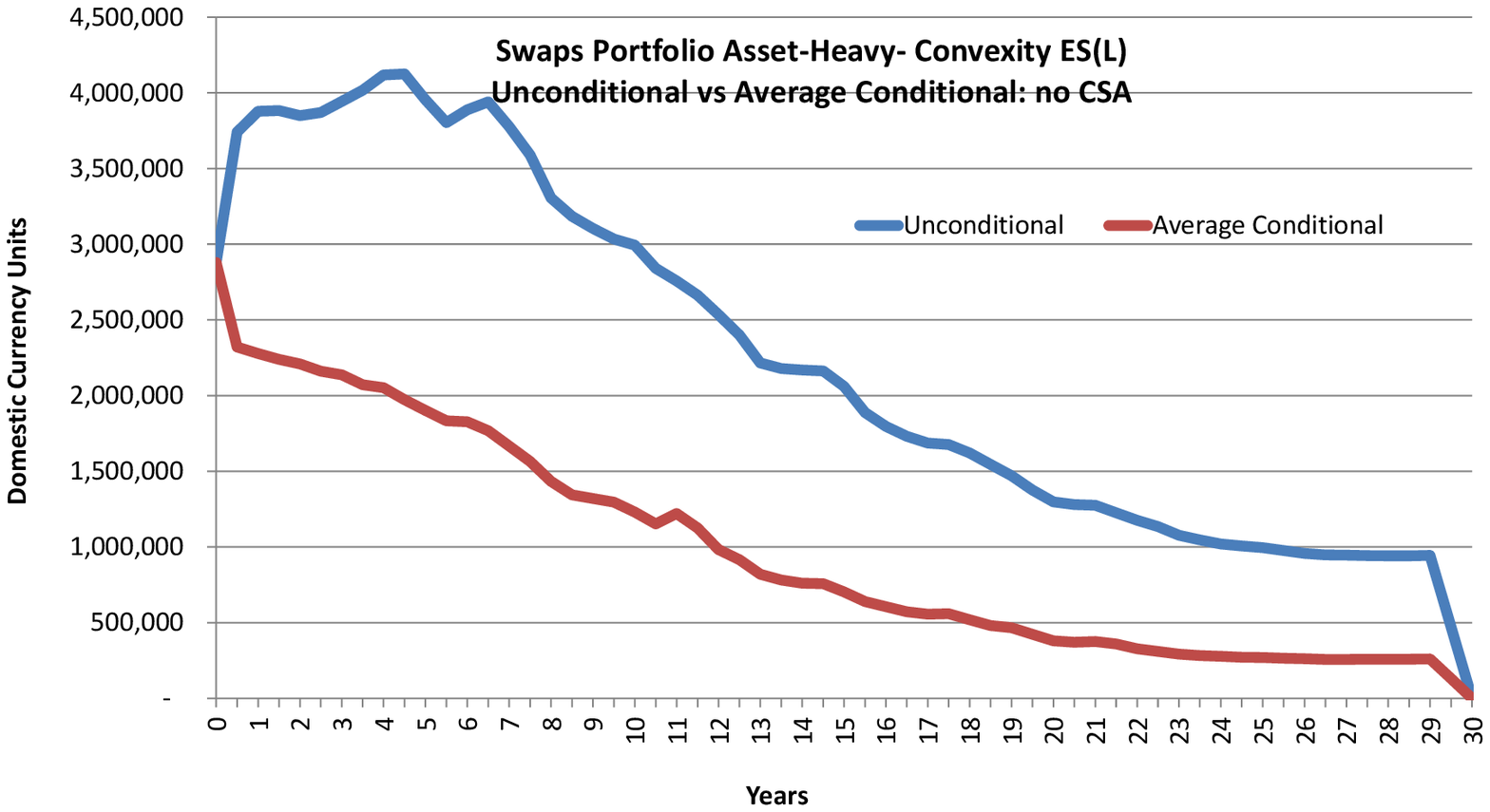} 
\caption{\textit{(Top left)} FVA ignoring the 
off-setting impact of reserve capital and capital at risk, cf.~Section \ref{rem:compl} (blue), FVA as per \qr{e:fvainit} accounting for the
off-setting impact of reserve capital but ignoring the one of capital at risk (green), 
refined FVA as per \qr{e:L} accounting for both impacts (red).
\textit{(Top right)} KVA ignoring the off-setting impact of the risk margin, i.e. with CR instead of ($\CR-\kva$) in \qr{e:kconsuite} (red), refined KVA as per \qr{e:ecpt}--\qr{e:kcon} (blue).
\textit{(Bottom left)} In the case of the asset-heavy portfolio under CSA, unconditional PIM profile, i.e. with $\VaR_t$  replaced by $\VaR$ in \qr{e:gapri} (blue), vs.~pathwise
PIM profile, i.e. mean of the
pathwise PIM process as per \qr{e:gapri} (red).
\textit{(Bottom right)} In the asset-heavy portfolio no CSA case, unconditional economic capital profile, i.e. EC profile ignoring the words ``time-$t$
conditional'' in Definition \ref{defi:ec} (blue), vs.~pathwise economic capital profile, i.e. mean of the
pathwise EC process as per Definition \ref{defi:ec} (red).}
\label{fig:AssetHeavyIMConvexityIMCSA}
\end{figure}

The above findings 
demonstrate the necessity of pathwise capital and margin calculations for accurate FVA, MVA, and KVA calculations.

\subsection{Trade Incremental XVA Profiles}


  
Next, we consider, on top of the previous portfolios, an incremental trade given as a par 30 year (receive fix or pay fix) swap with 100K notional.
Figure \ref{fig:LiabilityHeavyXVANoCSAIncrRec} shows the trade incremental XVA profiles produced by our deep learning approach.
Note that, for obtaining such smooth incremental profiles, it has been key to use
common random numbers, as much as possible, between the original portfolio XVA computations and the ones regarding the portfolio expanded with the new trade.
\begin{figure}[!htbp]
\centering
\includegraphics[viewport=50 250 580 550,width=6.625cm,clip]{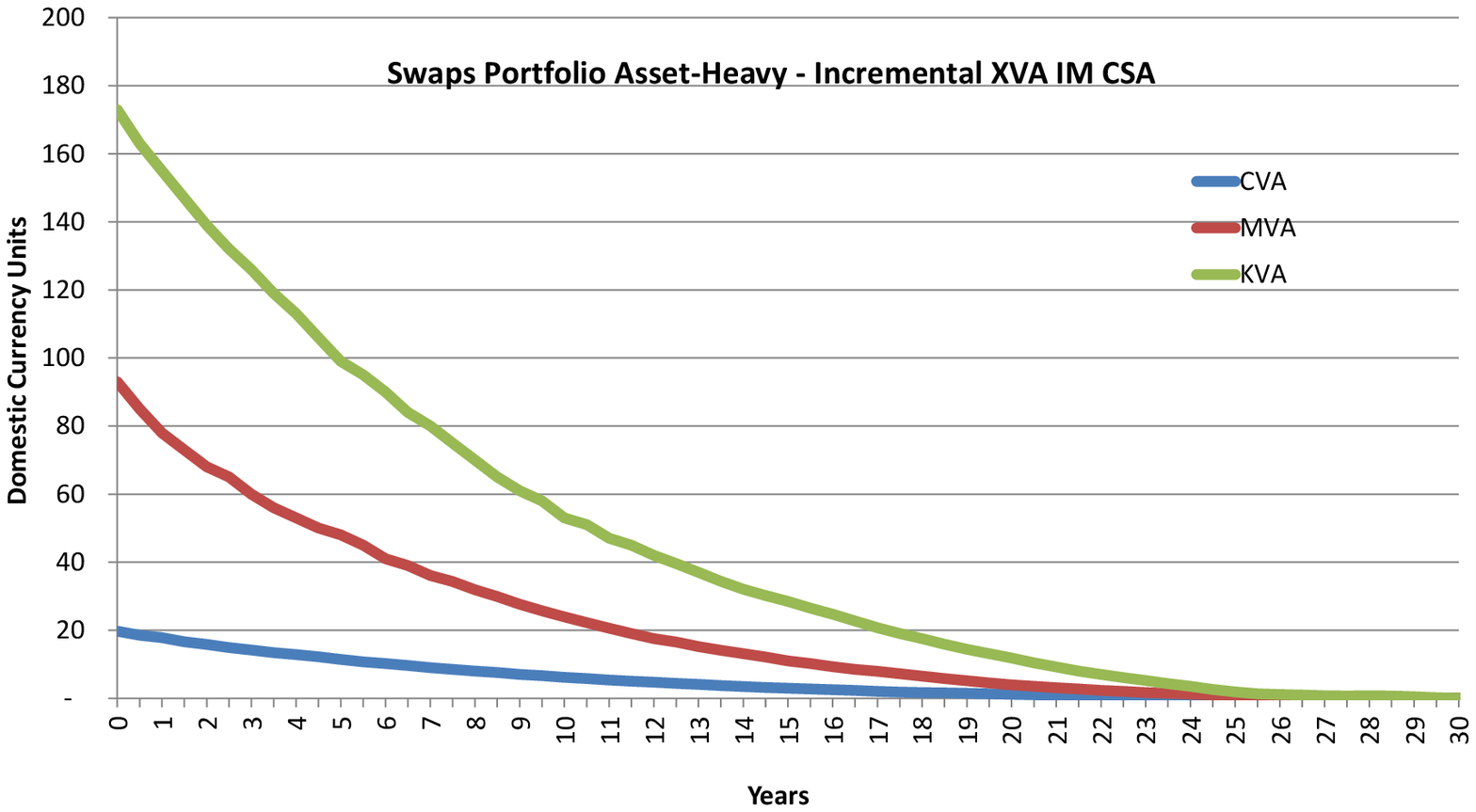}
\includegraphics[viewport=50 250 580 550,width=6.625cm,clip]{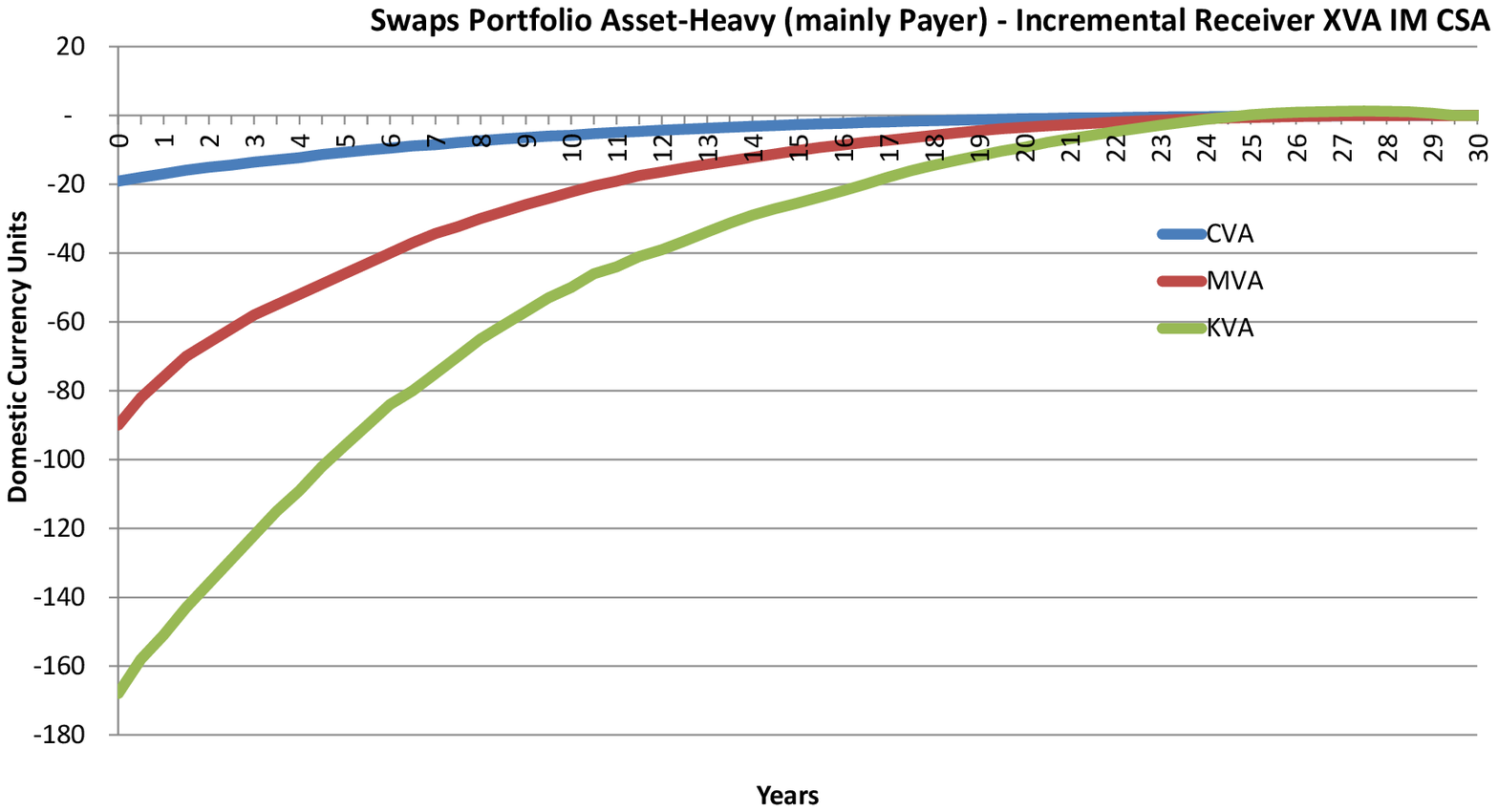}\\
\includegraphics[viewport=50 250 580 550,width=6.625cm,clip]{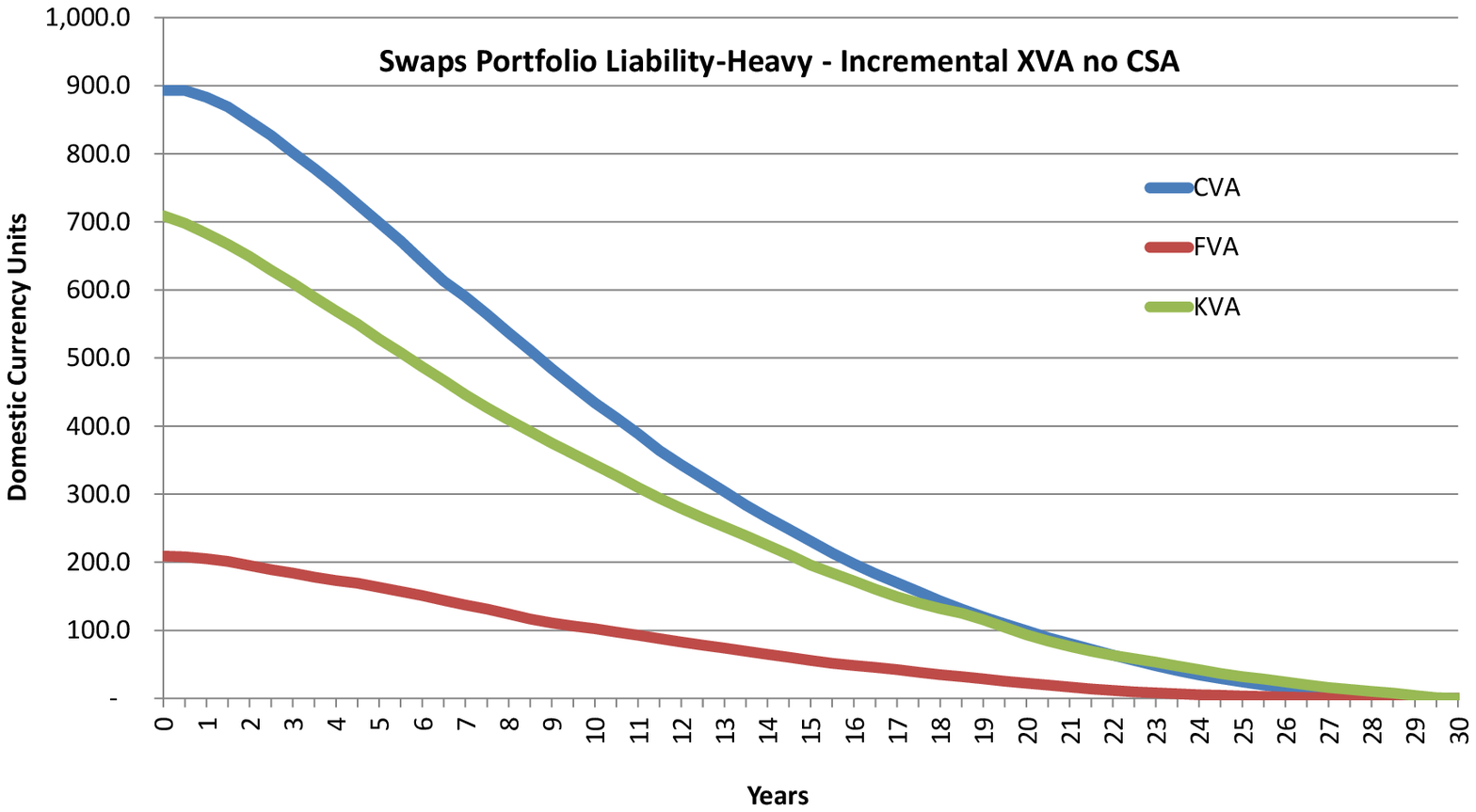}
\includegraphics[viewport=50 250 580 550,width=6.625cm,clip]{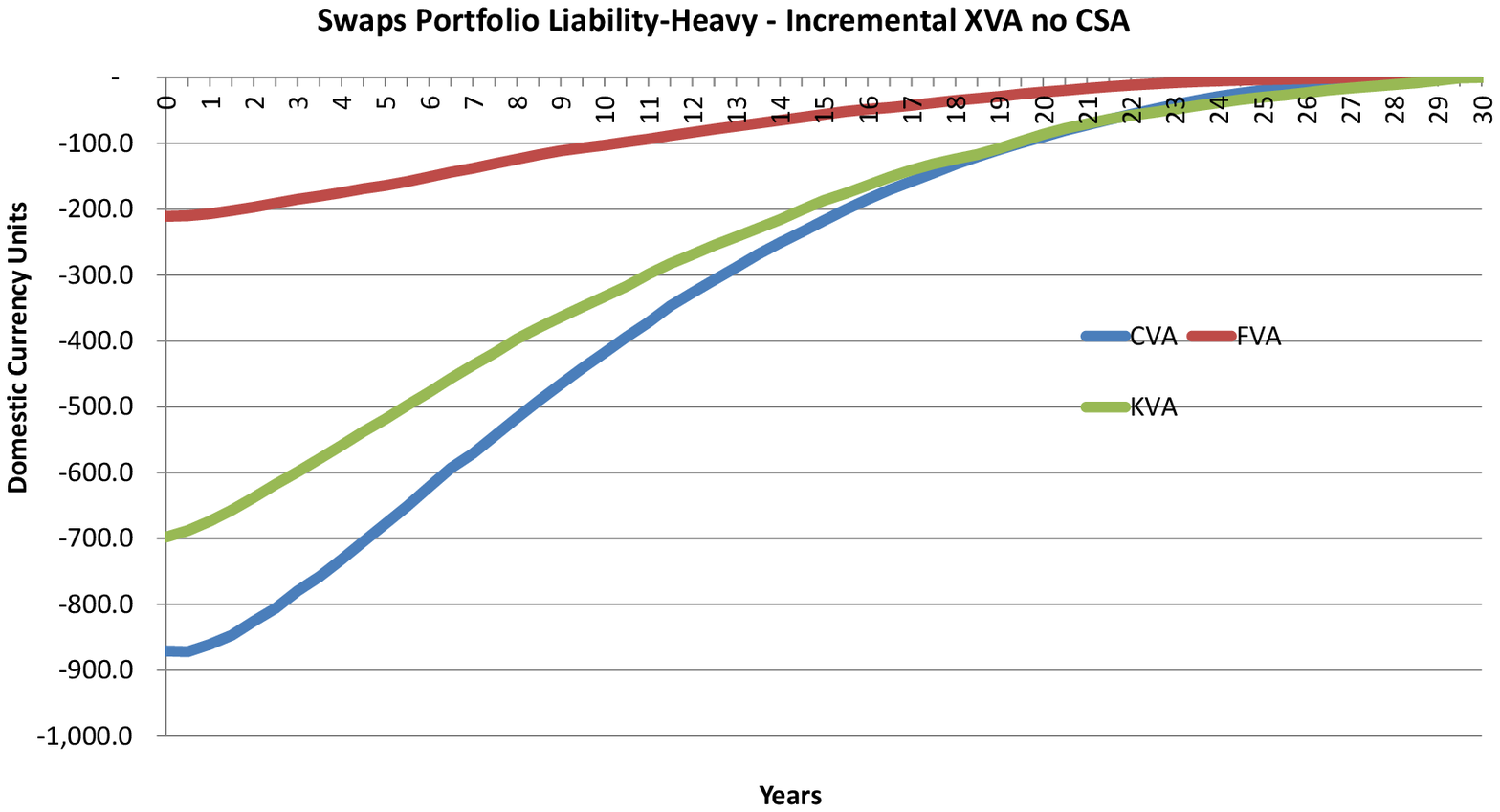}
\caption{\textit{(Top left)} Asset-heavy portfolio, no CSA. Incremental receive fix trade.  
\textit{(Top right)} Liability-heavy portfolio, no CSA. Incremental pay fix trade. 
\textit{(Bottom left)} Asset-heavy portfolio under CSA. Incremental Pay Fix Trade. 
\textit{(Bottom right)} Liability-heavy portfolio under CSA. Incremental receive fix trade.}
\label{fig:LiabilityHeavyXVANoCSAIncrRec}
\end{figure}
 
\subsection{Trade and Hedge Incremental XVA Profiles}\label{ss:deale}

Our model assumes the market risk of trades to be fully hedged (see the paragraph following Remark \ref{rem:int} and the proofs of Lemma \ref{l:cfoneper} and Proposition \ref{p:wtprel}). In the previous subsection, the new swap was implicitly meant to be hedged, in terms of market risk, by the clean desks, through an accordingly modified hedging loss process $\cH$ (see \sr{ss:runoffprel}).
Here we consider an alternative situation where the market risk of the new swap is back-to-back hedged via a financial, hedge counterparty.
Specifically, we deal with
\begin{itemize}
  \item 10 counterparties: 8 no CSA clients and 2 bilateral VM/IM CSA hedge counterparties, 
  \item portfolios of 5K randomly generated swap trades as before, plus 5K corresponding hedge trades,
  \item an incremental trade given as a par 30 year swap with 100K notional, along with the corresponding hedge trade.
\end{itemize}
In particular, $\mtm_0=0$ (cf.~\qr{e:mtm}), in both portfolios excluding or including the new swap. 
In case a client or hedge counterparty defaults, 
 the corresponding market hedge is assumed to be rewired through the clean desks via an accordingly modified hedging loss process $\cH$.  

The 8 no CSA counterparties are primarily asset or liability heavy.
One bilateral  CSA hedge counterparty is asset-heavy and one liability-heavy. 
Figure \ref{fig:XVAReduceTd} provides the {trade incremental XVA profiles} of the bilateral hedge alternatives in combination with those for the initial counterparty trade. \b{The main XVA impact of the  
hedge is then a corresponding incremental MVA term, which can contribute to make the global
FTP 
related to the trade+hedge package more or less positive or negative, depending on the data (cf.~the four panels in Figure \ref{fig:XVAReduceTd}), as can only be inferred by a refined XVA computation.} 
\begin{figure}[!htbp] 
\centering
\includegraphics[viewport=50 250 580 550,width=6.625cm,clip]{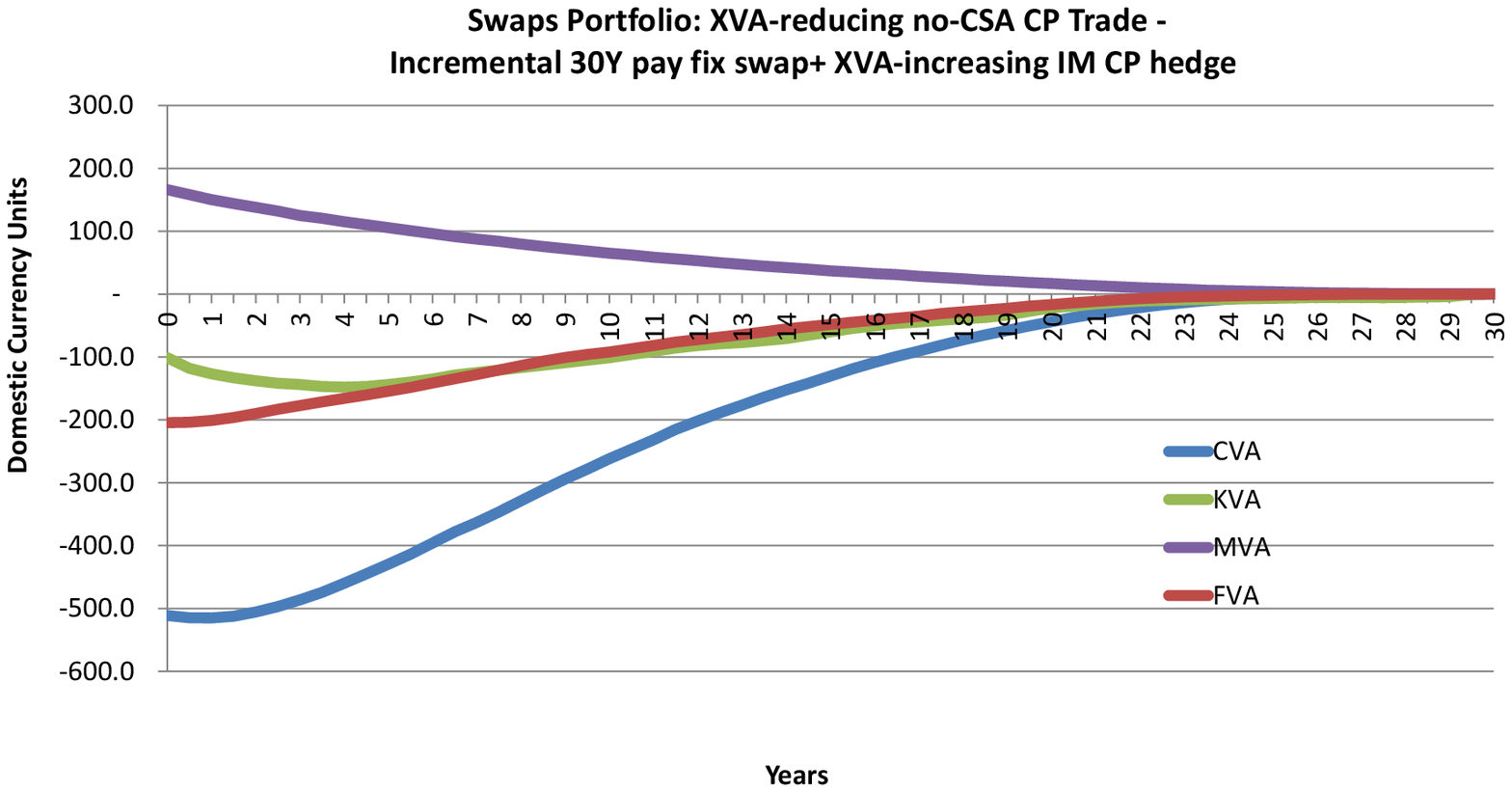}\includegraphics[viewport=50 250 580 550,width=6.625cm,clip]{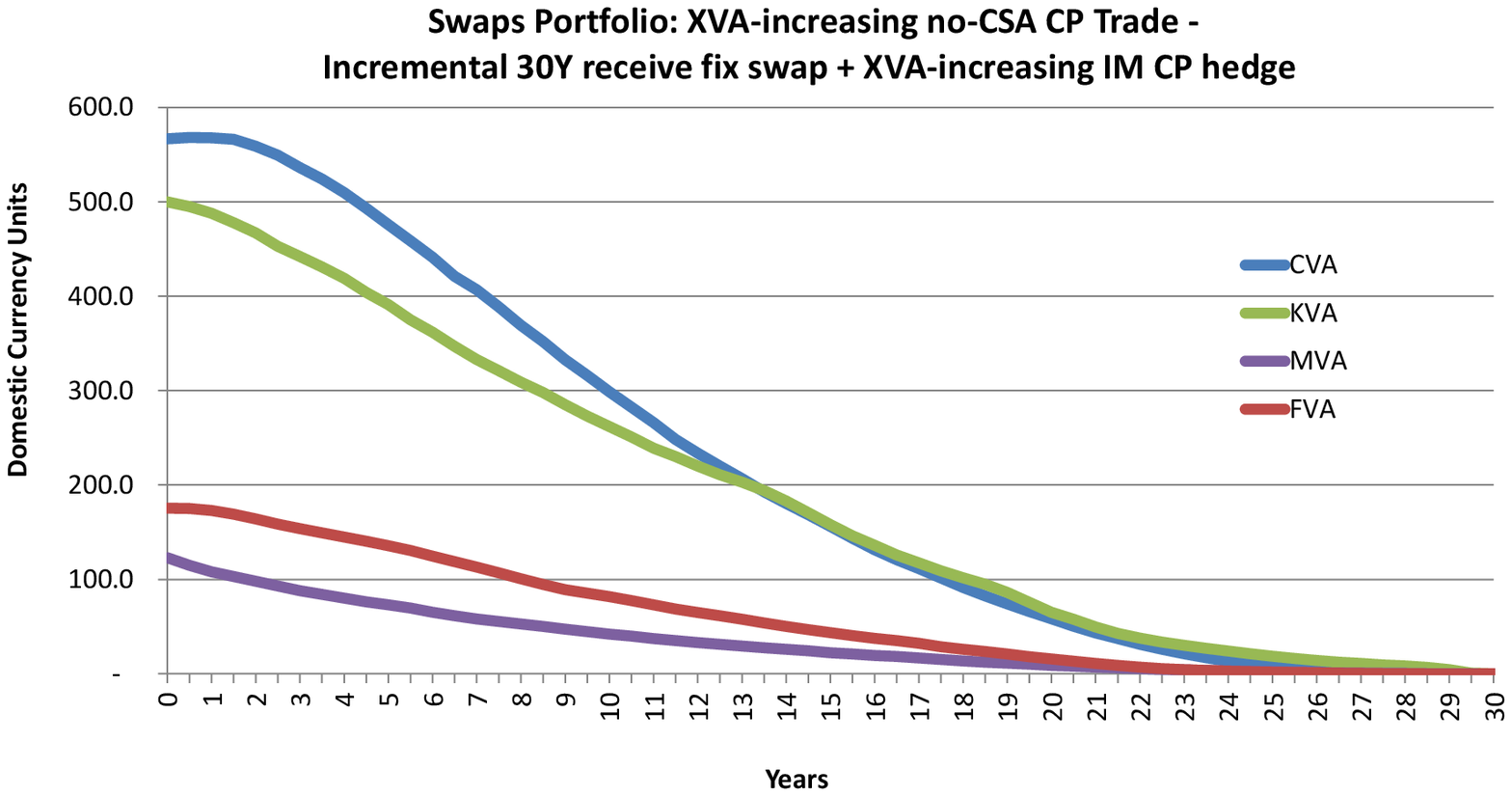}\\
\includegraphics[viewport=50 250 580 550,width=6.625cm,clip]{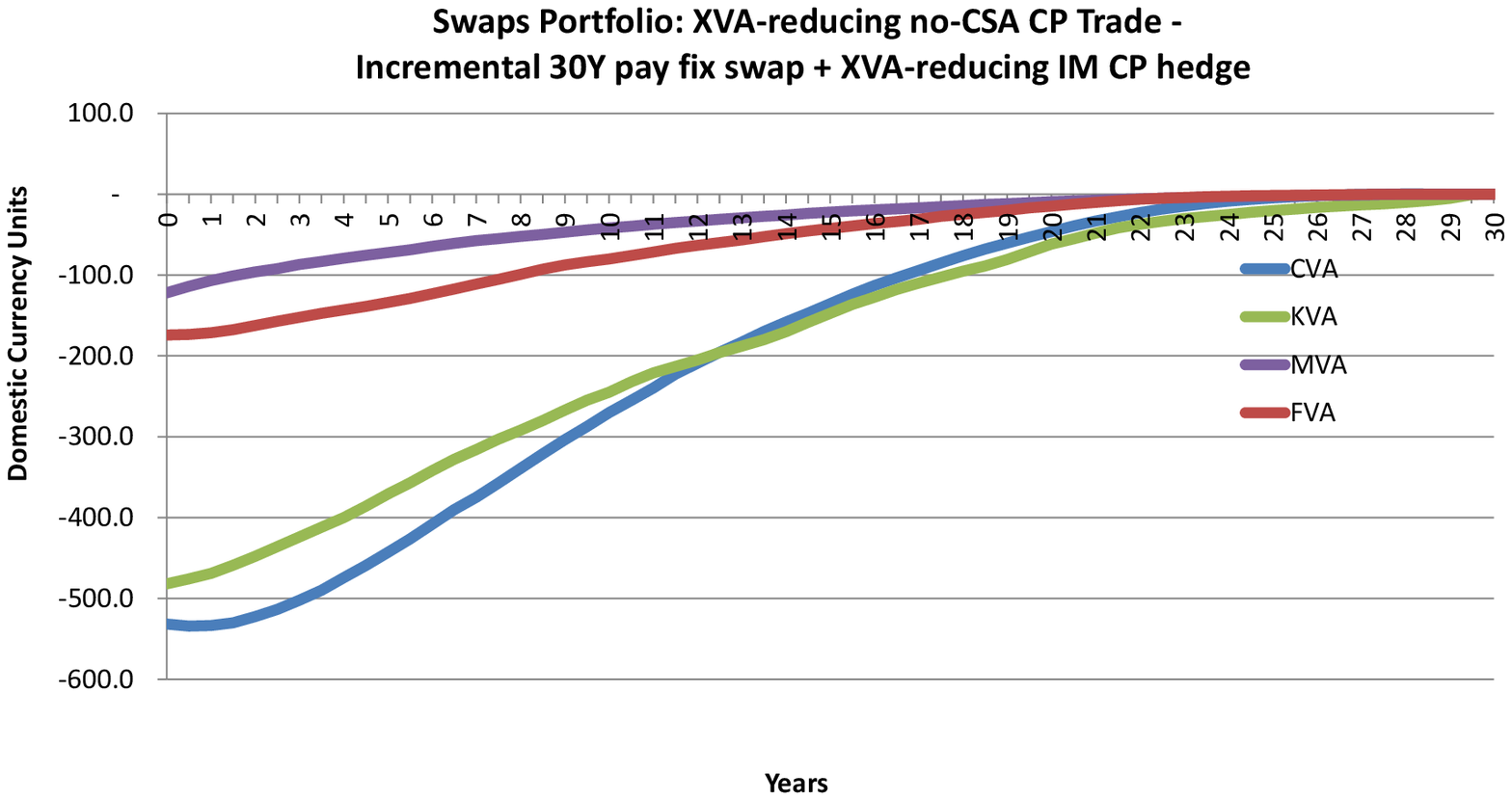}\includegraphics[viewport=50 250 580 550,width=6.625cm,clip]{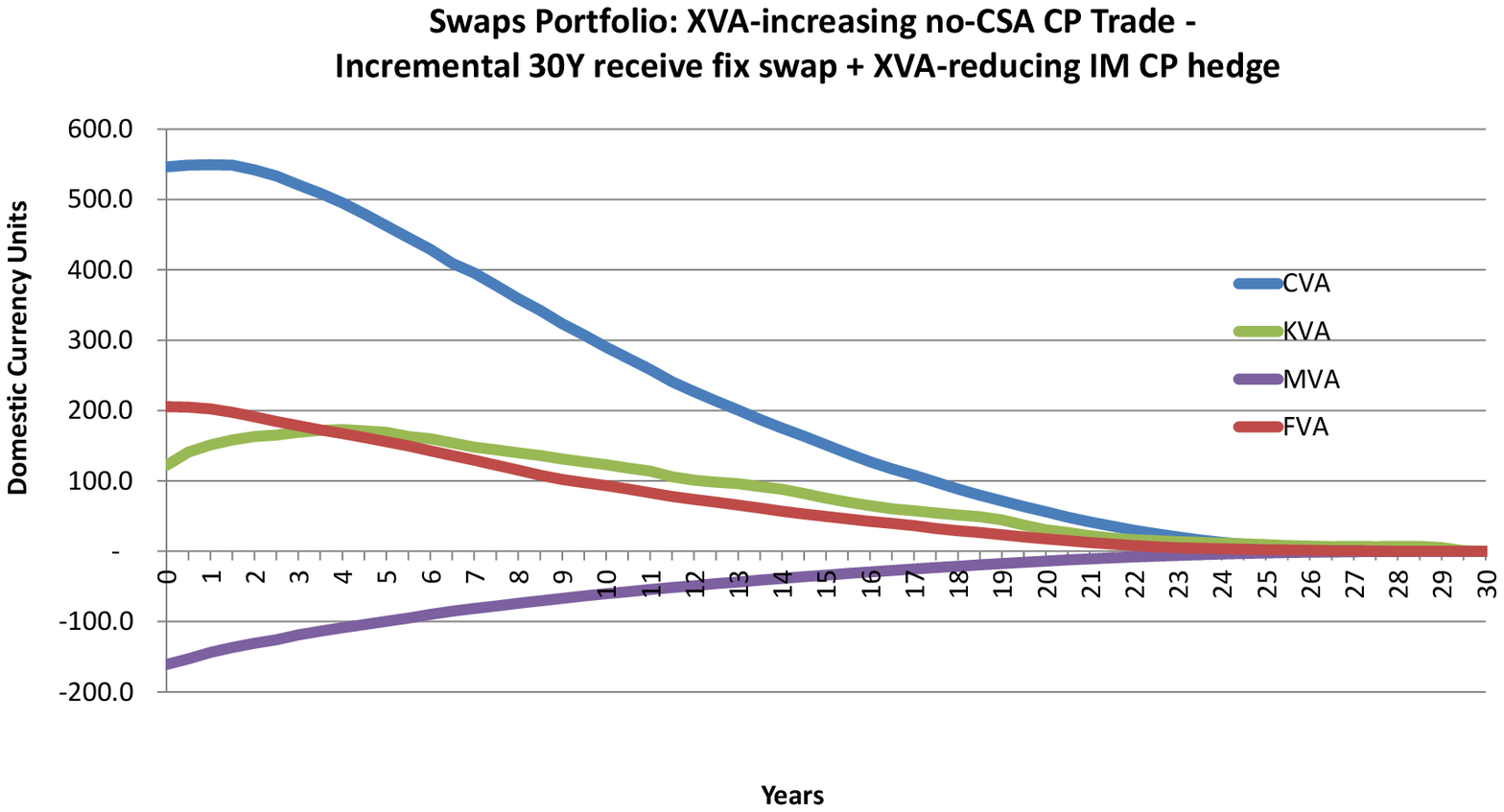}
\caption{\textit{(Top left)} {XVA-reducing trade + XVA-increasing bilateral hedge} \textit{(Top right)} XVA-increasing trade + XVA-increasing bilateral hedge.  \textit{(Bottom left)} {XVA-reducing trade + xva-reducing bilateral hedge} \textit{(Bottom right)} {XVA-increasing trade + XVA-reducing bilateral hedge.}} 
\label{fig:XVAReduceTd}
\end{figure}
  
\brem
In the above, we do not include the XVA costs/benefits of the bilateral hedge counterparty itself.
Given Remark \ref{rem:nodelta}, in different circumstances it may be possible to attribute them to client trades of the original or hedge bank.
Space is lacking for a fuller discussion of economics of XVA trading in different setups.
In particular, many hedge trades now face central instead of bilateral counterparties.
This occurs at additional XVA costs for the client of the initial swap
that can be computed the way explained in \citeN{ArmentiCrepey16}.~\erem

\subsection{Scalability}\label{s:timings}
 
Our deep learning XVA implementation uses CNTK, the Microsoft Cognitive Toolkit.
CNTK is written in core  C++/CUDA (with wrappers for Python, C\#, and Java). 
This is convenient for XVA applications, which are usually developed in C++: CNTK automatic differentiation in C++/CUDA enables C++ in-process training.
This allows embedding the deep learning task within XVA processing. 

Table \ref{tab:compTimes} sets out computation times, including additional results obtained by doubling the numbers of counterparties and risk factors (to 20 counterparties and 80 risk factors). 

\begin{table}[!htbp]
\small
  \centering
\begin{tabular}{|r|r|r|r|r|}
\hline
\multirow{2}{*}{} & \multicolumn{2}{|c|}{10 CP 40 risk factors} & \multicolumn{2}{|c|}{20 CP 80 risk factors}\\
\cline{2-5}
 &No CSA  &IM CSA  &No CSA  &IM CSA \\
    \hline
    Initial risk factor \& trade pricing simulation Cuda &        352 &                352 &        426 &        426 \bigstrut[t]\\
    Counterparty and bank level learning calculations &     4,529 &          13,466 &   19,154 &   59,342 \\
    Total initial batch &     4,881 &          13,818 &   19,580 &   59,768    \\\hline
    Re-simulate 1 counterparty trade pricing Cuda &          40 &                  40 &          51 &          51 \\
    Counterparty and bank level learning calculations &     2,785 &             2,736 &     7,694 &     6,628 \\
    Total incremental trade &     2,825 &             2,776 &     7,745 &     6,679 \bigstrut[b]\\
    \hline
\end{tabular}

  \caption{XVA deep learning computation timings (seconds).}
  \label{tab:compTimes}%
\end{table}%

All these results were based on 50K simulation paths, 32 time steps per year for risk factor simulation, and 16 time steps per year for all XVA calculations and deep learning.
They were computed on a Lenovo P52 laptop with NVidia Quadro P3200 GPU @ 5.5 Teraflops peak FP32 performance, and 14 streaming multiprocessors.

The computations for 20 counterparties took more than twice as long as those for 10 counterparties. However, our deep learning calculations
achieved around 80 to 90\% Cuda occupancy for 10 counterparties and at times fell to half that level for 20 counterparties.
Scaling to realistically high dimensions should be achievable, but acceptable trade incremental pricing performance in production would require server-grade GPU hardware, performance tuning for high GPU utilisation, and, possibly, caching computations.

\appendix 
\section{Continuous-Time XVA Equations}\label{s:eqns}

We recall from \citeN{CrepeyElie16} the continuous-time XVA equations for bilateral trade portfolios when capital at risk is deemed fungible with variation margin, also adding here initial margin and MVA as in the refined static setup of Section \ref{rem:compl}. 
 
We write $\boldsymbol\delta_\eta(dt)=d\indi{\eta\le t}$ for the Dirac measure at a random time $\eta$.

\subsection{Cash Flows}\label{s:cfs} 

\r{We suppose that the derivative portfolio of the bank is partitioned into bilateral netting sets of contracts which are jointly collateralized and liquidated upon bank or counterparties (whether these are clients or market hedge counterparties) 
 default.}
Given a netting set $c$ of the 
\r{bank}
portfolio, we denote by:
\begin{itemize}
\item $\cP^c$ 
and $P^c$, the corresponding  
contractually promised  cash flows and clean value processes;
\item 
 $\tau_c$, $J^c $, and $R_c$, the corresponding default times, survival indicators, and 
 recovery rates, whereas $\tau$, $J$, and $R$
are the analogous data regarding the bank itself, with bank credit spread process $\lambda =(1-R)\gamma $ taken as a proxy of its risky funding spread process\footnote{See
\citeN[Section 5]{ArmentiCrepey16}
 for the discussion of cheaper funding schemes for initial margin.};
\item $\tau^\delta_c= \tau_c+\delta$ and $\tau^\delta= \tau+\delta$, where $\delta$ is a positive margin period of risk, in the sense that
the liquidation of the netting set $c$ happens at time $\tau^\delta_c\wedge \tau^\delta$;
\item $\VM^c,$ the variation margin (re-hypothecable collateral) exchanged between the bank and 
\r{counterparty}
$c$, counted positively when received by the bank;
\item $\PIM^c$ and $\RIM^c$, the related initial margin (segregated collateral) posted and received by the bank;
\item $\RC$ and $\CR$, the reserve capital and capital at risk of the bank.
\end{itemize}
 
The contractually promised  cash flows 
are supposed to be hedged out by the bank but one conservatively assumes no XVA hedge, so that the bank is left with the 
following 
trading cash flows $\cC$ and $\cF$
(cf.~\qr{e:clpinew} and see 
\b{\citeN[Lemmas 5.1 and 5.2]{AlbaneseCaenazzoCrepey16abx}}
for detailed derivations of analogous equations in a slightly simplified setup):
 \begin{itemize}
\item The (counterparty) credit cash flows  
{
\beql{e:cmmtm}&  
d\cC_t=\\&\sum_{ c; \tau_c \le  \tau^\delta }  (1-R_c) 
\Big( (P^c +\cP^c)_{\tau^\delta_c\wedge \tau^\delta}   -(\cP^c+ \VM^c+\RIM^c)_{ (\tau_c\wedge \tau)-} \Big)^+\boldsymbol\delta_{\tau^\delta_c\wedge \tau^\delta}(dt) 
\\&
-
(1-R) 
  \sum_{ c; \tau \le  \tau^\delta_c }  \Big( (P^c +\cP^c)_{\tau^\delta\wedge \tau^\delta_c} -( \cP^c 
 + \VM^c  
-\PIM^c)_{(\tau\wedge \tau_c)-}\Big)^- 
\boldsymbol\delta_{\tau^\delta\wedge \tau^\delta_c}(dt) ;
\eeql }
\item The (risky) funding cash flows
\beql{e:fundi}  
&d\cF_t= J_t\lambda_t\Big(\sum_c J^c  (P^c-\VM^c) -\RC {  -\CR  } \Big)_{t}^{+} dt
 \\& 
- (1-R)\Big(\sum_c J^c  (P^c-\VM^c)  -\RC {  -\CR  } \Big)_{\tau-}^{+} \boldsymbol\delta_{\tau}(dt)\\&
+J_t\tilde{\lambda}_t \sum_c J^c _t \PIM^c_{t}  dt
  - (1-\tilde{R}) \sum_c   J^c _{\tau-}   \PIM^c_{\tau-}  \boldsymbol\delta_{\tau}(dt),
\eeql
where the $\RC$ and $\CR$ terms account for the fungibility of reserve capital and capital at risk with variation margin.
\end{itemize}

 \subsection{Valuation}\label{ss:basics} 

 Here (as in our numerics) we distinguish between a (strict) FVA, 
 in the strict sense of the cost of raising variation margin, and 
an MVA for the cost of raising initial margin
(see Remark \ref{rem:mva}).
The (other than K)VA equations are then
\beql{e:rcfm}\RC=\ca=\cva+\fva+\mva,\eeql 
the so-called ``contra-assets valuation''  sourced  from the clients and deposited in the reserve capital account of the bank, where, for $t<\tau$,
\beql{e:L} 
& {\rm {CVA}}_t = \Ep_t
\sum_{t<\tau^\delta_c}  (1-R_c) 
\Big( (P^c +\cP^c)_{\tau^\delta_c }   -(\cP^c+ \VM^c+\RIM^c)_{ \tau_c-} \Big)^+  
 \\&{\rm {FVA}}_t=\Ep_t\int_t^{\tb} \lambdabar_s\Big(\sum_c J^c  (P^c-\VM^c) -{\rm CA} - {\rm CR}
\Big)_{s}^{+} 
 ds \\&{\rm {MVA}}_t=\Ep_t\int_t^{\tb} \tilde{\lambda}_s\sum_c J^c _s \PIM^c_{s}  ds.
\eeql 
The corresponding
trading loss and profit process $L$ of the bank is such that
\begin{eqnarray}\label{e:cva-fvabisl}\begin{aligned}
&  L_0=0 \mbox{ and, for $t<\tau,$}  \\
&    dL_t =
\sum_{c}  (1-R_c) 
\Big( (P^c +\cP^c)_{\tau^\delta_c }   -(\cP^c+ \VM^c+\RIM^c)_{ \tau_c-} \Big)^+ \boldsymbol\delta_{\tau^\delta_c  } (dt) 
\\ & + 
\lambdabar_t\Big(\sum_c J^c  (P^c-\VM^c) -{\rm CA}- {\rm CR}
 \Big)_{t}^{+} dt\\&
+\tilde{\lambda}_t \sum_c J^c _t \PIM^c_{t}  dt\\&
+   d \ca_t ,
\end{aligned}\end{eqnarray} 
so that $L$ is a $\Qp$ martingale, hence (by Lemma \ref{e:lemQps}) $\LOSS$ is a $\Qs$ martingale.

%
 

%
%
 

By the same rationale as Definitions \ref{d:ccrbis} and \ref{d:ccrter} in the static setup:
\begin{defi} \label{defi:ec}
\em$\EC_{t}$  
is the time-$t$
conditional 97.5\% expected shortfall of ($\LOSS_{t+1}-\LOSS_{t}$) \b{under $\Qp $}.~\finproof\end{defi}


\noindent
Given a positive target hurdle rate $h$:

\bd\label{e:crkva} We set
\beql{e:ecpt}
\CR=\max(\EC,\kva),\eeql
for 
a $\kva$ process  
such that, for $t<\tau$,
\beql{e:kcon}
&\kva_t 
=\Ep_t\Big[\int_t^T h \big( \CR_s -\kva_s\big)
ds
\Big].~\finproof\eeql 
\eds
Hence, for $t<\tau$,
\beql{e:kconsuite}
\kva_t & = \Ep_t\Big[\int_t^T h  e^{-h(s-t)} \CR_s ds\Big]
\\&=  \Ep_t\Big[\int_t^T h  e^{-h(s-t)} \max( \EC_s,\kva_s \big) ds \Big].
\eeql

\noindent
The next-to-last identity is the continuous-time analog of the risk margin formula under the Swiss solvency test cost of capital methodology: 
see {\citeN[Section 6, middle of page 86 and top of page 88]{FOPI06}}.

\subsection{The XVA Equations are Well-Posed} \label{ss:wp}

In view of 
\qr{e:rcfm}, the second line
in
\qr{e:L} 
 is in 
fact an $\fva$ \emph{equation}. Likewise,
the second line in \qr{e:kconsuite} is a $\kva$ equation. Moreover, as capital at risk is fungible with variation margin (cf.~\sr{rem:compl}), i.e.~in consideration of the $\CR$ term
in \qr{e:L}-\qr{e:cva-fvabisl}, where
$\CR=\max(\EC,\kva)$, we actually deal with an $(\fva, \kva)$ \emph{system},
and even, as EC depends on $L$ (cf.~Definition \ref{defi:ec}), with a forward backward system for the forward loss process $L$ and the backward pair $(\fva, \kva)$.

However, as in the refined static setup of Section \ref{rem:compl},
the coupling between $(\fva,\kva)$ and $L$ can be disentangled by the following Picard iteration:

\begin{itemize}
\item Let CVA and MVA be as in \qr{e:L},
$L^{(0)}=\kva^{(0)}=0$, and , for $t<\tau$,
\beql{e:fvainit}{\rm {FVA}}^{(0)}_t=  \Ep_t\int_t^{\tb} \lambdabar_s\Big(\sum_c J^c  (P^c-\VM^c) 
-{\rm CA}^{(0)}
\Big)_{s}^{+} ds , 
\eeql
where
${\rm \TRC}^{(0)} =\cva +\fva^{(0)}+\mva ;$

\item For $k\ge 1,$ writing explicitly $\EC=\EC(L)$ to emphasize the dependence of $\EC$ on $L$, let
$L^{(k)}_0= 0$ and, for $t<\tau$,
\begin{equation} \label{e:prat-csa}
\bal
&d L^{(k)}_t = 
\sum_{c}  (1-R_c) 
\Big( (P^c +\cP^c)_{\tau^\delta_c }   -(\cP^c+ \VM^c+\RIM^c)_{ \tau_c-} \Big)^+ \boldsymbol\delta_{\tau^\delta_c  } (dt) 
\\ & \qqq+ 
\lambdabar_t\Big(\sum_c J^c  (P^c-\VM^c) -{\rm CA}^{(k-1)} {-\max\big(\EC(L^{(k-1)}), {\rm KVA}^{(k-1)}\big)}  \Big)_{t}^{+} dt\\&\qqq
+\tilde{\lambda}_t \sum_c J^c _t \PIM^c_{t}  dt+  d {\ca^{(k-1)}_t },
\\&{\rm {KVA}}^{(k)}_t=  
h \Ep_t \int_t^\tb e^{-h (s-t)}
{\max\big({{\rm \EC}}_s (L^{(k)}), \kva^{(k)}_s\big)} 
ds,\\& {\rm \TRC}^{(k)}_t={\rm CVA}_t+{\rm {FVA}}^{(k)}_t  +{\rm MVA}_t\mbox{  where } {\rm {FVA}}^{(k)}_t= 
\\& \qqq  
\Ep_t\int_t^{\tb} \lambdabar_s\Big(\sum_c J^c  (P^c-\VM^c) -{\rm CA}^{(k)}  -{\max\big( \EC(L^{(k)}), {\rm KVA}^{(k)}\big) } \Big)_{s}^{+} ds 
.  \eal
\end{equation} 
 \end{itemize}

\noindent\textbf{Theorem 4.1 in \citeN{CrepeyElie16}}
{\em 
Assuming square integrable data, the XVA equations are well-posed within square integrable solution (including when one accounts for the fact that capital at risk can be used for funding variation margin). Moreover, the 
above Picard iteration
converges to the unique square integrable solution of the XVA equations.~\finproof\\
}

\subsection{Collateralization Schemes}\label{ss:collat}

We denote by $\Delta_{t}^{c}=\mathcal{P}^c_t - \mathcal{P}^c_{(t-\delta)-}$ 
the cumulative contractual cash flows with the \r{counterparty}
$c$ 
accumulated over a past period of length $\delta$. In our case study,
we consider  both ``no CSA'' netting sets $c$, with
$\VM=\RIM=\PIM=0$,
and  ``(VM/IM) CSA'' netting sets $c$, with
$\VM^c _t=P^c _t$ and,  for $t\leq \tau_c$,
\begin{equation}\label{e:gapri}
\RIM^c _t = \VaR
_t\Big(
(P^c _{t^\delta} +\Delta^c _{t^\delta})- 
P^{c}_{t}\Big)  \sp
\PIM^c _t = \VaR
_t \Big(-
(P^c _{t^\delta} +\Delta^c _{t^\delta})+
P^{c}_{t}\Big) 
,
\end{equation}
for some PIM and RIM quantile levels $\apim$ and $\arim$ (and $t^\delta=t+\delta$).

%

The following result can be derived by similar computations as the ones in \citeN[Section A]{ArmentiCrepey16}.

\bp\label{p:cva} In a common shock default model of the 
\r{counterparties}
\b{and the bank itself} (see the beginning of
\sr{s:num}), with \b{pre-}default intensity processes $\gamma^c$ of the \r{counterparties}
and $\gamma$ of the bank,
then
$\mathrm{CVA} =\mathrm{CVA}^{nocsa}+\mathrm{CVA}^{csa}$, where, for $t<\tau,$
\begin{eqnarray}
&&
 \mathrm{CVA}_{t}^{nocsa}=
 \sum_{c~nocsa} \mathds{1}_{t < \tau_{c}}(1-R_c)  \Ep_{t}
\int_{t}^{\Ts}  
 ( P_{s^\delta }^{c} + \Delta_{s^\delta }^{c}) 
^{+} 
\gamma^{c}_s e^{-\int_t^s \gamma^{c}_u  du} ds \nonumber\\\label{e:instrunocsa}
&&\qqq + 
\sum_{c~nosca}
 \mathds{1}_{\tau_c  < t <\td_{c}}(1-R_c) \Ep_{t}  
  ( P_{\td_{c} }^{c} + \Delta_{\td_{c} }^{c})^{+} ,\\
&&
\mathrm{CVA}_{t}^{csa}=\sum_{c~csa} \mathds{1}_{t < \tau_{c}} (1-R_c) (1-\arim)\times\nonumber\\&&\qqq\qqq \Ep_{t}
\int_{t}^{\Ts} 
  (\ES_{s} -  \VaR_s) \left( 
(P^c _{s^\delta} + \Delta^c _{s^\delta}) 
- 
 P^{c}_{s}\right)
\gamma^{c}_s e^{-\int_t^s \gamma^{c}_u  du} ds \label{e:instrucsa}\\\nonumber
&&\qqq +
\sum_{c~csa} 
\mathds{1}_{\tau_c  < t <\td_{c}} (1-R_c)   \Ep_{t}
 \left( 
( P_{\td_{c} }^{c} + \Delta_{\td_{c} }^{c}) 
-
 ( P_{\tau_{c}}^{c} + {\RIM}_{\tau_{c}}^{c})
\right)^{+},
\end{eqnarray}
where ($\ES_{s} -  \VaR_s$) in \qr{e:instrucsa} is computed at the $\arim$ confidence level.
Assuming its posted initial margin borrowed unsecured by the bank, then
  $\mathrm{MVA} =\mathrm{MVA}^{csa}$, where, for $t<\tau,$
\begin{eqnarray}  \label{e:instrumva}
&&
{\rm {MVA}}^{csa}_t=\sum_{c~csa} J^c_t \Ep_t\int_t^{\tb}
(1-R) \b{\gamma_s }
 \PIM^c_{s}    e^{-\int_t^s \gamma^{c}_u  du} ds.~\finproof
\end{eqnarray}
\ep \noindent

\eee


\begin{thebibliography}{}

\bibitem[\protect\citeauthoryear{Abbas-Turki, Diallo, and Cr\'epey}{Abbas-Turki
  et~al.}{2018}]{AbbasturkiCrepeyDiallo17}
Abbas-Turki, L., B.~Diallo, and S.~Cr\'epey (2018).
\newblock {XVA} principles, nested {Monte Carlo} strategies, and {GPU}
  optimizations.
\newblock {\em International Journal of Theoretical and Applied Finance\/}~{\em
  21}, 1850030.

\bibitem[\protect\citeauthoryear{Albanese, Armenti, and Cr\'epey}{Albanese
  et~al.}{2020}]{ArmentiCrepey16}
Albanese, C., Y.~Armenti, and S.~Cr\'epey (2020).
\newblock {XVA Metrics for CCP optimisation}.
\newblock {\em Statistics \& Risk Modeling\/}~{\em 37\/}(1-2), 25--53.

\bibitem[\protect\citeauthoryear{Albanese, Caenazzo, and Cr\'epey}{Albanese
  et~al.}{2017}]{AlbaneseCaenazzoCrepey17b}
Albanese, C., S.~Caenazzo, and S.~Cr\'epey (2017).
\newblock Credit, funding, margin, and capital valuation adjustments for
  bilateral portfolios.
\newblock {\em Probability, Uncertainty and Quantitative Risk\/}~{\em 2\/}(7),
  26 pages.

\bibitem[\protect\citeauthoryear{Albanese and Cr\'epey}{Albanese and
  Cr\'epey}{2020}]{AlbaneseCaenazzoCrepey16abx}
Albanese, C. and S.~Cr\'epey (2020).
\newblock The cost-of-capital {XVA} approach in continuous time.
\newblock Working paper available on https://math.maths.univ-evry.fr/crepey.

\bibitem[\protect\citeauthoryear{Andersen, Duffie, and Song}{Andersen
  et~al.}{2019}]{AndersenDuffieSong2016}
Andersen, L., D.~Duffie, and Y.~Song (2019).
\newblock Funding value adjustments.
\newblock {\em Journal of Finance\/}~{\em 74\/}(1), 145--192.

\bibitem[\protect\citeauthoryear{Artzner, Eisele, and Schmidt}{Artzner
  et~al.}{2020}]{ArtznerEiseleSchmidt20}
Artzner, P., K.-T. Eisele, and T.~Schmidt (2020).
\newblock No arbitrage in insurance and the {QP}-rule.
\newblock {Working paper available as arXiv:2005.11022}.

\bibitem[\protect\citeauthoryear{Barrera, Cr\'epey, Diallo, Fort, Gobet, and
  Stazhynski}{Barrera et~al.}{2019}]{BarreraCrepeyDialloFortGobetStazhynski17}
Barrera, D., S.~Cr\'epey, B.~Diallo, G.~Fort, E.~Gobet, and U.~Stazhynski
  (2019).
\newblock Stochastic approximation schemes for economic capital and risk margin
  computations.
\newblock {\em ESAIM: Proceedings and Surveys\/}~{\em 65}, 182--218.

\bibitem[\protect\citeauthoryear{Beck, Becker, Cheridito, Jentzen, and
  Neufeld}{Beck et~al.}{2019}]{BeckBeckerCheriditoJentzenNeufeld19}
Beck, C., S.~Becker, P.~Cheridito, A.~Jentzen, and A.~Neufeld (2019).
\newblock {Deep splitting method for parabolic PDEs}.
\newblock arXiv:1907.03452.

\bibitem[\protect\citeauthoryear{Bichuch, Capponi, and Sturm}{Bichuch
  et~al.}{2018}]{BichuchCapponiSturm16}
Bichuch, M., A.~Capponi, and S.~Sturm (2018).
\newblock Arbitrage-free {XVA}.
\newblock {\em Mathematical Finance\/}~{\em 28\/}(2), 582--620.

\bibitem[\protect\citeauthoryear{Bielecki and Rutkowski}{Bielecki and
  Rutkowski}{2002}]{BieleckiRutkowski2002}
Bielecki, T. and M.~Rutkowski (2002).
\newblock {\em Credit {R}isk: {M}odeling, {V}aluation and {H}edging}.
\newblock Springer Finance, Berlin.

\bibitem[\protect\citeauthoryear{Bielecki and Rutkowski}{Bielecki and
  Rutkowski}{2015}]{BieleckiCrepeyRutkowski11}
Bielecki, T.~R. and M.~Rutkowski (2015).
\newblock Valuation and hedging of contracts with funding costs and
  collateralization.
\newblock {\em SIAM Journal on Financial Mathematics\/}~{\em 6}, 594--655.

\bibitem[\protect\citeauthoryear{Brigo and Capponi}{Brigo and
  Capponi}{2010}]{BrigoCapponi2010}
Brigo, D. and A.~Capponi (2010).
\newblock Bilateral counterparty risk with application to {C}{D}{S}s.
\newblock {\em Risk Magazine\/}, March 85--90.
\newblock Preprint version available at https://arxiv.org/abs/0812.3705.

\bibitem[\protect\citeauthoryear{Brigo and Pallavicini}{Brigo and
  Pallavicini}{2014}]{PallaviciniBrigo13bprel}
Brigo, D. and A.~Pallavicini (2014).
\newblock Nonlinear consistent valuation of {CCP} cleared or {CSA} bilateral
  trades with initial margins under credit, funding and wrong-way risks.
\newblock {\em Journal of Financial Engineering\/}~{\em 1}, 1--60.

\bibitem[\protect\citeauthoryear{Burgard and Kjaer}{Burgard and
  Kjaer}{2011}]{BurgardKjaer11}
Burgard, C. and M.~Kjaer (2011).
\newblock {In the balance}.
\newblock {\em Risk Magazine\/}, October 72--75.

\bibitem[\protect\citeauthoryear{Burgard and Kjaer}{Burgard and
  Kjaer}{2013}]{BurgardKjaer14}
Burgard, C. and M.~Kjaer (2013).
\newblock Funding costs, funding strategies.
\newblock {\em Risk Magazine\/}, December 82--87.
\newblock Preprint version available at https://ssrn.com/abstract=2027195.

\bibitem[\protect\citeauthoryear{Burgard and Kjaer}{Burgard and
  Kjaer}{2017}]{BurgardKjaer17}
Burgard, C. and M.~Kjaer (2017).
\newblock Derivatives funding, netting and accounting.
\newblock {\em Risk Magazine\/}, March 100--104.
\newblock Preprint version available at https://ssrn.com/abstract=2534011.

\bibitem[\protect\citeauthoryear{Castagna}{Castagna}{2014}]{Castagna2014}
Castagna, A. (2014).
\newblock Towards a theory of internal valuation and transfer pricing of
  products in a bank: Funding, credit risk and economic capital.
\newblock Available at http://ssrn.com/abstract=2392772.

\bibitem[\protect\citeauthoryear{Collin-Dufresne, Goldstein, and
  Hugonnier}{Collin-Dufresne
  et~al.}{2004}]{CollinDufresneGoldsteinHugonnier2004}
Collin-Dufresne, P., R.~Goldstein, and J.~Hugonnier (2004).
\newblock A general formula for valuing defaultable securities.
\newblock {\em Econometrica\/}~{\em 72\/}(5), 1377--1407.

\bibitem[\protect\citeauthoryear{{Committee~of~European~Insurance~and~Occupational~Pensions~Supervisors}}{{Committee~of~European~Insurance~and~Occupational~Pensions~Supervisors}}{2010}]{CEIOPS10}
{Committee~of~European~Insurance~and~Occupational~Pensions~Supervisors} (2010).
\newblock {QIS5} technical specifications.
\newblock
  https://eiopa.europa.eu/Publications/QIS/QIS5-technical\_specifications\_20100706.pdf.

\bibitem[\protect\citeauthoryear{Cr\'epey}{Cr\'epey}{2015}]{Crepey2012bc}
Cr\'epey, S. (2015).
\newblock {Bilateral counterparty risk under funding constraints. Part I:
  Pricing, followed by Part II: CVA.}
\newblock {\em Mathematical Finance\/}~{\em 25\/}(1), 1--22 and 23--50.
\newblock First published online on 12 December 2012.

\bibitem[\protect\citeauthoryear{Cr\'epey, Bielecki, and Brigo}{Cr\'epey
  et~al.}{2014}]{BieleckiBrigoCrepeyHerbertsson13}
Cr\'epey, S., T.~R. Bielecki, and D.~Brigo (2014).
\newblock {\em Counterparty Risk and Funding: A Tale of Two Puzzles}.
\newblock Chapman \& Hall/CRC Financial Mathematics Series.

\bibitem[\protect\citeauthoryear{Cr\'epey, Sabbagh, and Song}{Cr\'epey
  et~al.}{2020}]{CrepeyElie16}
Cr\'epey, S., W.~Sabbagh, and S.~Song (2020).
\newblock When capital is a funding source: The anticipated backward stochastic
  differential equations of {X}-{V}alue {A}djustments.
\newblock {\em SIAM Journal on Financial Mathematics\/}~{\em 11\/}(1), 99--130.

\bibitem[\protect\citeauthoryear{Cr\'epey and Song}{Cr\'epey and
  Song}{2016}]{CrepeySong15FS}
Cr\'epey, S. and S.~Song (2016).
\newblock Counterparty risk and funding: Immersion and beyond.
\newblock {\em Finance and Stochastics\/}~{\em 20\/}(4), 901--930.

\bibitem[\protect\citeauthoryear{Cr\'epey and Song}{Cr\'epey and
  Song}{2017}]{CrepeySong15c}
Cr\'epey, S. and S.~Song (2017).
\newblock Invariance times.
\newblock {\em The Annals of Probability\/}~{\em 45\/}(6B), 4632--4674.

\bibitem[\protect\citeauthoryear{Dimitriadis and Bayer}{Dimitriadis and
  Bayer}{2019}]{dimitriadis2017}
Dimitriadis, T. and S.~Bayer (2019).
\newblock A joint quantile and expected shortfall regression framework.
\newblock {\em Electronic Journal of Statistics\/}~{\em 13\/}(1), 1823--1871.

\bibitem[\protect\citeauthoryear{Duffie and Huang}{Duffie and
  Huang}{1996}]{DuffieHuang}
Duffie, D. and M.~Huang (1996).
\newblock Swap rates and credit quality.
\newblock {\em Journal of Finance\/}~{\em 51}, 921--950.

\bibitem[\protect\citeauthoryear{Duffie and Sharer}{Duffie and
  Sharer}{1986}]{DuffieSharer86}
Duffie, D. and W.~Sharer (1986).
\newblock Equilibrium and the role of the firm in incomplete market.
\newblock Stanford University, Working Paper No. 915, available at
  https://www.gsb.stanford.edu/faculty-research/working-papers/equilibrium-role-firm-incomplete-markets.

\bibitem[\protect\citeauthoryear{Dybvig}{Dybvig}{1992}]{Dybvig92}
Dybvig, P. (1992).
\newblock Hedging non-traded wealth: when is there separation of hedging and
  investment.
\newblock In S.~Hodges (Ed.), {\em Options: recent advances in theory and
  practice}, Volume~2, pp.\  13--24. Manchester University Press.

\bibitem[\protect\citeauthoryear{Elouerkhaoui}{Elouerkhaoui}{2007}]{Elouerkhaoui07}
Elouerkhaoui, Y. (2007).
\newblock Pricing and hedging in a dynamic credit model.
\newblock {\em International Journal of Theoretical and Applied Finance\/}~{\em
  10\/}(4), 703--731.

\bibitem[\protect\citeauthoryear{Elouerkhaoui}{Elouerkhaoui}{2017}]{Elouerkhaoui17}
Elouerkhaoui, Y. (2017).
\newblock {\em Credit Correlation: Theory and Practice}.
\newblock Palgrave Macmillan.

\bibitem[\protect\citeauthoryear{Fissler and Ziegel}{Fissler and
  Ziegel}{2016}]{fissler2016}
Fissler, T. and J.~Ziegel (2016).
\newblock Higher order elicitability and {O}sband's principle.
\newblock {\em The Annals of Statistics\/}~{\em 44\/}(4), 1680--1707.

\bibitem[\protect\citeauthoryear{Fissler, Ziegel, and Gneiting}{Fissler
  et~al.}{2016}]{fiss:zieg:gnei:15}
Fissler, T., J.~Ziegel, and T.~Gneiting (2016).
\newblock Expected {S}hortfall is jointly elicitable with {V}alue at
  {R}isk---{I}mplications for backtesting.
\newblock {\em Risk Magazine\/}, January.

\bibitem[\protect\citeauthoryear{F\"{o}llmer and Schied}{F\"{o}llmer and
  Schied}{2016}]{foellmerSchied2016}
F\"{o}llmer, H. and A.~Schied (2016).
\newblock {\em Stochastic Finance: An Introduction in Discrete Time\/} (4th
  ed.).
\newblock De Gruyter Graduate.

\bibitem[\protect\citeauthoryear{Goodfellow, Bengio, and Courville}{Goodfellow
  et~al.}{2016}]{GoodfellowBengioCourville2017}
Goodfellow, I., Y.~Bengio, and A.~Courville (2016).
\newblock {\em Deep Learning}.
\newblock MIT Press.

\bibitem[\protect\citeauthoryear{Gottardi}{Gottardi}{1995}]{Gottardi95}
Gottardi, P. (1995).
\newblock {An analysis of the conditions for the validity of Modigliani-Miller
  Theorem with incomplete markets}.
\newblock {\em Economic Theory\/}~{\em 5}, 191--207.

\bibitem[\protect\citeauthoryear{Green, Kenyon, and Dennis}{Green
  et~al.}{2014}]{GreenKenyonDennis14}
Green, A., C.~Kenyon, and C.~Dennis (2014).
\newblock {KVA}: capital valuation adjustment by replication.
\newblock {\em Risk Magazine\/}, December 82--87.
\newblock Preprint version ``{KVA}: capital valuation adjustment'' available at
  ssrn.2400324.

\bibitem[\protect\citeauthoryear{Hoeting, Madigan, Raftery, and
  Volinsky}{Hoeting et~al.}{1999}]{Hoeting99bayesianmodel}
Hoeting, J.~A., D.~Madigan, A.~E. Raftery, and C.~T. Volinsky (1999).
\newblock Bayesian model averaging: A tutorial.
\newblock {\em Statistical Science\/}~{\em 14\/}(4), 382--417.

\bibitem[\protect\citeauthoryear{Hull and White}{Hull and
  White}{2012}]{HullWhite13de}
Hull, J. and A.~White (2012).
\newblock The {FVA} debate, followed by {The} {FVA} debate continued.
\newblock {\em Risk Magazine\/}, July 83--85 and October 52.

\bibitem[\protect\citeauthoryear{Hur\'e, Pham, and Warin}{Hur\'e
  et~al.}{2020}]{HurePhamWarin19}
Hur\'e, C., H.~Pham, and C.~Warin (2020).
\newblock {Some machine learning schemes for high-dimensional nonlinear PDEs}.
\newblock {\em Mathematics of Computation\/}~{\em 89\/}(324), 1547--1580.

\bibitem[\protect\citeauthoryear{{International~Financial~Reporting~Standards}}{{International~Financial~Reporting~Standards}}{2013}]{IFRS4Phase2ED}
{International~Financial~Reporting~Standards} (2013).
\newblock I{F}{R}{S} 4 insurance contracts exposure draft.

\bibitem[\protect\citeauthoryear{Kjaer}{Kjaer}{2019}]{Kjaer2019}
Kjaer, M. (2019).
\newblock In the balance redux.
\newblock {\em Risk Magazine\/}~(November).

\bibitem[\protect\citeauthoryear{Longstaff and Schwartz}{Longstaff and
  Schwartz}{2001}]{long}
Longstaff, F.~A. and E.~S. Schwartz (2001).
\newblock {Valuing {American} options by simulation: A simple least-squares
  approach.}
\newblock {\em The Review of Financial Studies\/}~{\em 14\/}(1), 113--147.

\bibitem[\protect\citeauthoryear{Merton}{Merton}{1974}]{Merton1974}
Merton, R. (1974).
\newblock On the pricing of corporate debt: the risk structure of interest
  rates.
\newblock {\em The Journal of Finance\/}~{\em 29}, 449--470.

\bibitem[\protect\citeauthoryear{Modigliani and Miller}{Modigliani and
  Miller}{1958}]{MM1958}
Modigliani, F. and M.~Miller (1958).
\newblock The cost of capital, corporation finance and the theory of
  investment.
\newblock {\em Economic Review\/}~{\em 48}, 261--297.

\bibitem[\protect\citeauthoryear{Myers}{Myers}{1977}]{Myers77}
Myers, S. (1977).
\newblock Determinants of corporate borrowing.
\newblock {\em Journal of Financial Economics\/}~{\em 5}, 147--175.

\bibitem[\protect\citeauthoryear{Piterbarg}{Piterbarg}{2010}]{Piterbarg10}
Piterbarg, V. (2010).
\newblock {Funding beyond discounting: collateral agreements and derivatives
  pricing}.
\newblock {\em Risk Magazine\/}, August 57--63.

\bibitem[\protect\citeauthoryear{Sch\"{o}nbucher}{Sch\"{o}nbucher}{2004}]{Schoenbucher04}
Sch\"{o}nbucher, P. (2004).
\newblock A measure of survival.
\newblock {\em Risk Magazine\/}~{\em 17\/}(8), 79--85.

\bibitem[\protect\citeauthoryear{{Swiss~Federal~Office~of~Private~Insurance}}{{Swiss~Federal~Office~of~Private~Insurance}}{2006}]{FOPI06}
{Swiss~Federal~Office~of~Private~Insurance} (2006).
\newblock Technical document on the {S}wiss solvency test.
\newblock
  https://www.finma.ch/FinmaArchiv/bpv/download/e/\\SST\_techDok\_061002\_E\_wo\_Li\_20070118.pdf.

\end{thebibliography}
